\documentclass[journal]{IEEEtran}
\usepackage{amsmath}
\usepackage{cite}
\usepackage{graphicx}
\usepackage{subfigure}
\usepackage{float}
\usepackage{stfloats}
\usepackage{subfloat}
\usepackage{multirow}
\usepackage{array}
\usepackage{lineno}
\usepackage{color}  %
\usepackage{amsfonts}
\begin{document}

\title{Image Separation with Side Information: A Connected Auto-Encoders Based Approach}

\author{Wei Pu,
Barak Sober,
Nathan Daly,
Zahra Sabetsarvestani, 
Catherine Higgitt,
Ingrid Daubechies, and
Miguel R.D. Rodrigues.

\thanks{
W. Pu, Z. Sabetsarvestani and M. Rodrigues are with the Department of Electronic and Electrical Engineering, University College London, UK.
Z. Sabetsarvestani is also with the American International Group, UK.
B. Sober is with the Department of Mathematics and Rhodes Information Initiative, Duke University, US.
N. Daly and C. Higgitt are with the Scientific Department, National Gallery, London, UK.
I. Daubechies is with the Department of Electrical and Computer Engineering, Department of Mathematics, and Rhodes Information Initiative, Duke University, US.
}}

\maketitle
\begin{abstract}
X-radiography (X-ray imaging) is a widely used imaging technique in art investigation. It can provide information about the condition of a painting as well as insights into an artist's techniques and working methods, often revealing hidden information invisible to the naked eye. In this paper, we deal with the problem of separating mixed X-ray images originating from the radiography of double-sided paintings. Using the visible color images (RGB images) from each side of the painting, we propose a new Neural Network architecture, based upon ’connected’ auto-encoders, designed to separate the mixed X-ray image into two simulated X-ray images corresponding to each side. In this proposed architecture, the convolutional auto encoders extract features from the RGB images. These features are then used to  (1) reproduce both of the original RGB images, (2) reconstruct the hypothetical separated X-ray images, and (3) regenerate the mixed X-ray image. The algorithm operates in a totally self-supervised fashion without requiring a sample set that contains  both the mixed X-ray images and the separated ones. The methodology was tested on images from the double-sided wing panels of the \textsl{Ghent Altarpiece}, painted in 1432 by the brothers Hubert and Jan van Eyck. These tests show that the proposed approach outperforms other state-of-the-art X-ray image separation methods for art investigation applications.
\end{abstract}

\begin{IEEEkeywords}
Image separation, image unmixing, deep neural networks, convolutional neural networks, auto-encoders, side information
\end{IEEEkeywords}

\section{Introduction}
\label{sec:intro}

\IEEEPARstart{O}{ld} Master paintings – precious objects illuminating Europe’s rich cultural heritage and history – are often the subject of detailed technical examination, whether to investigate an artist’s materials and technique or in support of conservation or restoration treatments in order to preserve them for future generations. These processes have traditionally relied on X-ray radiography (or X-ray imaging)\cite{A}, infrared reflectography\cite{B} or micro-sample analysis\cite{C} – an invasive and destructive process – in order to understand the materials present within specific features of a painting\cite{D, E}.

More recently, to complement these traditional  approaches to the technical study of works of art, the cultural heritage sector has been gradually witnessing the increased use of non-invasive and non-destructive, cutting-edge analytical and imaging techniques and generating large and typically multi-dimensional datasets associated with artwork\cite{F, G, H}. Such techniques include macro X-ray fluorescence (MA-XRF) scanning\cite{I, J, K} and hyperspectral imaging\cite{L, M, N, O}. Sophisticated multimodal image and data processing tools have been developed to exploit these new datasets and the increasingly high-resolution digital images now available using more traditional imaging techniques (e.g. X-ray imaging, infrared reflectography and various forms of multispectral imaging\cite{M}) to tackle various challenging tasks arising in the field of art investigation\cite{P, Q}, such as crack detection\cite{R, S}, material identification\cite{T, U, V, W}, brush stroke style analysis\cite{X, Y, Z, AA}, canvas pattern or stretcher bar removal\cite{BB, CC, DD}, automated canvas weave analysis\cite{EE, FF}, and improved visualization of concealed features or under-drawing\cite{L, GG, HH, II}.

Due to the ability of X-rays to penetrate deep into a painting’s stratigraphy, X-radiographs (X-ray images’) are especially important during the examination and restoration of paintings\cite{A, JJ, KK}. They can help to establish the condition of a painting (e.g., losses and damages not be apparent at the surface) and the status of different paint passages (e.g., to identify retouchings, fills or other conservation interventions). X-ray images can also provide insights into an artist’s technique and working methods, for example revealing the painting’s stratigraphy (the buildup of the different paint layers which may include concealed earlier designs or pentimenti), and information about the painting support (e.g., type of canvas or the construction of a canvas or panel) or even some indication of the pigments used.  However, the X-ray image of a painting — particularly those with design changes, areas of damage, hidden paintings, or paintings on both the front and reverse sides of the support — will inevitably contain a mix or blend of these various features, making it difficult for experts to interpret. Features within the structure of the painting may also appear in the image (e.g., nails and battens, wood grain, stretcher bars etc). Therefore, it is relevant to devise approaches that can separate a mixed X-ray image into its (hypothetical) constituent images, corresponding to the various individual features.

The task of separating mixed signals has been studied extensively in the blind source separation (BSS) and the informed source separation (ISS) literature. Among the approaches designed to tackle this challenge, we can mention independent component analysis (ICA)\cite{ICA1,ICA2}, robust principal component analysis (PCA)\cite{RPCA1,RPCA2,RPCA3} and morphological component analysis (MCA)\cite{MCA1,MCA2,MCA3}. These methods often rely on some strong prior assumptions including independence, sparsity, low-rankness and so on. However, the implementation of such techniques in art investigation applications – including the separation of mixed X-ray images from double-sided paintings into separate X-ray images of corresponding (hypothetical) single-sided paintings – is generally problematic because such typical prior assumptions adopted by other methods do not always hold.

Recently, deep learning architectures have also been successfully applied to various signal and image separation challenges\cite{SeparateSupervised1,SeparateSupervised2,SeparateSemiSupervised,SeparateUnSupervised}. Such approaches typically fall into three different categories: unsupervised, semi-supervised and supervised approaches. The supervised case typically assumes one has access to a training dataset containing a number of examples of mixed and associated component signals that can be used to train a deep neural network carrying out the separation task\cite{SeparateSupervised1,SeparateSupervised2}. In contrast, in the unsupervised case one does not have access to such a training dataset; instead, the sources are typically separated by minimizing joint adverse and remix losses as in \cite{SeparateUnSupervised}. Finally, in the semi-supervised case one may have access to samples of one individual source but not other sources; a neural egg separation (NES) method \cite{SeparateSemiSupervised} integrated with generative adversarial networks (GANs)\cite{GAN} has been recently proposed to tackle the semi-supervised source separation challenge. However, again, the application of these approaches to challenges associated with the unmixing of an X-ray image into its constituents can also be problematic since  the data one typically has access to prevents the use of supervised or semi-supervised approaches.

\begin{figure}[h]
\centering
\includegraphics[width=0.48\textwidth]{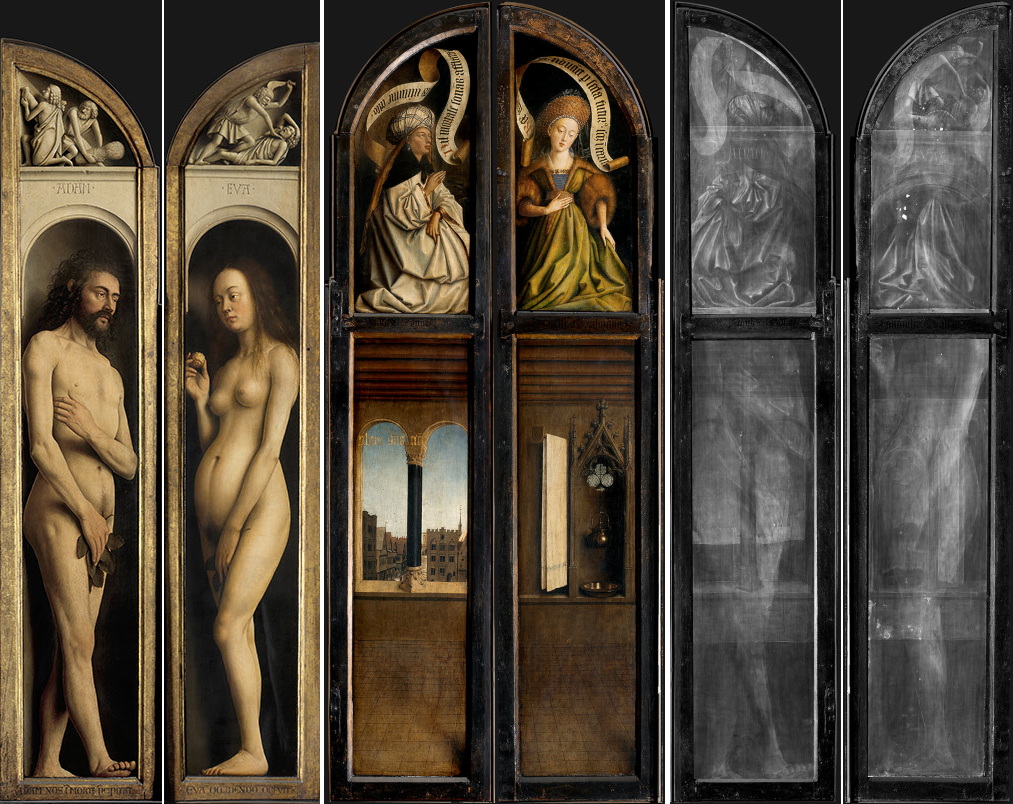}
\caption{Two double-sided panels from the \textsl{Ghent Altarpiece}\cite{Ghent1}: (left) visible RGB image of the front side, (centre) visible RGB image of the back side, (right) mixed X-ray image.}\label{Fig1}
\end{figure}

There are, however, various cases where one has access to both mixed X-ray images along with additional images that can potentially be exploited to aid in the image separation task. For example, in the separation of mixed X-ray images associated with double-sided paintings – such as the outer wing panels of the large polyptych \textsl{The Adoration of theMystic Lamb}, painted by Hubert and Jan Van Eyck and more commonly known as the \textsl{Ghent Altarpiece}\cite{Ghent1,Ghent2,Ghent3}, shown in Fig. \ref{Fig1} – one can also potentially use the RGB (or visible) images associated with both sides of the outer panels, in order to understand traits such as general contours, paint passages and losses to improve the separation.

A number of image processing approaches based on sparsity\cite{IS1}, Gaussian mixture models \cite{IS3} or deep learning \cite{IS2} have been proposed to address such a challenge in the context of double-sided paintings. The approaches proposed \cite{IS1,IS3} have been partially successful, whereas the self-supervised approach in \cite{IS2} has led to significantly better results.

In this paper, we propose another self-supervised learning  approach to perform the separation of mixed X-ray images originating from double-sided paintings given the visible images associated with each side of the painting. We show that our approach outperforms the state-of-the-art approaches designed to tackle this specific problem \cite{IS2}. 

Our main contributions can be summarized as follows:
\begin{itemize}

\item
First, we propose an entirely new self-supervised learning approach based on the use of “connected” auto-encoders that extract features from the RGB images in order to (1) reproduce both of the original RGB images, (2) reconstruct the associated separated X-ray images, and (3) regenerate the mixed X-ray image. This approach – akin to \cite{IS2} – allows us to carry out the image separation task without the need for labelled data.
\item
Second, we propose methods to tune these auto-encoders based on the use of a composite loss function involving reconstruction losses, energy losses and dis-correlation losses. This composite loss function allows us to improve further on the image separation task.
\item
Third, we also offer a detailed analysis of the effect of various hyper-parameters associated with our separation method on performance.
\item
Finally, we apply our proposed approach to a real dataset, showcasing state-of-the-art results over competing methods. The dataset relates to images taken from the double- sided wing panels of the \textsl{Ghent Altarpiece}, shown in Fig. \ref{Fig1}. 

\end{itemize}

The remainder of the paper is organized as follows:  section II formulates the image separation problem with side information; section III describes our proposed connected auto- encoder model to carry out image separation tasks with side information; section IV presents the selection of hyper-parameters along with an evaluation of the proposed algorithm;  section V draws the conclusions. 

\section{Problem formulation}
\label{sec:format}

The focus of this paper is the separation of mixed X-ray images, arising from X-radiography of double-sided paintings. Thus, the input data available is the mixed X-ray image accompanied by the visible images of each of the two sides of the painting. Our goal is to separate the mixed X-ray image into two components – where one component would contain features associated with the image on the front panel and the other component would contain features associated with the image on the rear panel – by leveraging the availability of the RGB images pertaining to the front and reverse of the painting.

We carry out this task by dividing these images into several smaller patches that overlap with respect to the vertical and horizontal dimensions of the image. In particular, suppose x denotes a mixed X-ray image patch and let x1 and x2 denote the hypothetical, separated X-ray image patches corresponding to the front and rear sides of the painting respectively. We then assume that the mixed X-ray patch x can be expressed in terms of the individual X-ray patches x1 and x2 as follows:
\begin{align}
\label{e1}
x \approx x_1 + x_2 .
\end{align}
This linear mixing assumption is motivated by the fact that paintings (and panels) can be quite thin so that higher-order attenuation effects can be neglected.

We further assume that there is a mapping $\cal F$ that is approximately able to convert an image patch in the RGB domain into an image patch in the X-ray domain so that
\begin{align}
\label{e2}
x \approx {\cal F}(r_1) + {\cal F}(r_2).
\end{align}
We can then cast the X-ray image separation problem with side information as the task of learning the mapping function $\cal F$.
In fact, we note that this idea has also been explored in \cite{IS2} where the mapping function $\cal F$ has been modelled via a 7-layer convolutional neural network (CNN). This mapping function was then learnt by minimizing the error between the sum of the two separated X-ray image patches and the original mixed X-ray image patch as follows:

\begin{align}
\label{e3}
{\mathop {\min }\limits_{{\cal F}}} \left \| x -{\cal F}(r_1) -{\cal F}(r_2) \right \|_F .
\end{align}
where $\| \cdot \|_F$ denotes the Frobenius norm of the argument.
However, due to a lack of constraints on the structure of $x_1$ and $x_2$, the individual X-ray images obtained using the algorithm in \cite{IS2} can be highly related to the corresponding RGB images. We therefore propose a different approach to learn such a mapping function as described below.

\section{Proposed Approach} 
\subsection{Connected auto-encoder structure} 
\begin{figure}[h]
\centering
\includegraphics[width=0.48\textwidth]{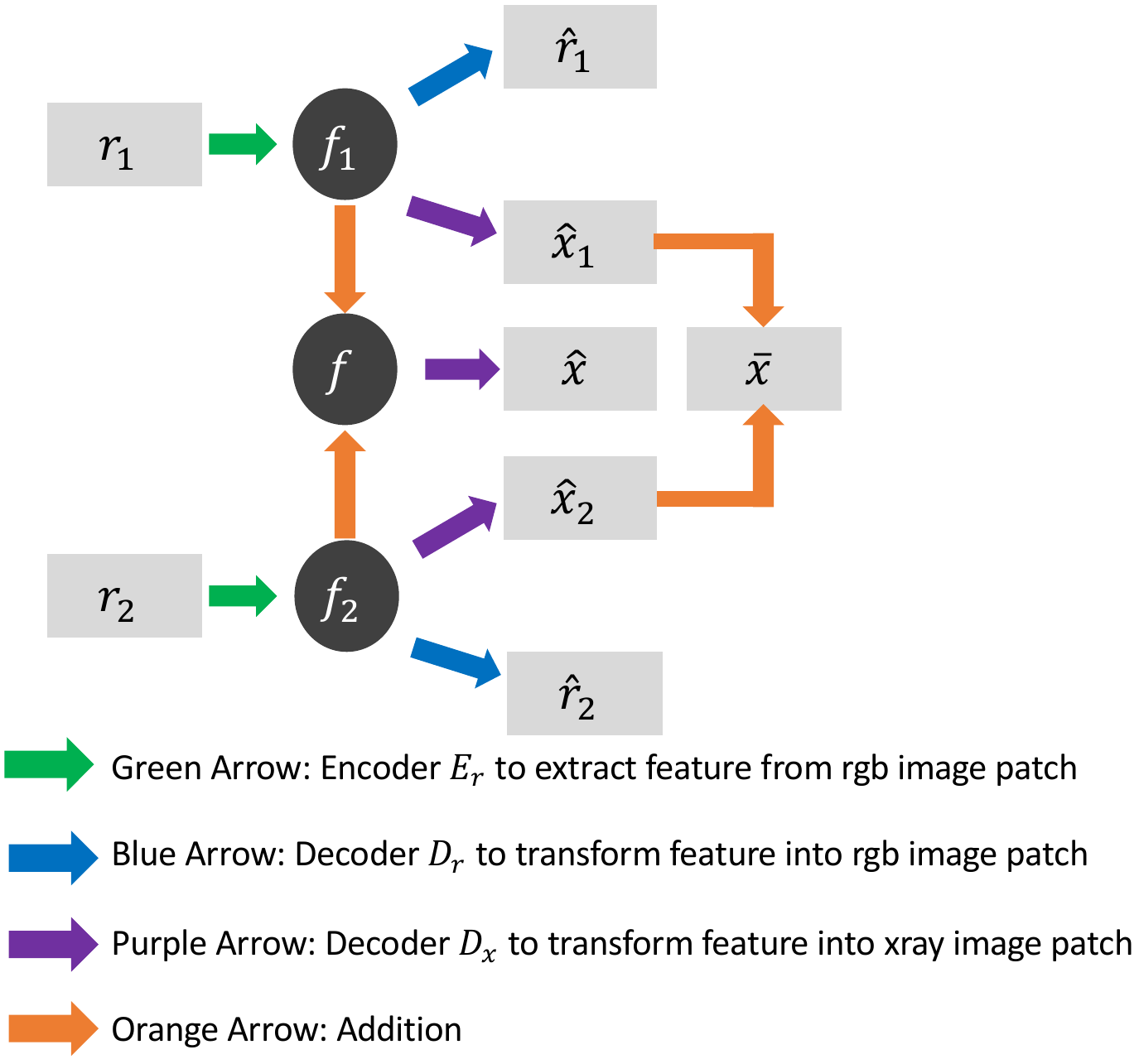}
\caption{Block diagram of the proposed method. 
}\label{BlockDiagram}
\end{figure}

Our mixed X-ray separation approach is based on the use of auto-encoders. Fig. \ref{BlockDiagram} depicts the main constituent blocks of our proposed approach where 
\begin{itemize}
    
    \item Encoder $E_r$ (represented by the green arrows) is used to extract features $f_1$ and $f_2$ from the RGB image patches $r_1$ and $r_2$, respectively;
    
    \item Decoder $D_r$ (represented by the blue arrows) is used to convert the features $f_1$ and $f_2$ onto an estimate of the RGB image patches $\hat{r}_1$ and $\hat{r}_2$, respectively;
    
    \item Decoder $D_x$ (represented by the purple arrows) is also used to convert the features $f_1$, $f_2$ and $f$ onto an estimate of the X-ray image patches $\hat{x}_1$, $\hat{x}_2$, and $\hat{x}$ respectively, where $f$ denotes a feature vector associated with the mixed X-ray image patch $x$;
    
    \item One 'addition' process (represented by orange arrows) is used in the feature and X-ray domain to get another version of the mixed X-ray $\bar{x} = \hat{x}_1 + \hat{x}_2$ and the corresponding feature map $f = f_1 + f_2$.
    
\end{itemize}

Our proposed approach is therefore based on some implicit assumptions:

\begin{itemize}
    
    \item \textit{Shared-Features Assumption\cite{Translation}}: There is a set of features underlying both an RGB image patch and the associated X-ray image that  make the mutual conversion between RGB and X-ray images possible. In particular, for a certain pair of X-ray image patch $x_1$ and its corresponding RGB image patch $r_1$, one postulates there is some latent feature vector $f_1$ such that $x_1 = D_x(f_1) $ and $r_1 = E_r(f_1)$ (and likewise for rear panel patches).

    \item \textit{Linear-Feature Assumption}: The features underlying a mixed X-ray image correspond to the sum of the individual features underlying the individual X-ray components. That is, the feature map $f$ associated with a mixed X-ray patch $x$ corresponds to the sum of the feature map $f_1$ associated with X-ray image of patch $x_1$ and feature map $f_2$ associated with X-ray image of patch $x_2$. Furthermore, the feature map $f$ underlying the mixed X-ray can be used to reconstruct $\hat x$, an estimate of the mixed X-ray image patch $x$ using decoder $D_x$; i.e., $\hat{x}=D_x(f)$.

\end{itemize}

One can of course argue that such assumptions may not always hold, but as will be shown below this proposed method leads to state-of-the-art image separation results. The details of our proposed architecture are described further – including the encoder and decoder models and the learning algorithms – in the following section.

\subsection{Encoder and Decoder Models}

Our encoder $E_r$ and decoder $D_r$ and $D_x$ models are based on 3-layer CNN networks. This choice is due to the fact that CNNs are very suited to image processing tasks.

In addition, for $E_r$, each CNN layer is followed by batch normalization (BN), ReLU activation as well as average pooling (AP) layers, while for $D_r$ and $D_x$, each CNN layer is followed by batch normalization, ReLU activation and up sampling (US) layers.

\begin{figure*}
\centering
\includegraphics[width=0.73\textwidth]{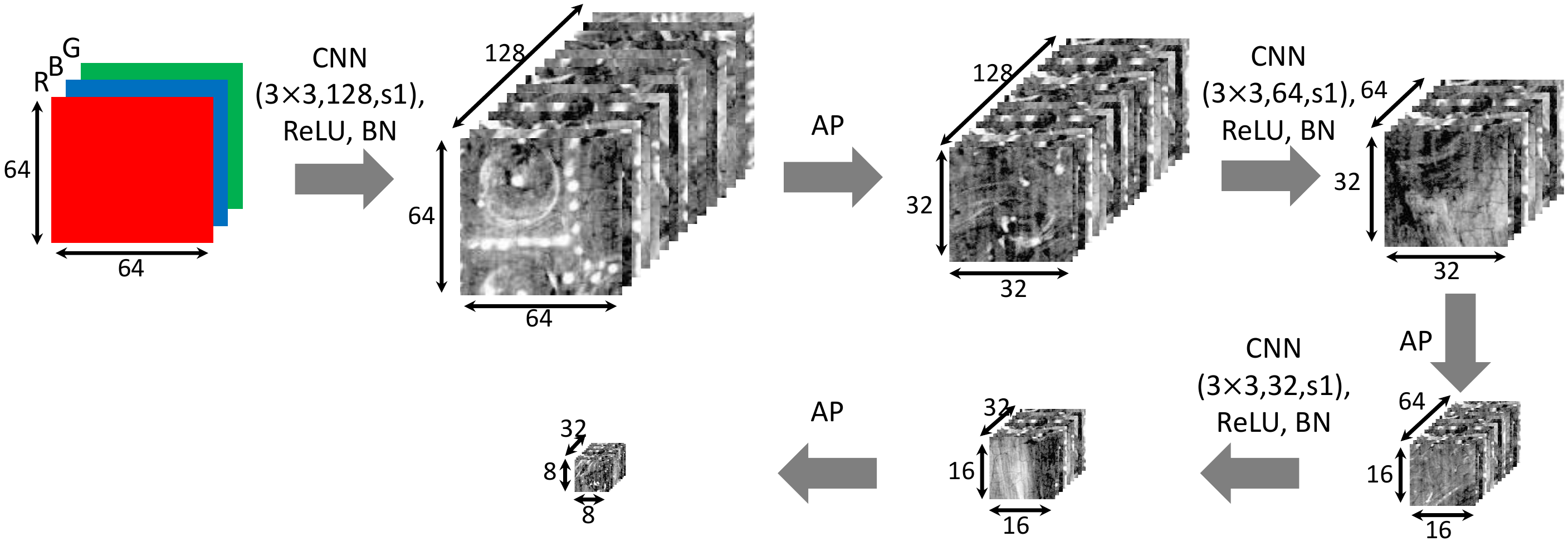}
\caption{Encoder $E_r$ is modelled as 3-layer 2-dimensional CNNs , wherein each CNN layer is followed by batch normalization (BN), ReLU activation as well as average pooling (AP) layers.}\label{Encoder}
\end{figure*}

\begin{figure*}
\centering
\includegraphics[width=0.7\textwidth]{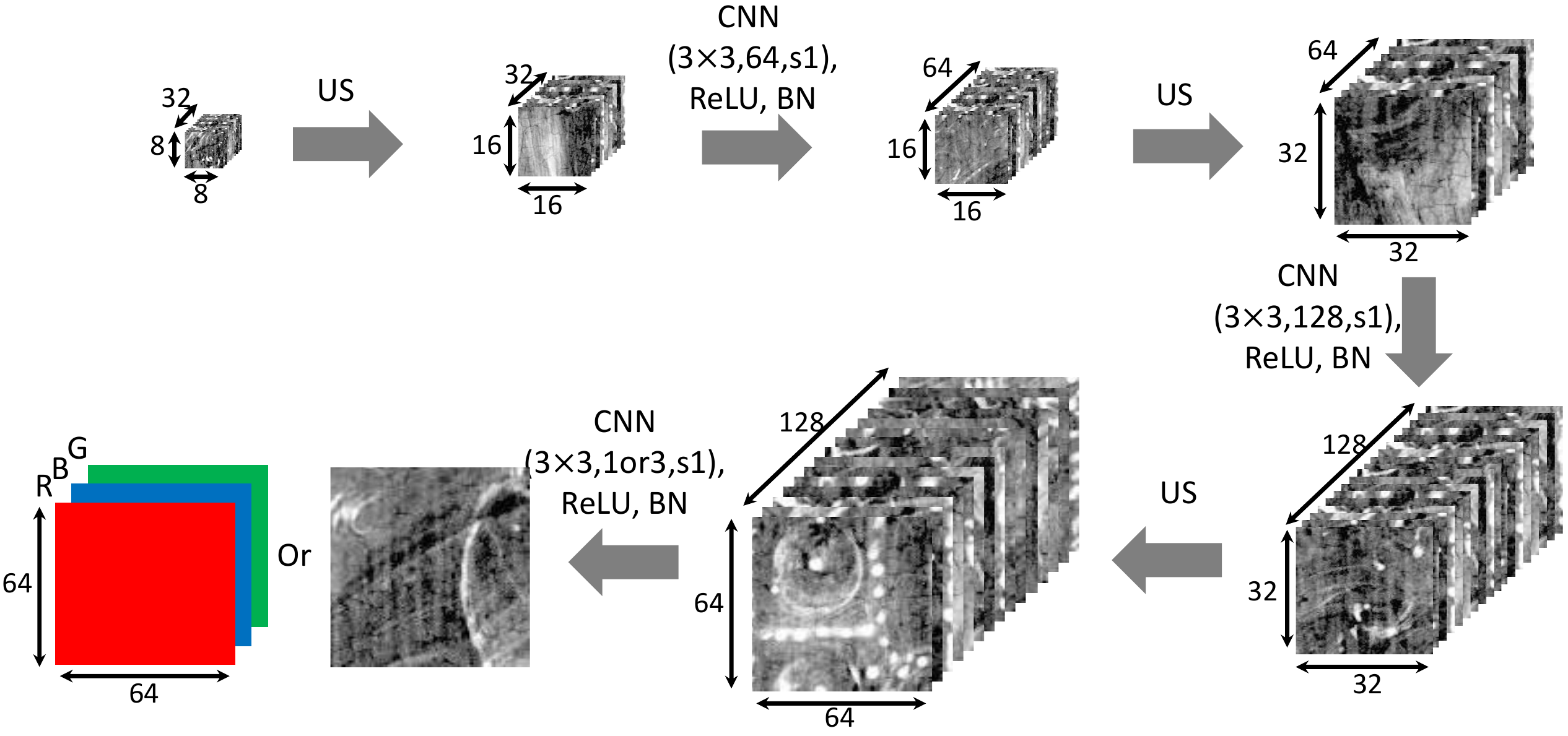}
\caption{Decoders $D_x$ and $D_r$ are modelled as 3-layer CNNs, wherein each CNN layer is followed by batch normalization (BN), ReLU activation as well as upsampling (US) layers.}\label{Decoder}
\end{figure*}

The architectures of the various encoder and decoders appear in Fig. \ref{Encoder} and \ref{Decoder}, respectively. In Fig. \ref{Encoder} and \ref{Decoder}, “3$\times$3, 128, s1” denotes 3$\times$3 filters, 128 feature maps and stride 1 for this convolutional layer.

We have adopted these architectures of the encoder and decoders after experimenting with various structures and observing that they serve the purpose of X-ray reconstruction with high performance.

Finally, we note that the encoder $E_r$ used to generate the features $f_1$ from $r_1$ is exactly the same as the encoder $E_r$ used to generate features $f_2$ from $r_2$ (both have same architecture, weights and biases). The same applies to the decoders $D_r$ and $D_x$. This represents another set of constraints that enforces the learning process to lock onto general features rather than get into an overfit.

\subsection{Learning Algorithm}

Our strategy for training the various encoders and decoders is based on various considerations:
    \begin{itemize}
        
        \item First, we want to minimize the difference  between the reconstructed and original RGB images given by
        \begin{align}
        \label{e4}
        L_1 =\left \| r_1 - \hat r_1 \right \|_F + \left \| r_2 - \hat r_2 \right \|_F .
        \end{align}
        
        \item Second, we also want to minimize the difference between the reconstructed mixed X-ray image patch $\hat x$ and the original mixed X-ray image patch $x$ given by
        \begin{align}
        \label{e5}
        L_2 = \left \| x - \hat x \right \|_F ,
        \end{align}
        where $\hat{x} = D_x (E_r(r_1) + E_r(r_2))$.
        
        \item Third, we also desire that the difference between the sum of two separated X-ray image patch $\bar{x}$ and original mixed X-ray $x$ should be minimized as
        \begin{align}
        \label{e6}
        L_3 = \left \| x - \bar x \right \|_F ,
        \end{align}
        where $\bar{x} = \hat x_1 +\hat x_2 = D_x (E_r(r_1)) + D_x(E_r(r_2))$.
    \end{itemize}

We have also noted that these individual losses by themselves do not entirely promote reasonable results in view of the fact that:
    \begin{enumerate}
        \item It is possible to obtain degenerate results such as  $\hat x_1 \approx x$ and $\hat x_2 \approx 0$ or $\hat x_1 \approx 0$ and $\hat x_2 \approx x$ by using these loss functions alone.   
        \item It is also possible to obtain results where a portion of the content of the X-ray image from one side appears in the X-ray image of the other side (and vice versa).
    \end{enumerate}

In particular, using a loss function $L = L_1 + \lambda_1 L_2 + \lambda_2 L_3$, can often lead to such degenerate results depending on the exact random initializations of the encoder and decoder model parameters.
Therefore, we also introduce two additional losses:
\begin{itemize}
    \item To address the first issue, we introduce an energy penalty function given by
    \begin{align}
    \label{e7}
    L_4 =  \left\| \hat {x}_1\right\|^2 _{ F }  + \left\| \hat {x} _2 \right\| ^2_{ F } .
    \end{align}
    This loss $L_4$ promotes non-zero separated X-ray image patches $\hat x_1$ and $\hat x_2$. In particular, under the constraint in $L_3$, loss $L_4$ lead to an outcome where the energy of x is approximately evenly divided into $\hat x_1$ and $\hat x_2$. 

    \item To address the second issue, we also introduce another loss function capturing the correlation between the feature maps given by
    \begin{align}
    \label{e8}
    L_5 =  C^2(f_1,f_2) ,
    \end{align}
    where $C(f_1,f_2)$ denotes the Pearson correlation coefficient between $f_1$ and $f_2$ given by
	\begin{equation}
	\label{e9}
	C(f_1,f_2)=\frac {  \sum (f_{v1}-\mu_1)(f_{v2}-\mu_2)}{\sqrt{\sum (f_{v1}-\mu_1)^2\sum (f_{v2}-\mu_2)^2}} .
	\end{equation}
    In (\ref{e9}), $\mu_1$ and $\mu_1$ denote the mean of $f_1$ and $f_2$, respectively, $f_{v1}={\rm vec} (f_1 )$ and $f_{v2}={\rm vec} (f_2)$, where $\rm vec$ refers to vectorization operation.
     This loss $L_5$ attempts to make sure the features associated with the separated X-ray image patches $\hat x_1$ and $\hat x_2$ are as different as possible. 
     We apply this loss exclusively on the feature domain because we prefer a dis-correlation between individual separated X-ray images in terms of overall shape, border and content rather than detail.

\end{itemize}

Therefore, we eventually learn the various encoders and decoders -- including the decoder delivering an estimate of the unmixed X-ray signals -- by using the composite loss function
\begin{align}
\label{e10}
 L_{ total }=L_{ 1 }+\lambda_1 \cdot L_{ 2 }+\lambda_2 \cdot L_{ 3 }+\lambda_3 \cdot L_{ 4 } + \lambda_4 \cdot L_{ 5 },
\end{align}
where $\lambda_1$, $\lambda_2$, $\lambda_3$ and $\lambda_4$ are the hyper-parameters corresponding to the losses $L_2$, $L_3$, $L_4$ and $L_5$, respectively.

Finally, the learning process of the various encoders and decoders is done by using a stochastic gradient descent (SGD) algorithm with the ADAM optimization strategy with learning rate 0.0001. In particular, the encoders and decoders $E_r$, $D_r$ and $D_x$ are learned simultaneously during the process, trying to minimize the total loss $L_{total}$.

We note that our unmixing approach shown in Fig.  \ref{BlockDiagram}  is such that – via our assumptions – the different auto-encoders are effectively connected, enabling us to learn an unmixing  strategy in a totally unsupervised (or, more appropriately, self-supervised) fashion.

\section{Experimental Results}

We conducted a number of experiments to assess the effectiveness of our proposed X-ray separation approach. These involved:
\begin{itemize}

\item an analysis of the effect of the various hyper-parameters associated with our approach on X-ray separation performance;

\item an analysis of the effectiveness of our approach in relation to the state-of-the-art, both on synthetically mixed X-ray images and real mixed X-ray images.

\end{itemize}

\subsection{Datasets}

\begin{figure}[h]
    \centering
    \subfigure[]{\includegraphics[width=0.23\textwidth]{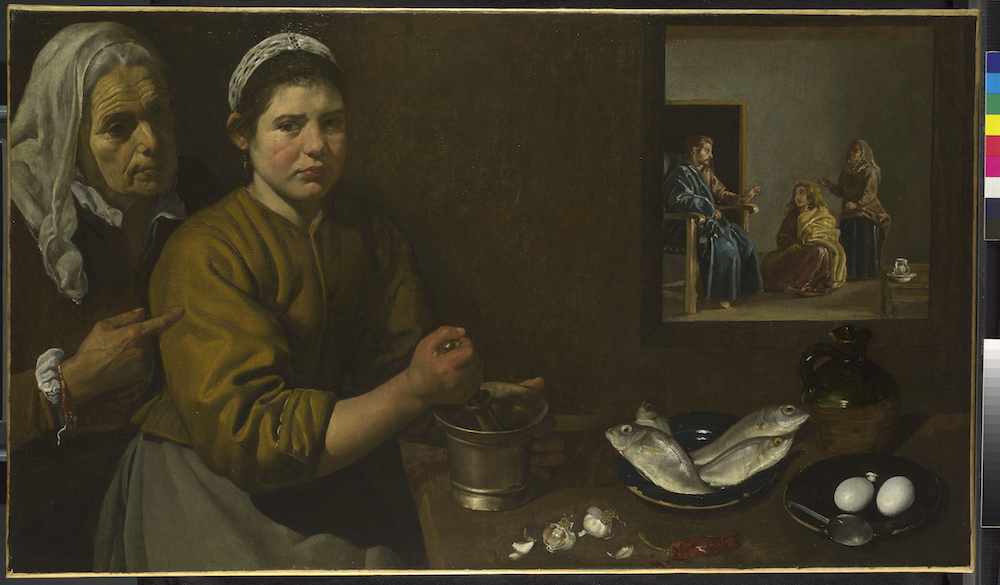}}
    \hfil
    \subfigure[]{\includegraphics[width=0.23\textwidth]{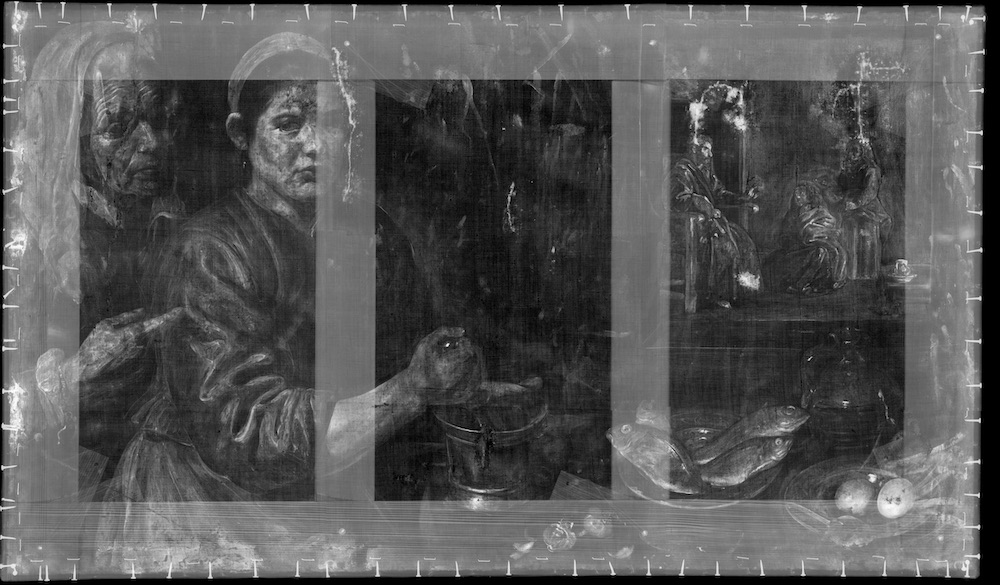}}
    \caption{ Diego Velázquez, Kitchen Scene with Christ in the House of Martha and Mary (NG1375), probably 1618. Oil on canvas © The National Gallery, London. (a). RGB image. (b). X-ray image.}\label{figs1}
\end{figure}

\begin{figure}[h]
\centering
  \subfigure[]{\includegraphics[width=0.23\textwidth]{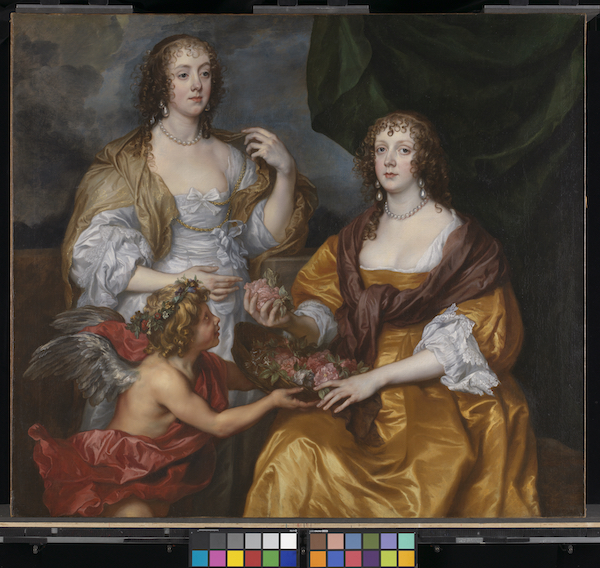}}
  \hfil
  \subfigure[]{\includegraphics[width=0.23\textwidth]{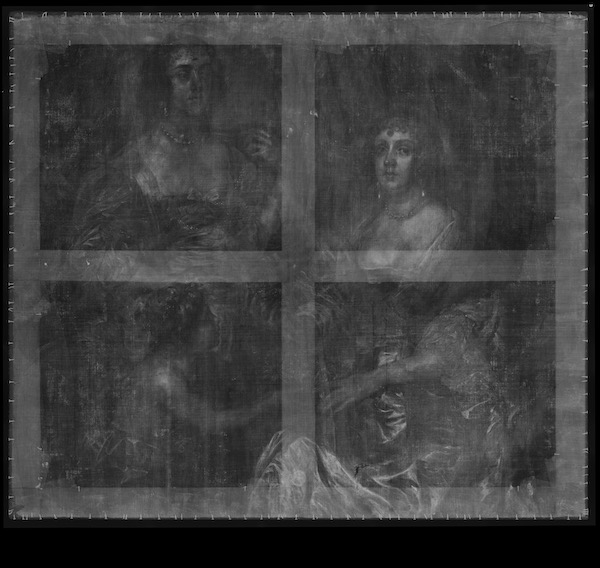}}
    \caption{ Anthony van Dyck, \textsl{Lady Elizabeth Thimbelby and Dorothy}, Viscountess Andover (NG6437), about 1635. Oil on canvas © The National Gallery, Londo. (a). RGB image. (b). X-ray image.}\label{figp1}
\end{figure}

Our experiments rely on a number of datasets associated with real paintings, including:

\begin{itemize}

\item \textsl{Ghent Altarpiece} by Hubert and Jan van Eyck. This large, complex 15th-century polyptych altarpiece comprises a series of panels – including panels with a
composition on both sides (see Fig. \ref{Fig1}) – that we use to showcase the
performance of our algorithm on real mixed X-ray data.

\item \textsl{Kitchen Scene with Christ in the House of Martha and Mary} by Diego Velázquez (Fig. \ref{figs1}). This one-sided canvas painting was used to showcase the performance of our algorithm on synthetically mixed X-ray data.

\item \textsl{Lady Elizabeth Thimbelby and Dorothy, Viscountess Andover} by Anthony Van Dyck (Fig. \ref{figp1}). This canvas painting, also one-sided, was used to design a number of experiments allowing us to understand the impact of the various hyper-parameters associated with our separation approach.

\end{itemize}

We next describe in detail our various experiments, starting with our hyper-parameter selection protocol.

\subsection{Hyper-parameter Selection Protocol}

We start be evaluating the effect of the hyper-parameters $\lambda_1$-- $\lambda_4$ appearing in (\ref{e10}) on the X-ray separation performance. In view of the fact that it is not practical to visualize the optimal combination of these parameters, our experiments involve two phases:

\begin{itemize}

\item First, we report results for the optimal values for the hyper-parameters $\lambda_1$ and $\lambda_2$ with the hyper-parameters $\lambda_3$ and $\lambda_4$ set to be equal to zero.

\item Second, we report the results for the optimal values for the hyper-parameters $\lambda_3$ and $\lambda_4$ with the hyper-parameters $\lambda_1$ and $\lambda_2$ set to be equal to their optimal values from the first optimization step.

\end{itemize}

We also compared this selection approach to an approach involving the selection of the optimal tuple $\lambda_1$ -- $\lambda_4$ simultaneously on various datasets. However, both approaches result in similar separation performance.

\subsubsection{Experiment set-up}

\begin{figure*}[h]
    \centering
    \subfigure[]{\includegraphics[width=0.18\textwidth]{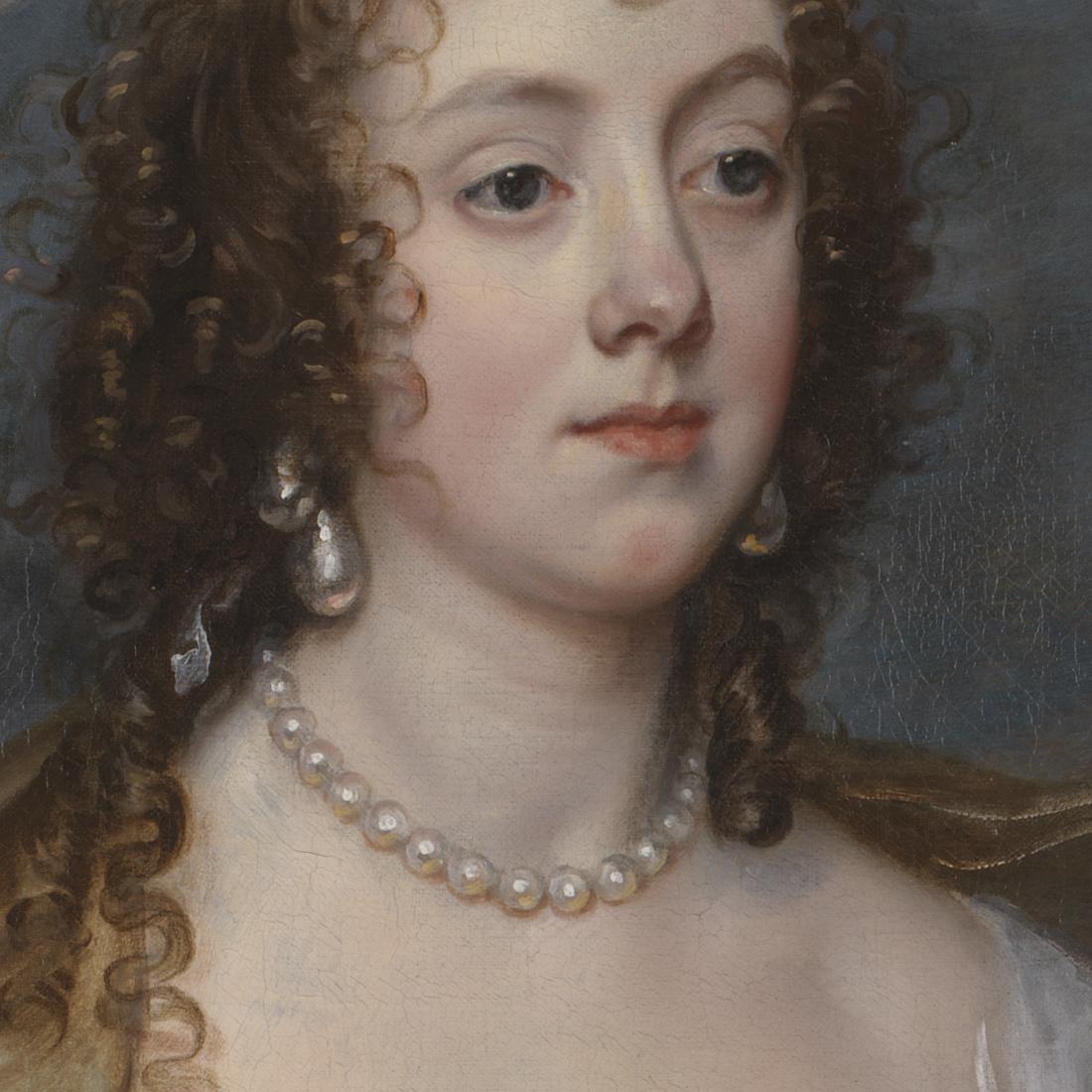}}
    \hfil
    \subfigure[]{\includegraphics[width=0.18\textwidth]{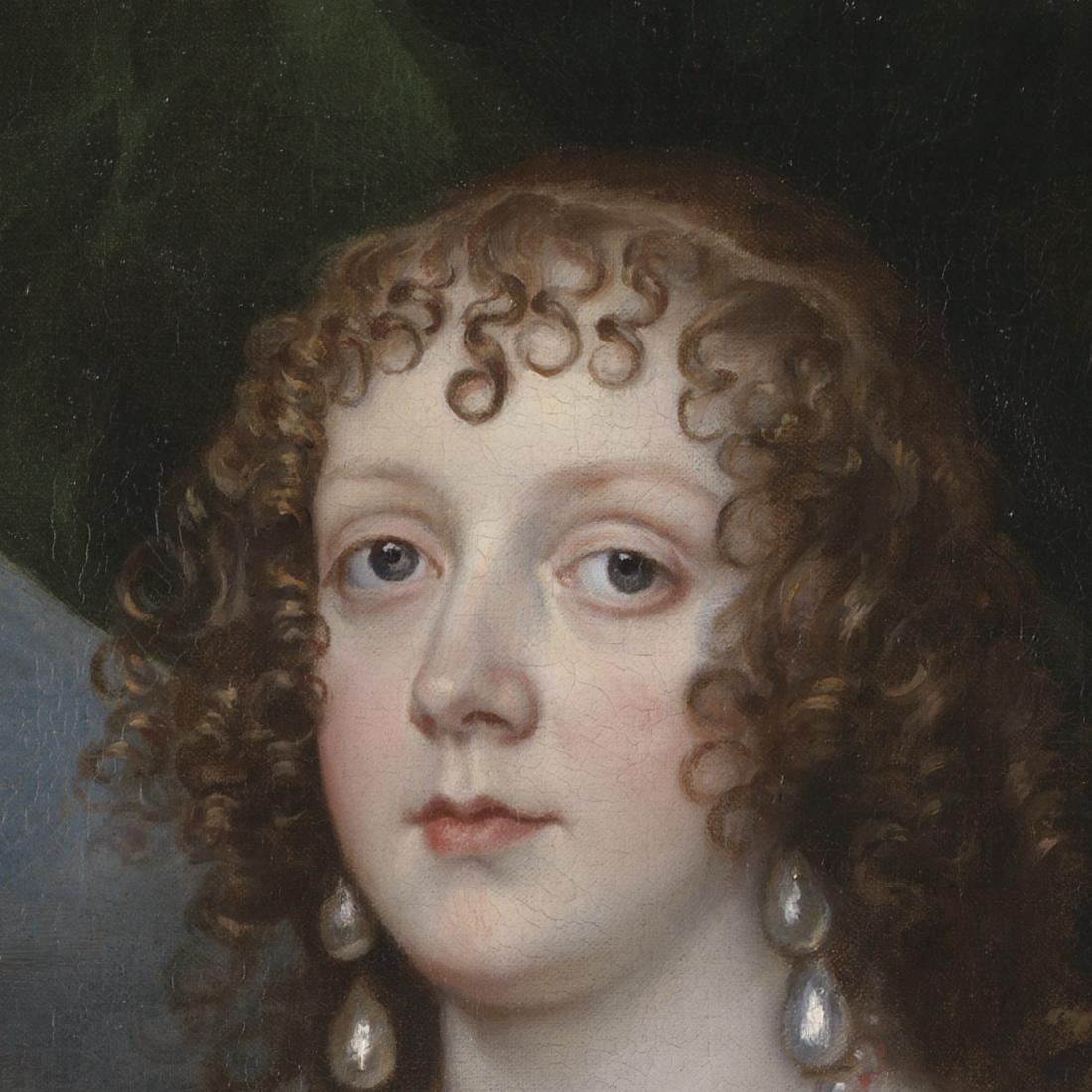}}
    \hfil
    \subfigure[]{\includegraphics[width=0.18\textwidth]{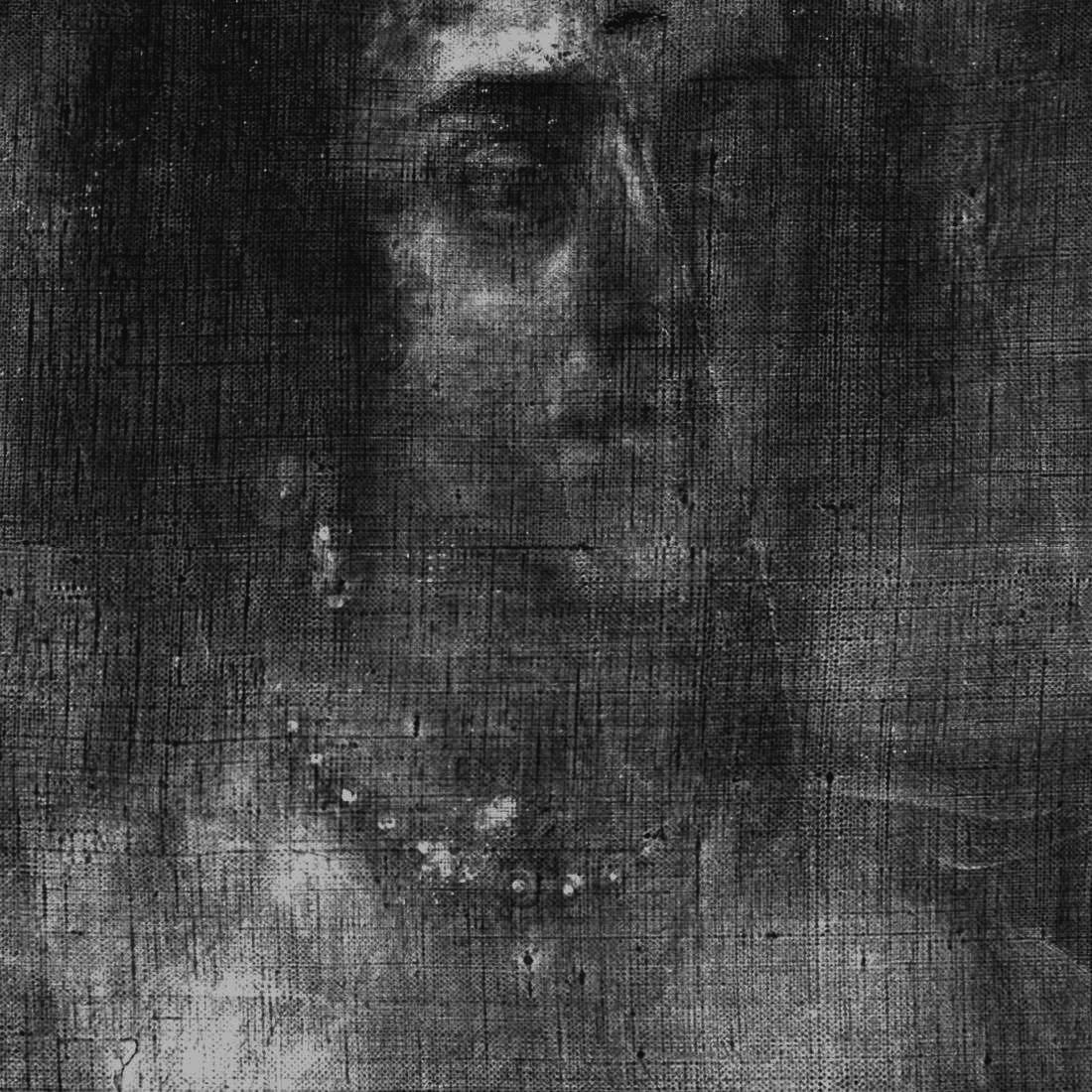}}
    \hfil
    \subfigure[]{\includegraphics[width=0.18\textwidth]{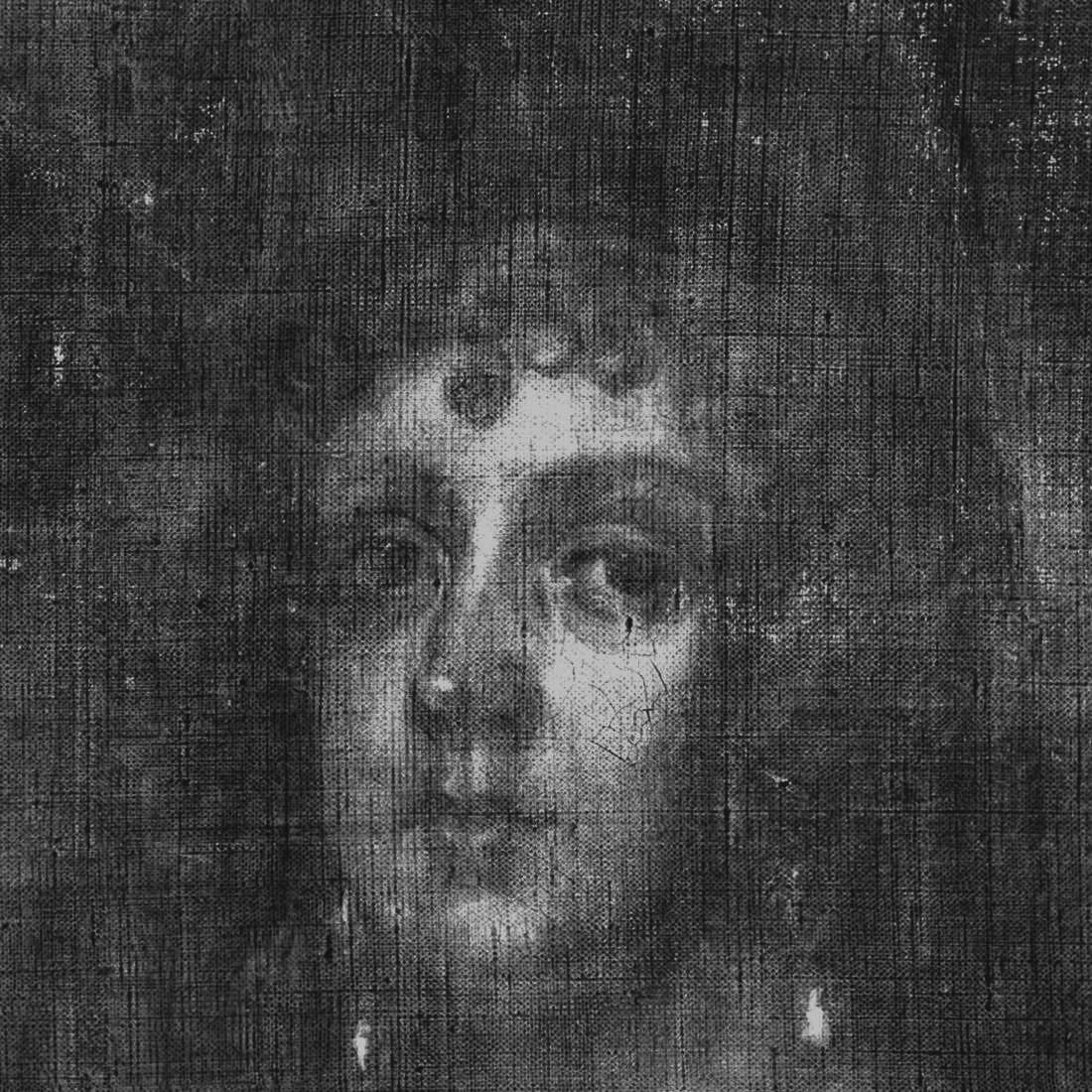}}
    \hfil
    \subfigure[]{\includegraphics[width=0.18\textwidth]{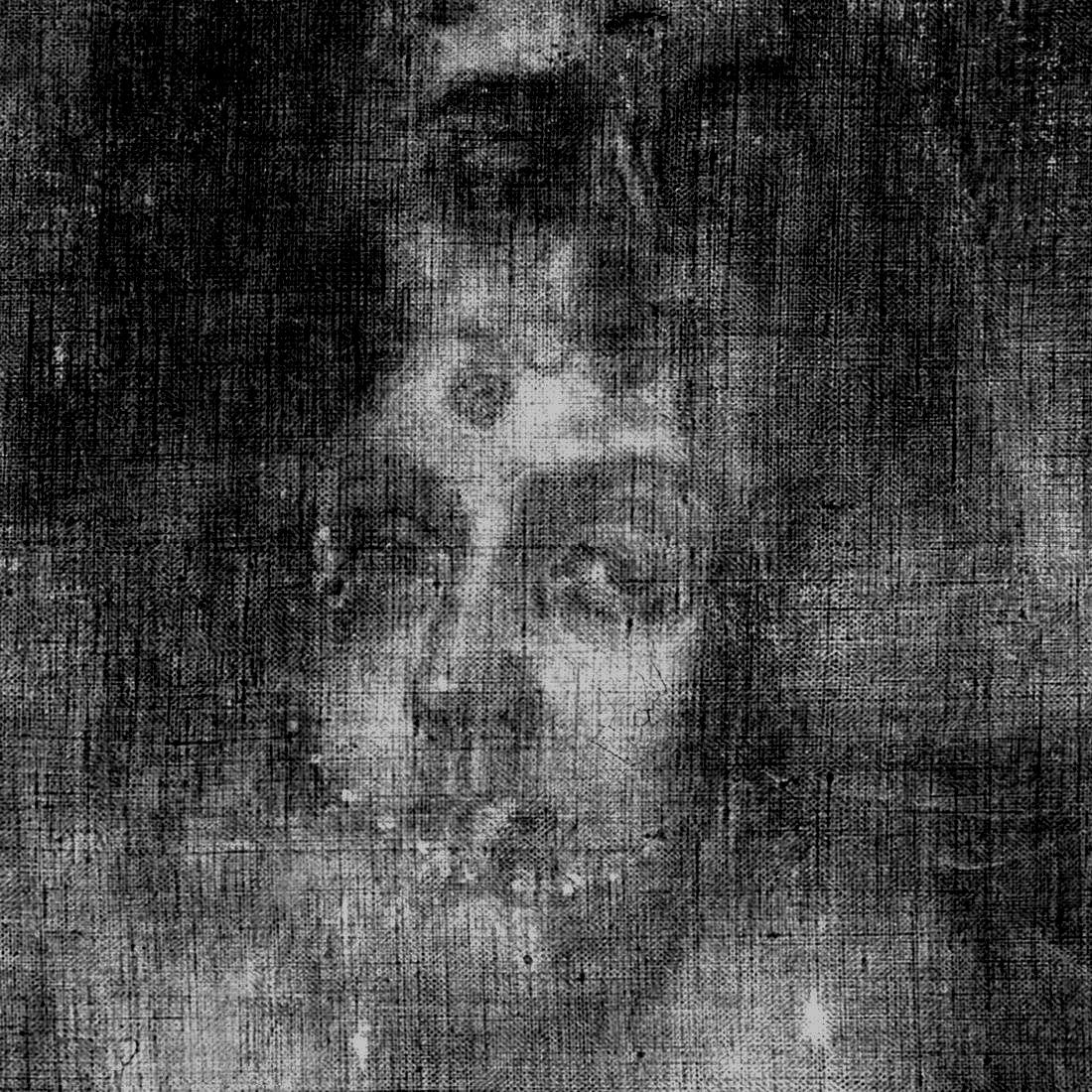}}
    \caption{Images used for hyper-parameter selection. (a). First RGB image. (b). Second RGB image. (c). X-ray image corresponding to first RGB image. (d). X-ray image corresponding to second RGB image. (e). Synthetically mixed X-ray image.}\label{figp2}
\end{figure*}

We use two small areas with the same size from the oil painting \textsl{Lady Elizabeth Thimbelby and Dorothy, Viscountess Andover} by Anthony Van Dyck, one associated with the face of lady Elizabeth Thimbelby and the other with the face of her sister in the portrait, in order to create a synthetically mixed X-ray image. The corresponding RGB images, X-ray images, and synthetically mixed X-ray image are shown in Fig. \ref{figp2}.
The synthetically mixed X-ray image in Fig. \ref{figp2} (e) is obtained by adding the X-ray images shown as Fig \ref{figp2} (c) and Fig \ref{figp2} (d).

Each such image is of size $1100 \times 1100$ pixels. These images were then further divided into patches of size 64$\times$64 pixels  with 56 pixels overlap (both in the horizontal and vertical direction), resulting in 11,236 patches. Each patch associated with the synthetically mixed X-ray was separated independently; the various patches associated with the individual separated X-rays are then put together by placing various patches in the original order and averaging the overlap portions.

We carried out the separation experiments over a number of trials associated with different random initializations of the auto-encoders in our method. We then assessed the separation performance by reporting on the average Mean Squared Error (MSE) given by:
\begin{align}
\label{e17}
MSE = \frac{1}{2R} \sum _{r=1}^{R}\left ( || X_1 -\hat X^r_1||_F + || X_2 -\hat X^r_1||_F\right ),
\end{align}
where $R$ corresponds to the number of trials, $\hat{X}_1^r$ corresponds to one of the separated X-ray images on trial $r$, $\hat{X}_2^r$ corresponds to the other separated X-ray image on trial $r$ and $X_1$ and $X_2$ correspond to the ground truth of the individual X-ray images. We set the number of trials $R$ to be equal to 50 in our experiments.

This experimental procedure was carried out for different combinations of $\lambda_1$ and $\lambda_2$ along with different combinations of $\lambda_3$ and $\lambda_4$. We restricted these hyper-parameters to lie in the interval $\lambda_1 \in [0,10], \lambda_2 \in [0,10], \lambda_3 \in [1,10]$ and $\lambda_4 \in [0.1,0.5]$, we also selected instances of the hyper-parameters from this interval in steps of $0.2, 0.2, 0.2$ and $0.02$, respectively.

\subsubsection{Effect of hyper-parameters $\lambda_1$ and $\lambda_2$}

\begin{figure}[h]
\centering
\includegraphics[width=0.35\textwidth]{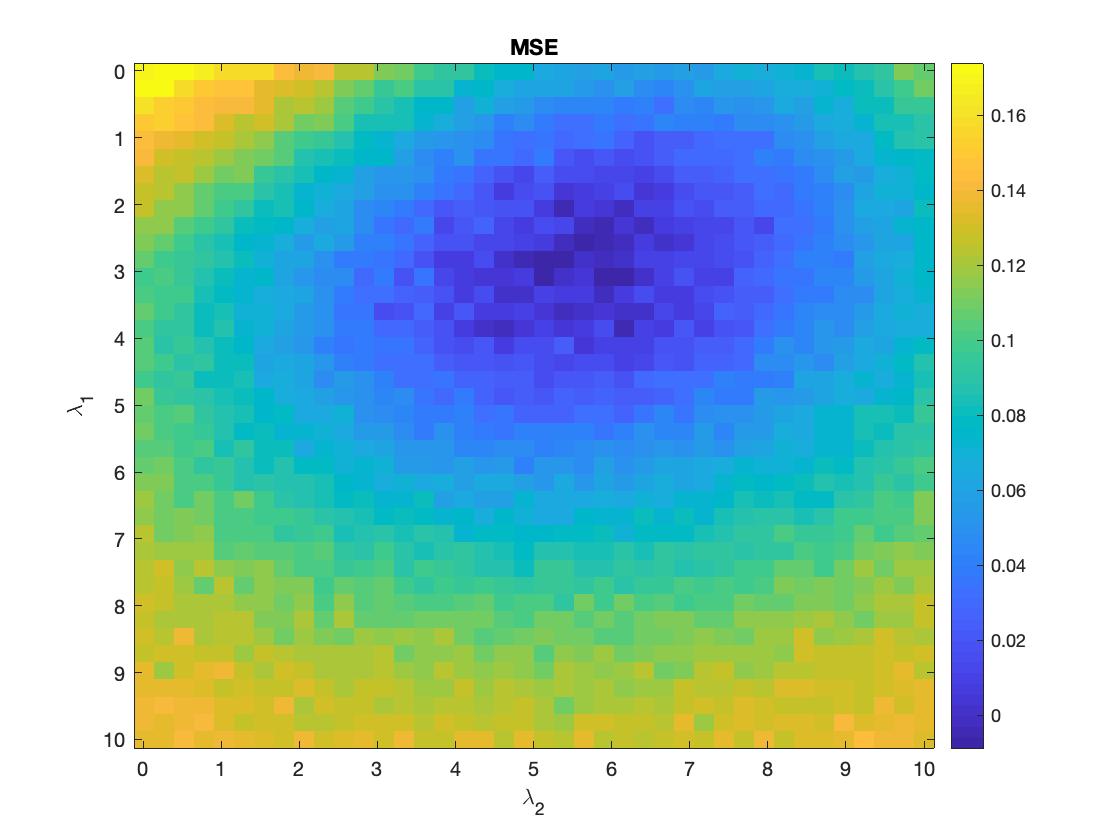}
\caption{Average MSE in (\ref{e10}) as a function of $\lambda_1$ and $\lambda_2$ with $\lambda_3$ and $\lambda_4$ set to be equal to zero.}\label{figp3}
\end{figure}

\begin{figure}[h]
\centering
  \subfigure[]{\includegraphics[width=0.2\textwidth]{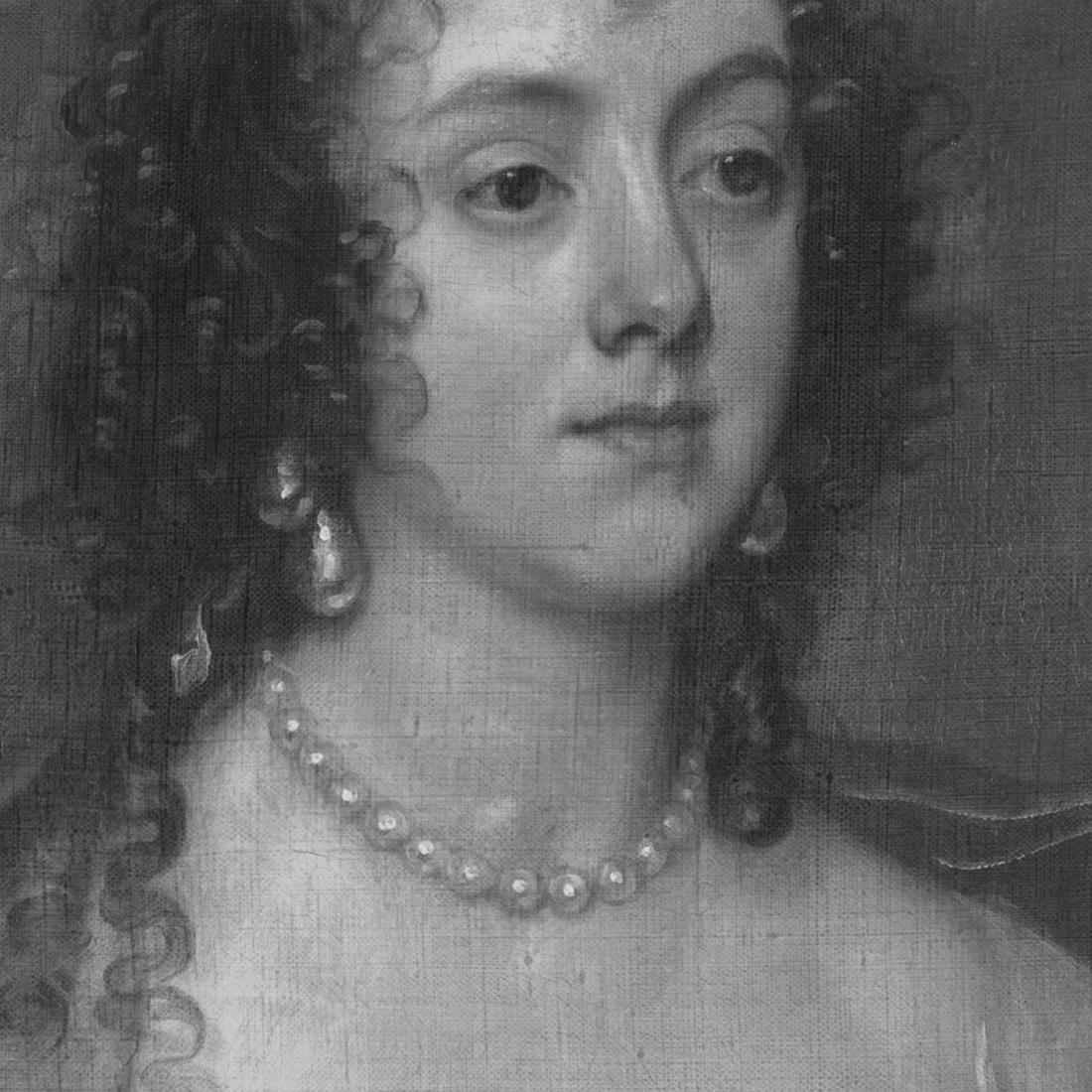}}
  \hfil
  \subfigure[]{\includegraphics[width=0.2\textwidth]{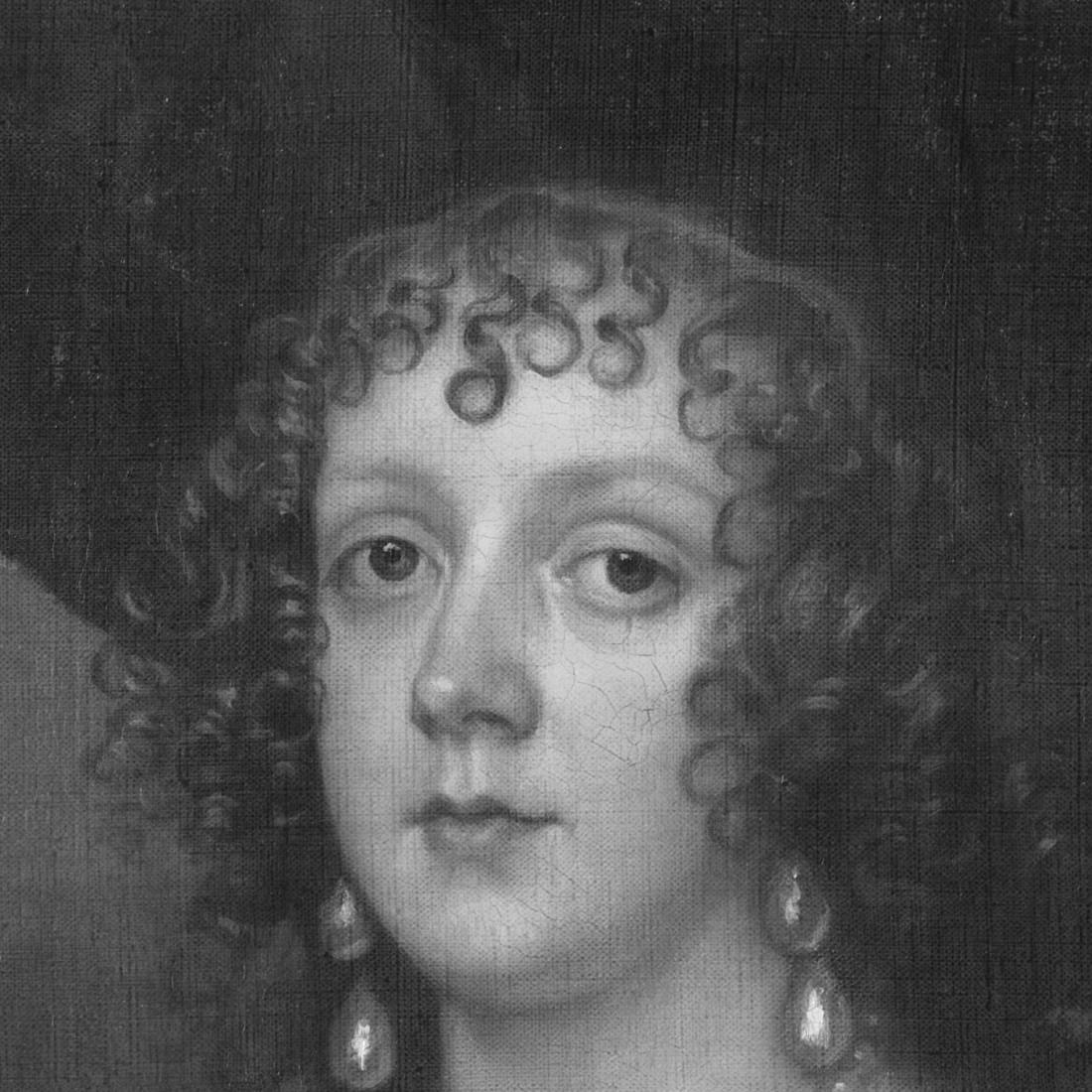}}

  \subfigure[]{\includegraphics[width=0.2\textwidth]{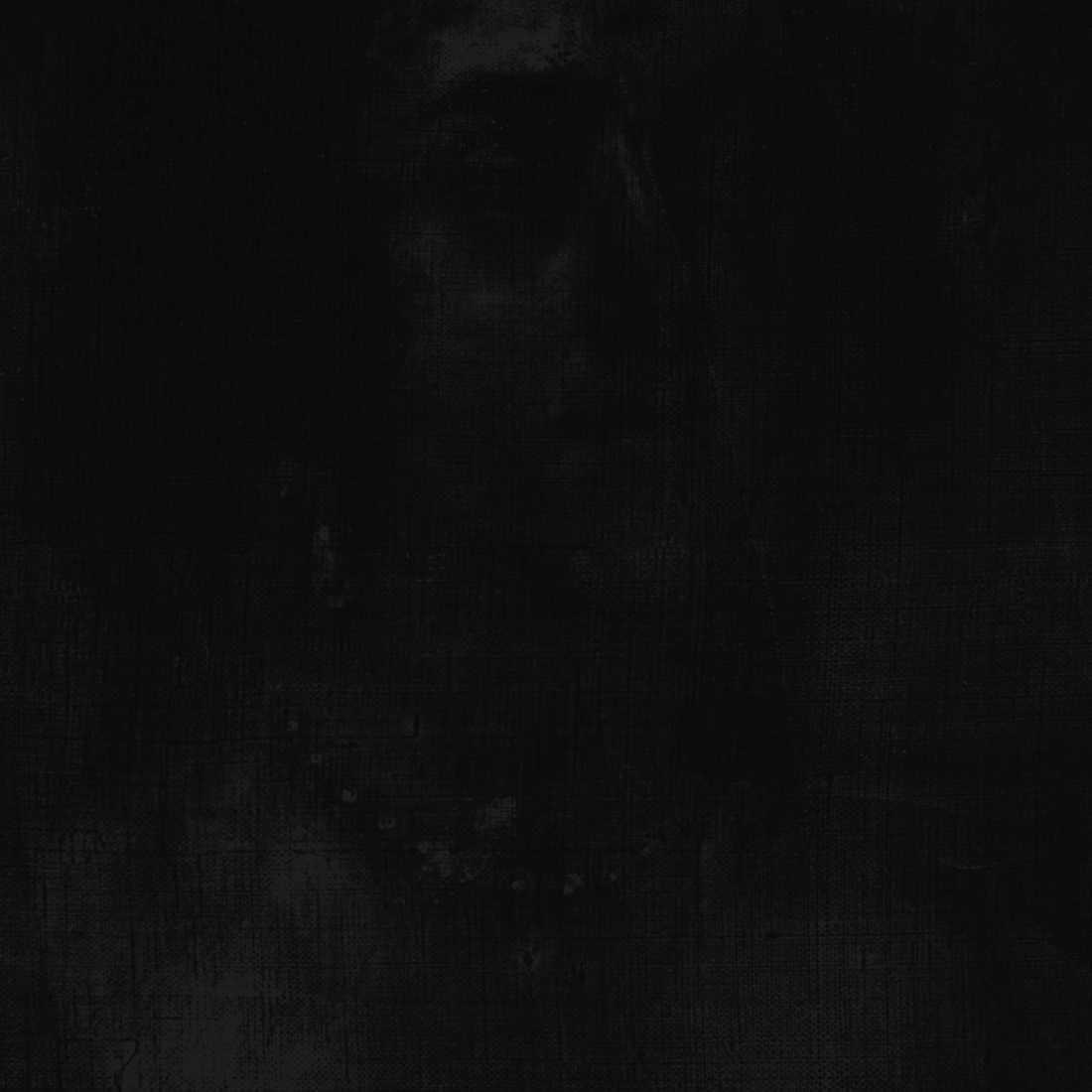}}
  \hfil
  \subfigure[]{\includegraphics[width=0.2\textwidth]{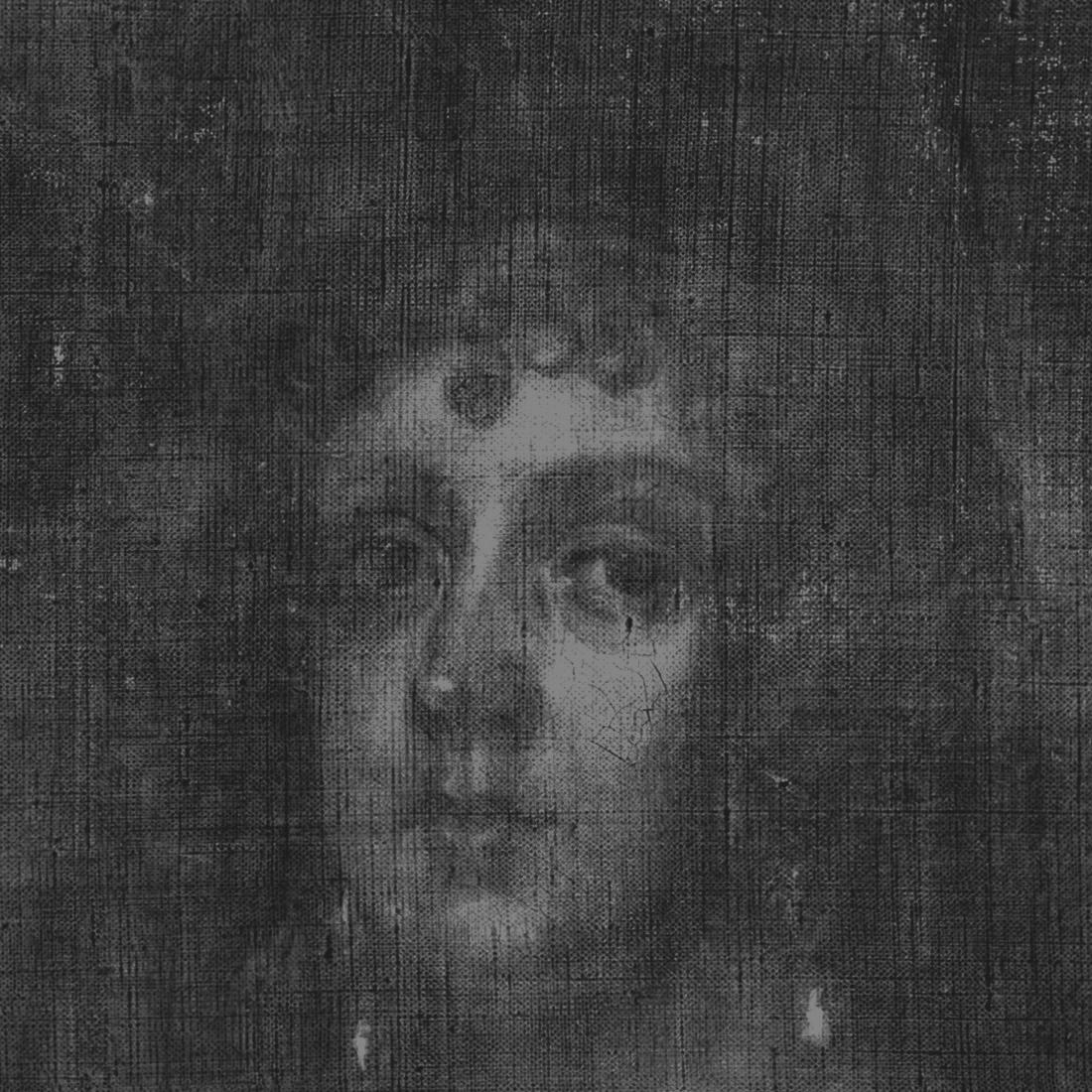}}
  \hfil
  \subfigure[]{\includegraphics[width=0.2\textwidth]{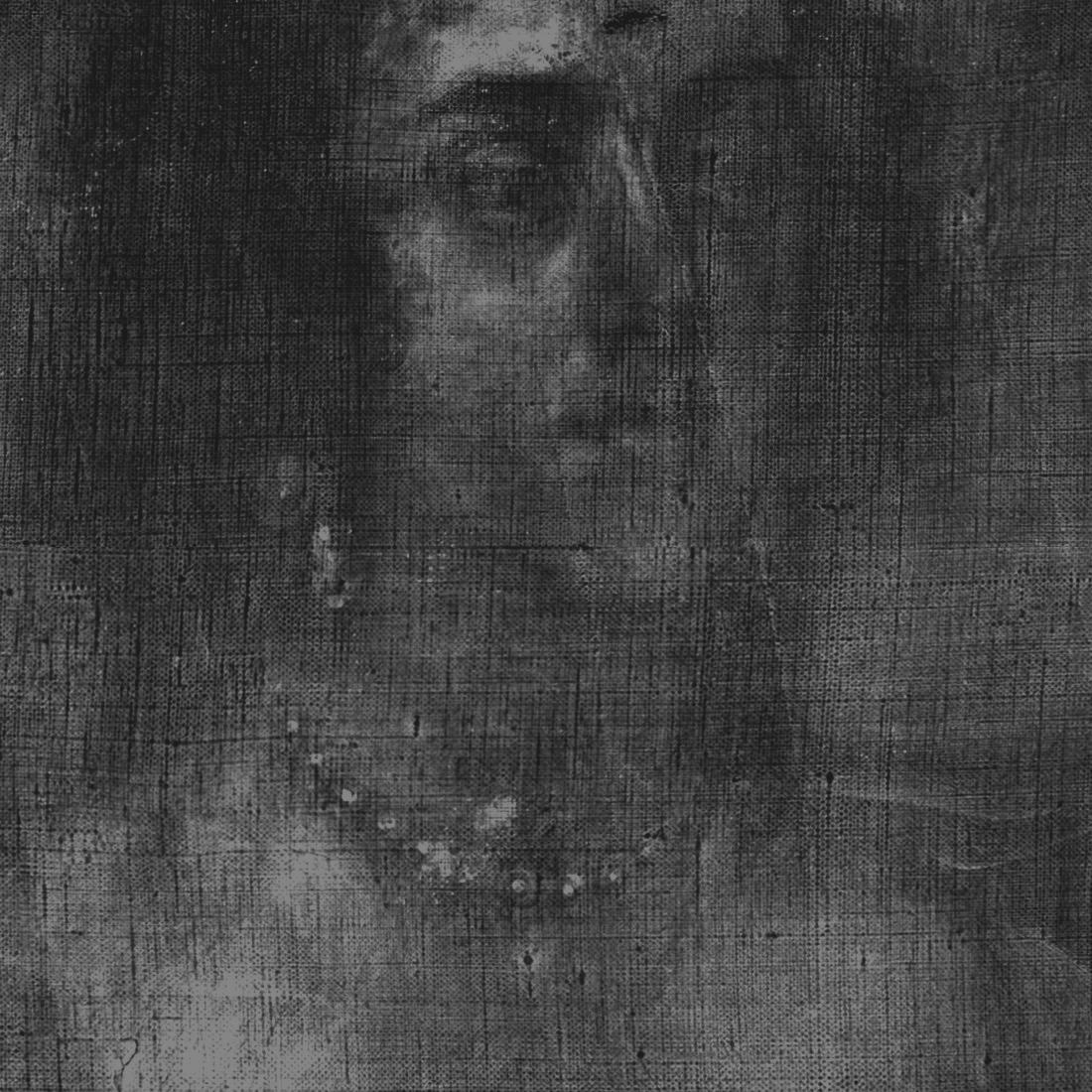}}
  \hfil
  \subfigure[]{\includegraphics[width=0.2\textwidth]{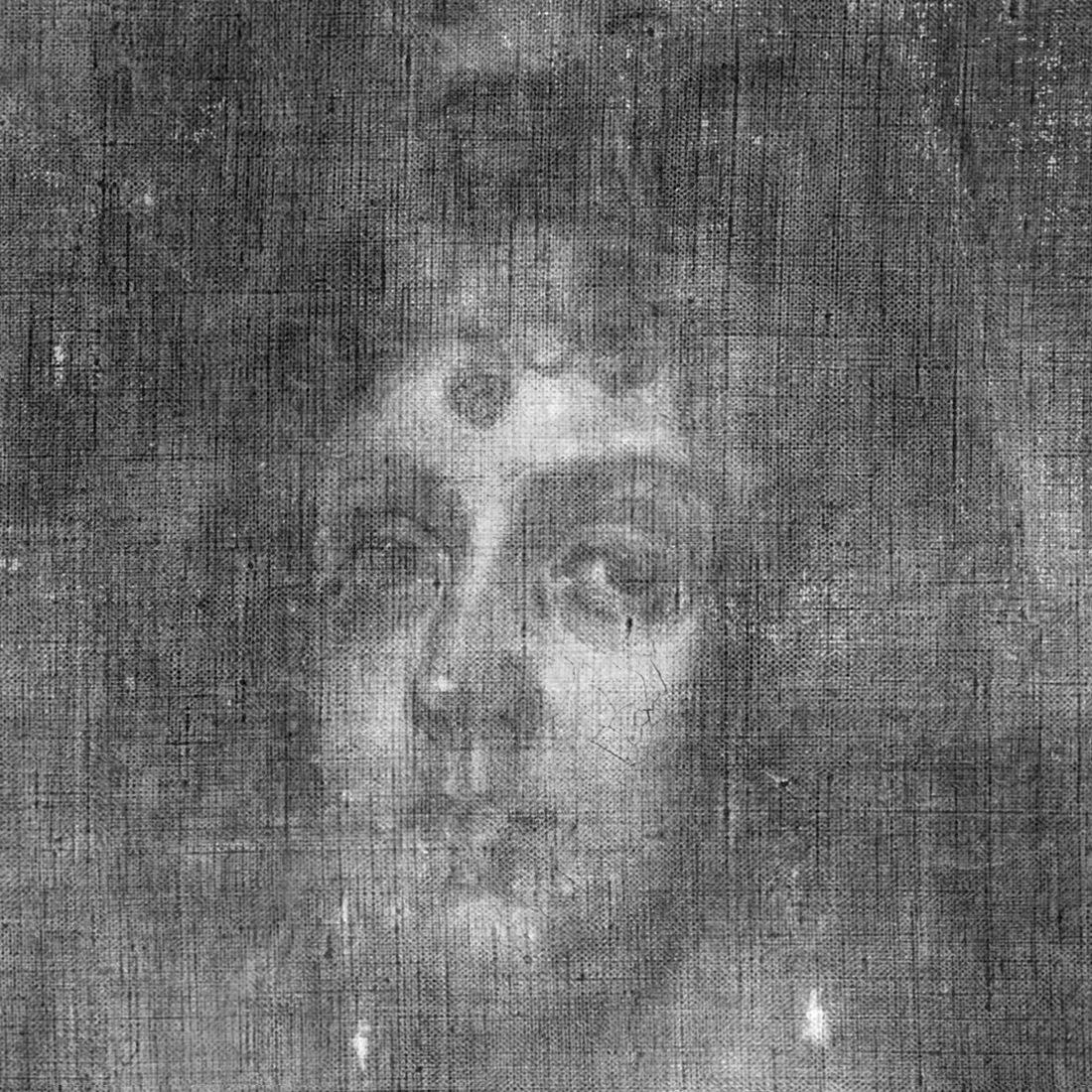}} 
\caption{ X-ray separation results: (a) and (b) Separated X-ray images for $\lambda_1 = 0.1$, $\lambda_2 = 0.1$, and $\lambda_3=\lambda_4=0$; (c) and (d) Separated X-ray images for $\lambda_1 = 10$, $\lambda_2 = 1$, and $\lambda_3=\lambda_4=0$; (e) and (f) Separated X-ray images for $\lambda_1 = 1$, $\lambda_2 = 10$, and $\lambda_3=\lambda_4=0$.}\label{figp4}
\end{figure}

Fig. \ref{figp3} depicts the average MSE in eq. (\ref{e10}) as a function of the hyper-parameters $\lambda_1$ and $\lambda_2$ with $\lambda_3$ and $\lambda_4$ set to be equal to zero. It is clear that different hyper-parameter values result in different separation performances. For example,

\begin{itemize}

\item With $\lambda_1 = 0.2$ and $\lambda_2 = 0.2$ the loss function component $L_1$ dominates over components $L_2$ and $L_3$ implying one tends to promote fidelity of the reconstruction of the individual RGB images. Fig. \ref{figp3} suggests that this may result in a relatively high average separation MSE and Fig. \ref{figp4} (a) and (b) also  confirm  that  the  separated X-ray images are very similar to the corresponding grayscale  versions of the visible RGB images (thereby losing information present in X-ray images associated with subsurface design features such as previous compositions and underdrawing, concealed areas of damage or structural features such as the wood grain (for paintings on panel) or canvas weave and wooden stretcher bars (for paintings on canvas).

\item With $\lambda_1 = 10$ and $\lambda_2 = 1$ the loss function component $L_2$ dominates over components $L_1$ and $L_3$ implying one tends to promote fidelity of the reconstruction of the mixed X-ray image. Fig. \ref{figp3} also suggests that this may result in a relatively high average separation MSE and Fig. \ref{figp4} (c) and (d) also confirm that one of the separated X-ray images becomes close to the wanted individual X-ray image while the other separated X-ray image becomes close to zero. 

\item Finally, with $\lambda_1 = 1$ and $\lambda_2 = 10$ the loss function component $L_3$ dominates over components $L_1$ and $L_2$. This situation seems to lead to a much better average MSE (Fig. \ref{figp3}) along with much better separation results (Fig. \ref{figp4} (e) and (f)) though we can still visualize content in one separated X-ray image that should be present in the other and vice-versa (e.g., the lady's face in Fig. \ref{figp4} (e) appears in Fig. \ref{figp4} (f)).

\end{itemize}

\begin{figure}[h]
\centering
  \subfigure[]{\includegraphics[width=0.2\textwidth]{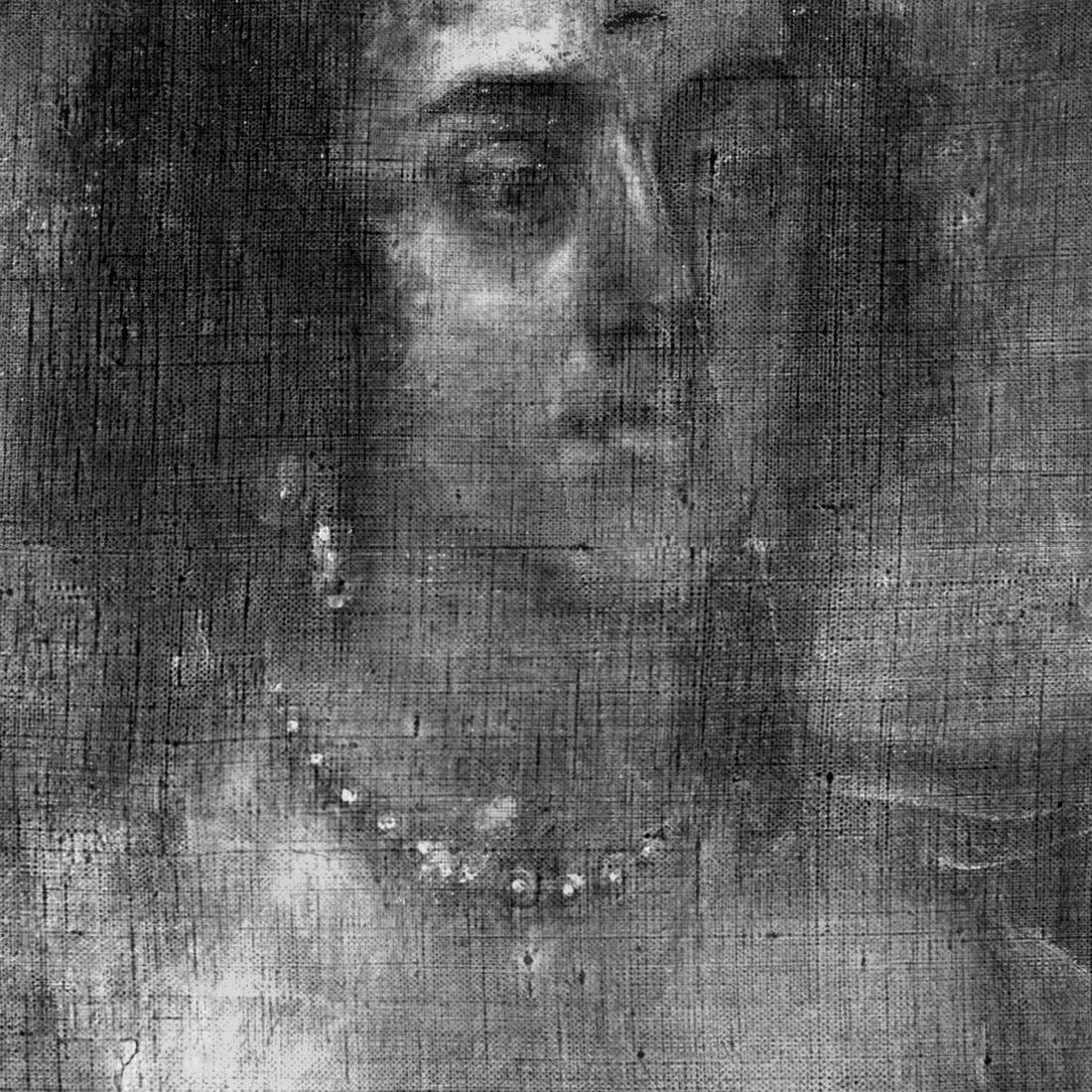}}
  \hfil
  \subfigure[]{\includegraphics[width=0.2\textwidth]{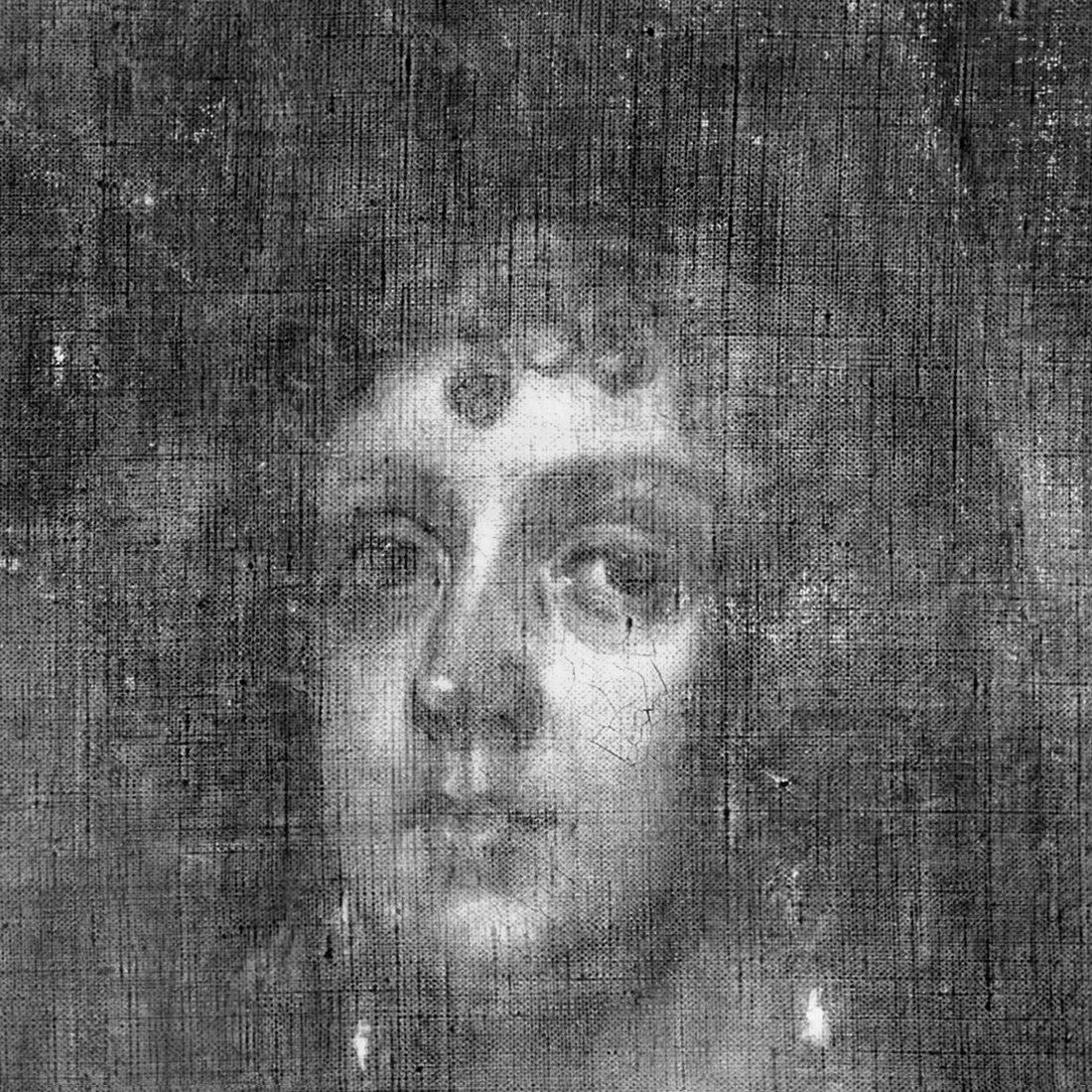}}

  \subfigure[]{\includegraphics[width=0.2\textwidth]{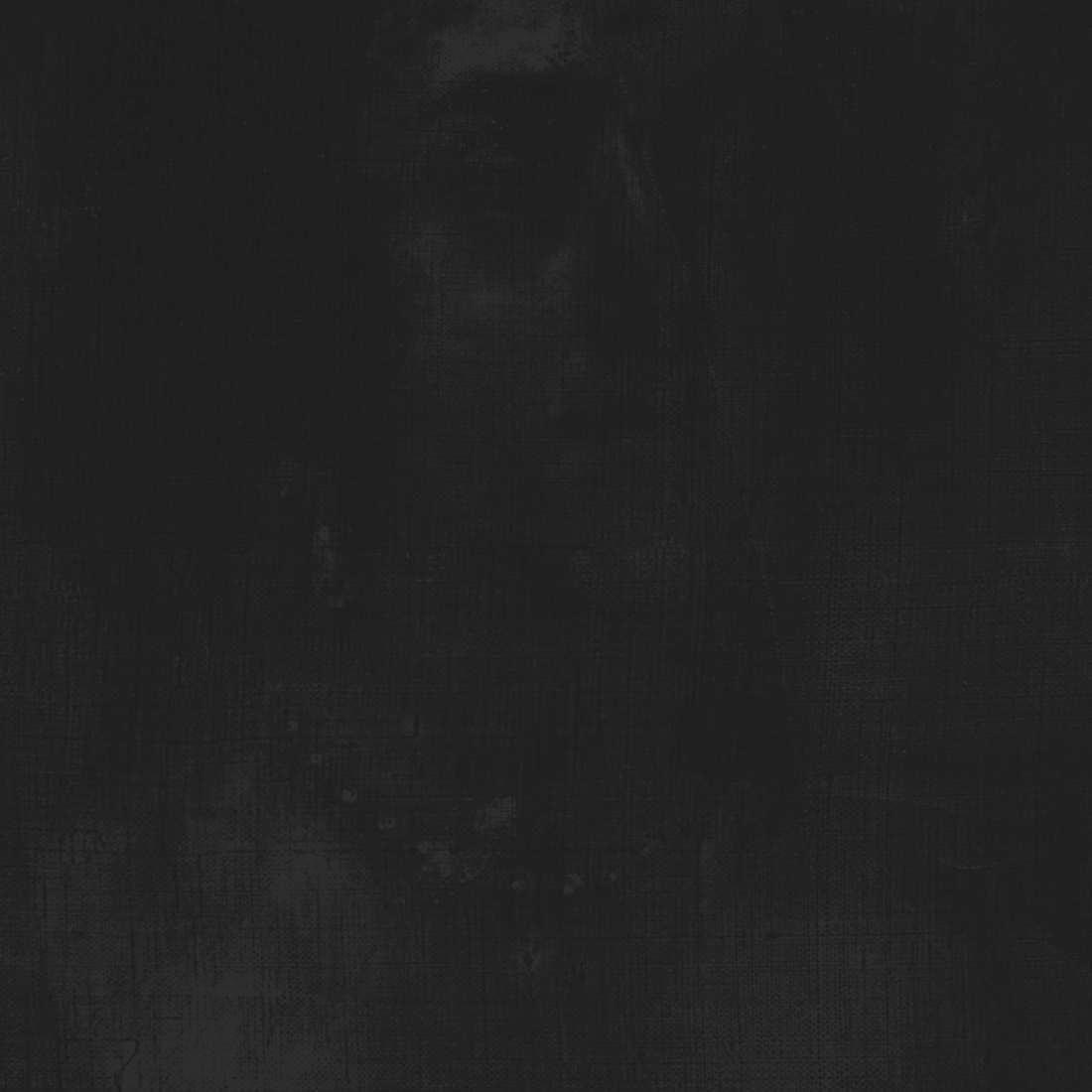}}
  \hfil
  \subfigure[]{\includegraphics[width=0.2\textwidth]{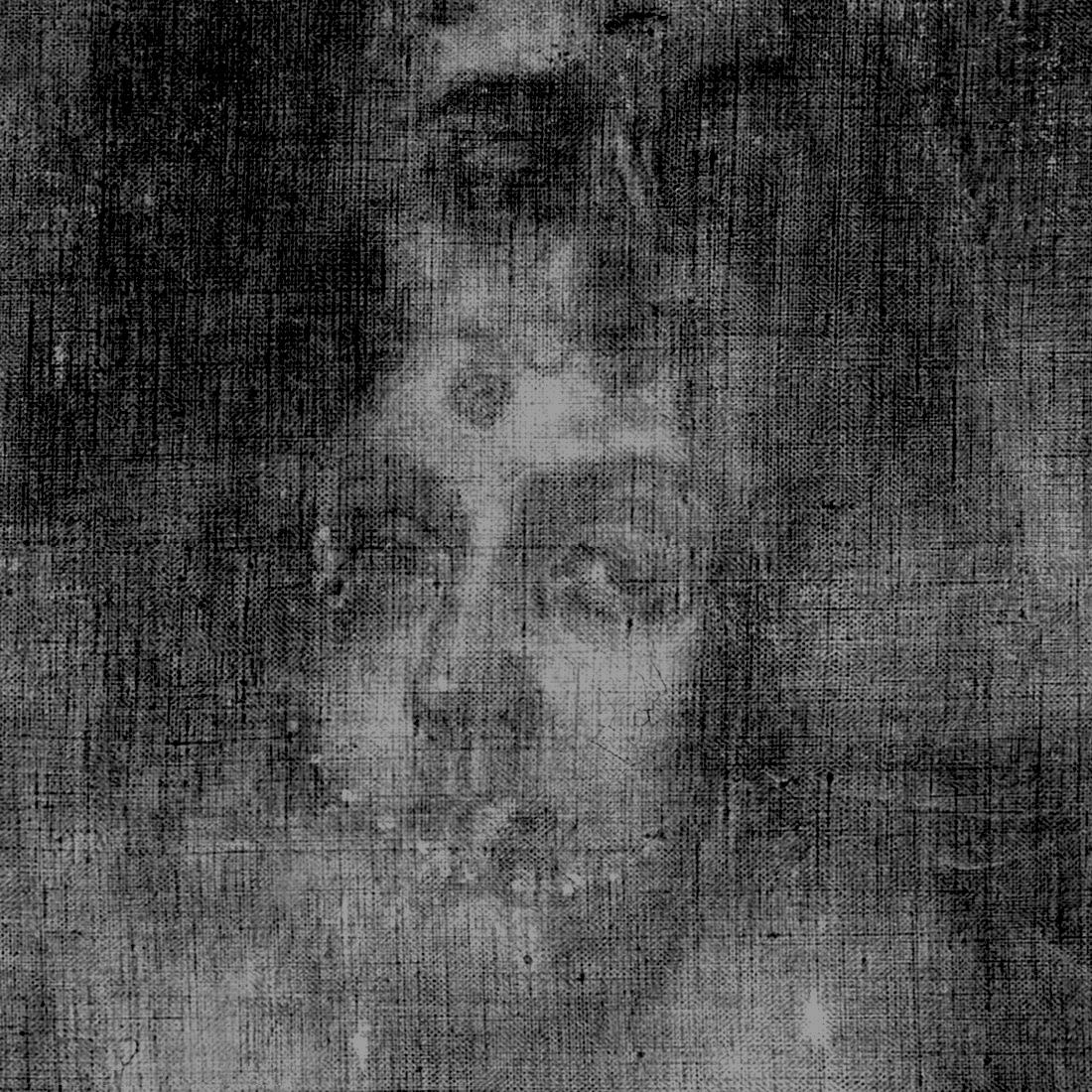}}

  \subfigure[]{\includegraphics[width=0.2\textwidth]{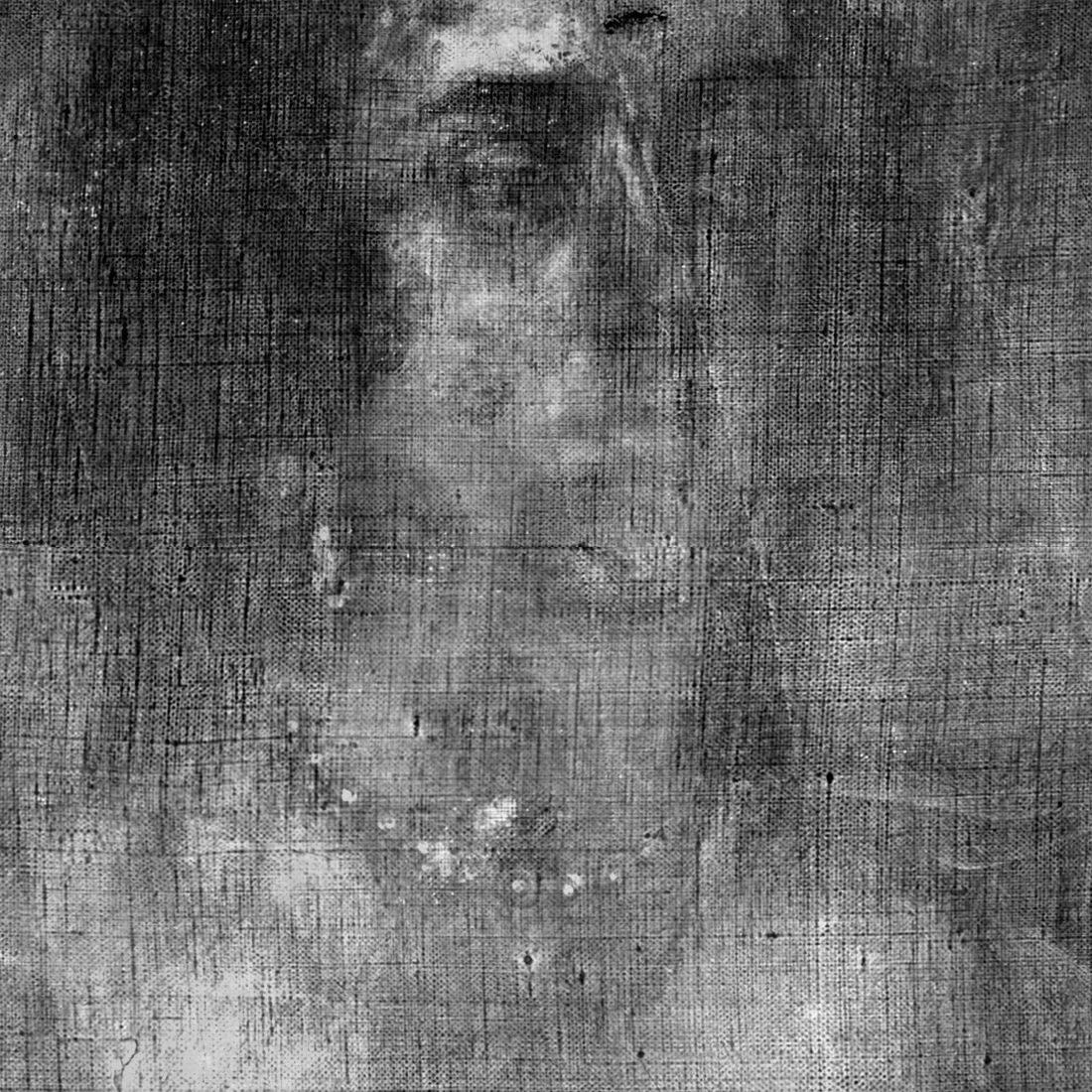}}
  \hfil
  \subfigure[]{\includegraphics[width=0.2\textwidth]{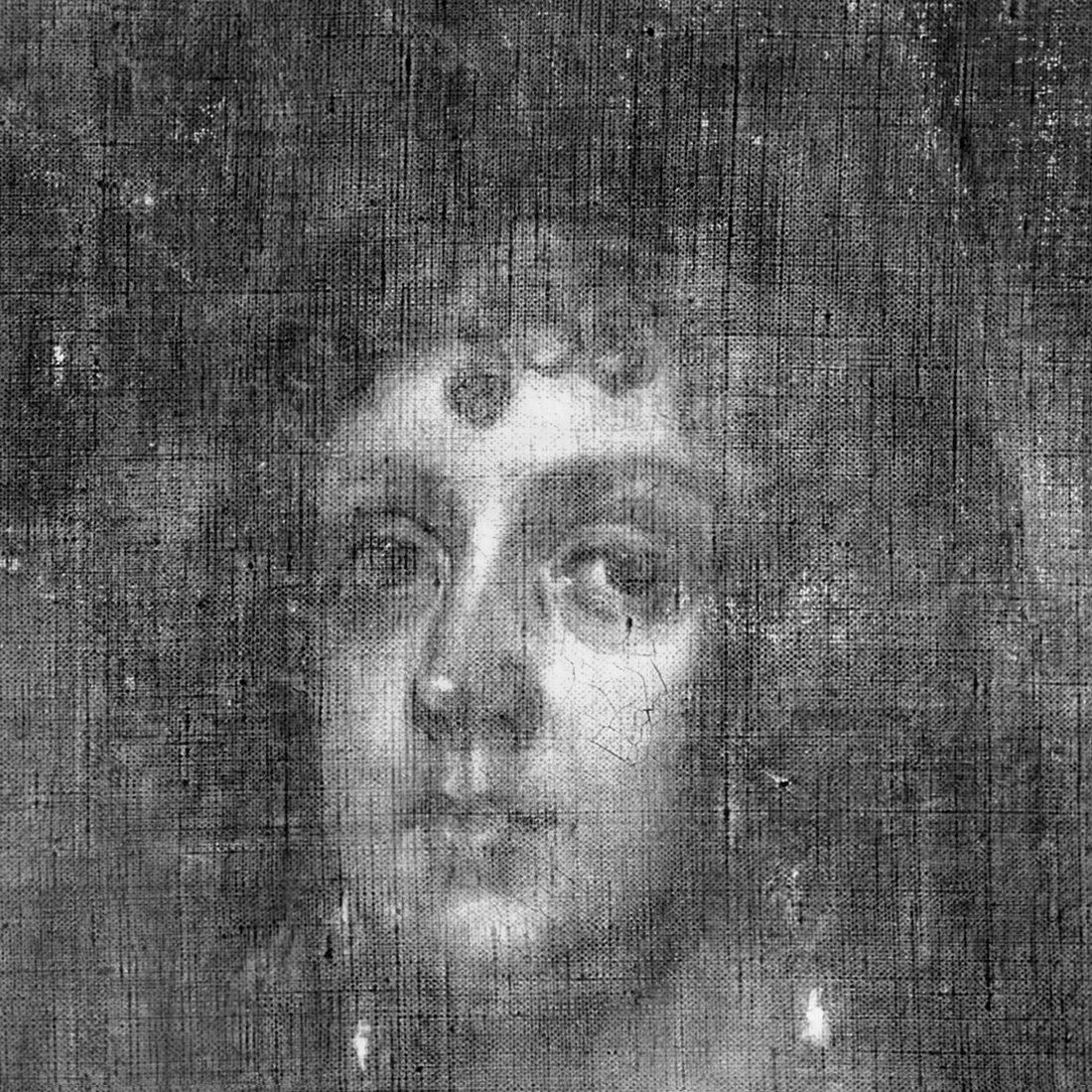}} 
\caption{ X-ray separation results: (a) and (b) Case i: Separated X-ray images for $\lambda_1 = 3$, $\lambda_2 = 5$, and $\lambda_3=\lambda_4=0$; (c) and (d) Case ii: Separated X-ray images for $\lambda_1 = 3$, $\lambda_2 = 5$, and $\lambda_3=\lambda_4=0$; (e) and (f) Case iii: Separated X-ray images for $\lambda_1 = 3$, $\lambda_2 = 5$, and $\lambda_3=\lambda_4=0$.}\label{figp5}
\end{figure}

Fig. \ref{figp3} suggests that the hyper-parameter values leading to the best average MSE performance are $\lambda_1=3$ and $\lambda_2=5$. However, depending on the exact random initialization of the auto-encoders in our method, the learning procedure can still lead to degenerate results due to poor convergence of the learning algorithm For example, there are initialization instances where 
(case i) we obtain the desired separation results $\hat{x}_1 \approx x_1$ and $\hat{x}_2 \approx x_2$   (see Fig. \ref{figp5} (a) and (b));
(case ii) we obtain a degenerate result where $\hat{x}_1 \approx x$ and $\hat{x}_2 \approx 0$ and vice-versa (see Fig. \ref{figp5} (c) and (d)); and 
(case iii) one separated X-ray image can contain features from the other X-ray image (see Fig. \ref{figp5} (e) and (f)), e.g., the lady's face in Fig. \ref{figp5} (f) appears in Fig. \ref{figp5} (e). 
Our experiments suggest that these cases occur with probability 64.6$\%$, 19.8$\%$ and 16.6$\%$, respectively.

We set our hyper-parameter values $\lambda_1$ and $\lambda_2$ to be equal to 3 and 5, respectively, but we remedy these degenerate cases by augmenting our loss function with two additional component losses $L_4$ and $L_5$ weighted by two additional hyper-parameters $\lambda_3$ and $\lambda_4$ as argued in the previous section.

\subsubsection{Effect of hyper-parameters $\lambda_3$ and $\lambda_4$}

\begin{figure}[h]
\centering
\includegraphics[width=0.35\textwidth]{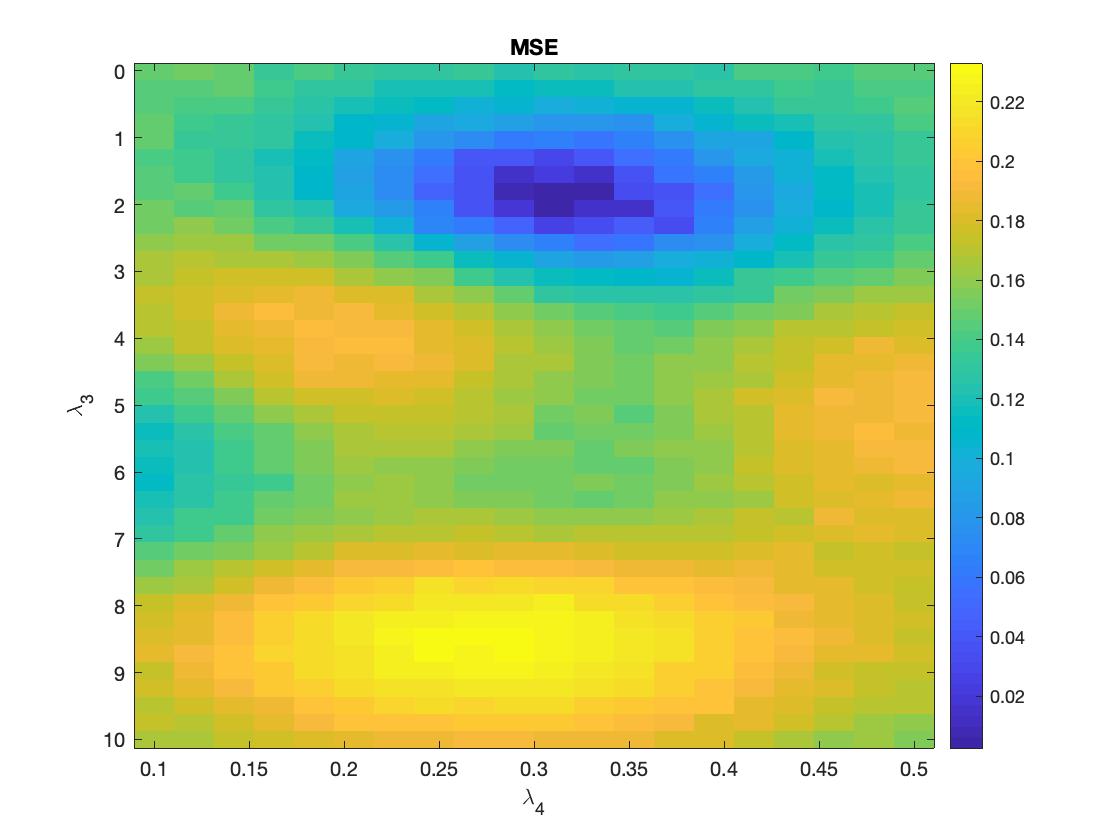}
\caption{Average MSE in eq. (\ref{e10}) as a function of $\lambda_3$ and $\lambda_4$ with $\lambda_1=3$ and $\lambda_2=5$.}\label{figp6}
\end{figure}

Fig. \ref{figp6} depicts the average MSE in (\ref{e10}) as a function of the hyper-parameters $\lambda_3$ and $\lambda_4$ with $\lambda_1$ and $\lambda_2$ set to be equal to 3 and 5 respectively. It is also clear that the hyper-parameters $\lambda_3$ and $\lambda_4$ can influence substantially the separation performance, with $\lambda_3 = 2$ and $\lambda_4 = 0.3$ leading to the best results.

\begin{figure}[h]
    \centering
    \subfigure[]{\includegraphics[width=0.23\textwidth]{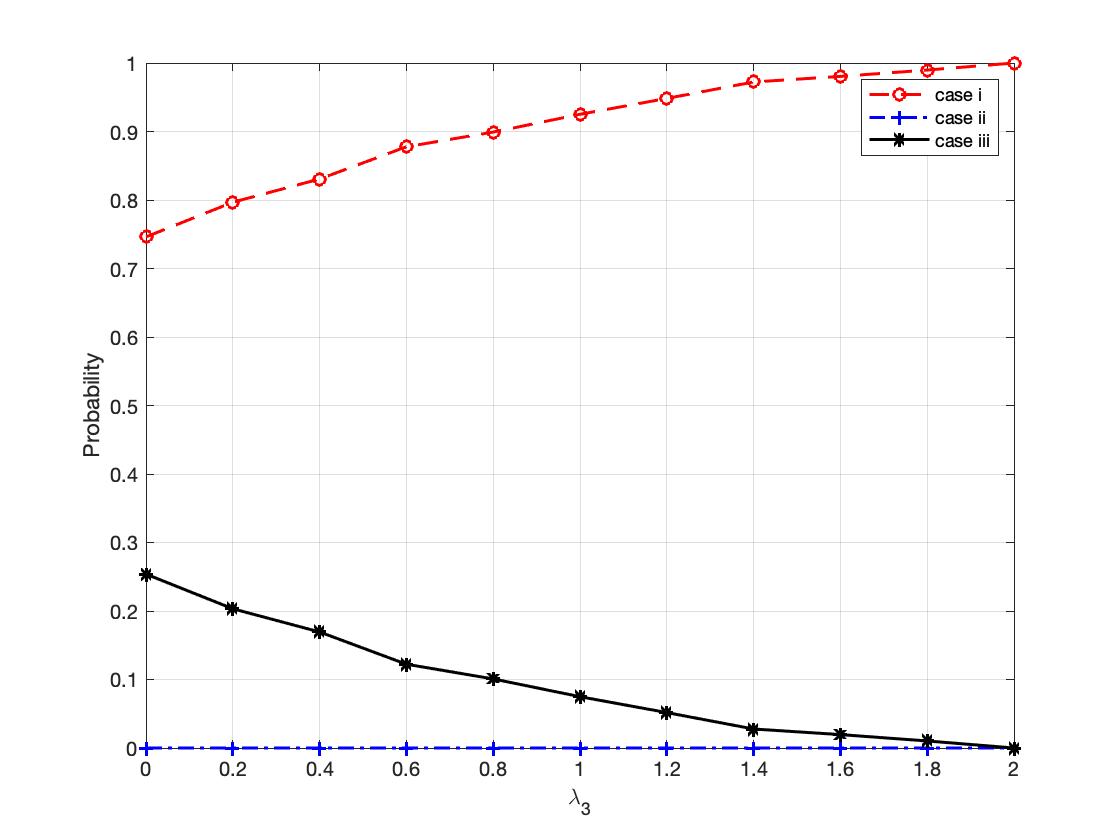}}
    \hfil
    \subfigure[]{\includegraphics[width=0.23\textwidth]{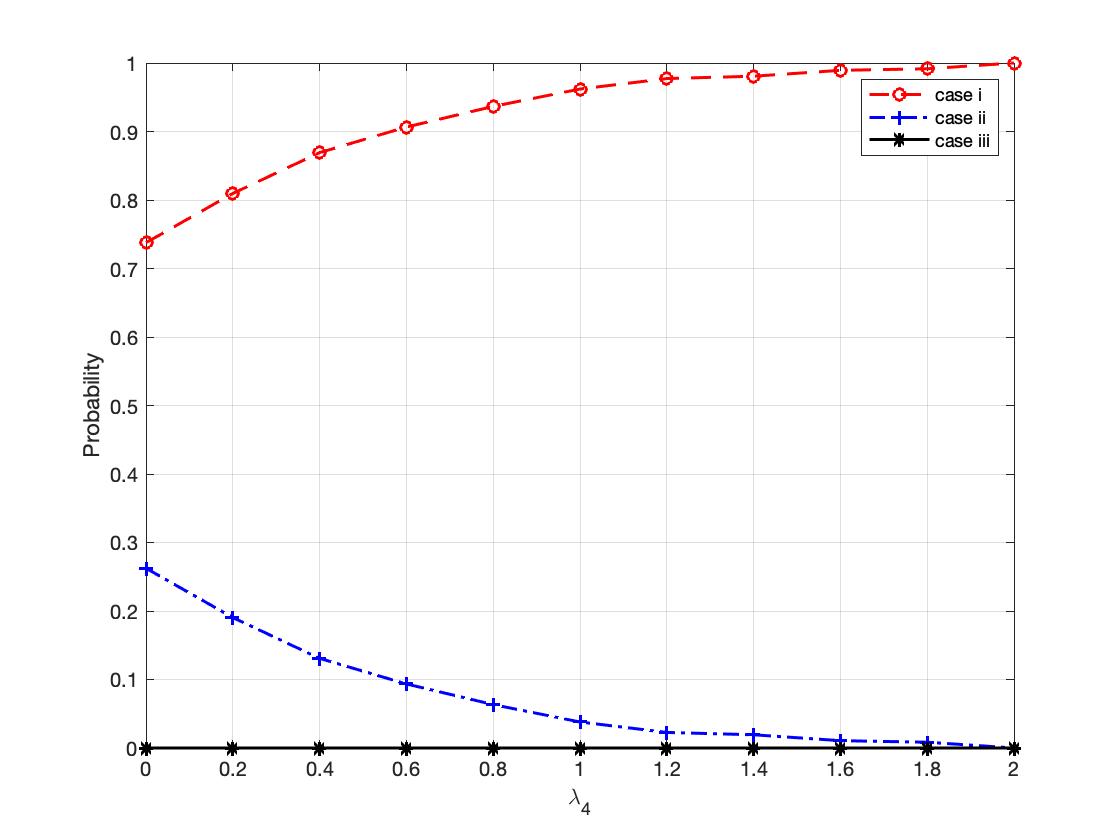}}
    \caption{ Probability of ocurrence of cases i, ii and iii as a function of hyper-parameters $\lambda_3$ and $\lambda_4$. (a). Probability vs. $\lambda_3$ with $\lambda_1=3$, $\lambda_2=5$, and $\lambda_4=0.3$. (b). Probability vs. $\lambda_4$ with $\lambda_1=3$, $\lambda_2=5$, $\lambda_3=2$.}\label{figp7}
\end{figure}

Two additional experiments  illustrate  how the additional loss functions address the issues highlighted in Fig. \ref{figp5}. These experiments involve quantifying the probability of the occurrence of the different cases i, ii, and iii -- where the probability is calculated over different random initializations of our models -- as a function of the hyper-parameters $\lambda_3$ and $\lambda_4$ associated with the component loss functions $L_4$ and $L_5$. Fig. \ref{figp7} suggests that the introduction of the extra loss functions leads indeed to a marked decrease in the probability of ocurrence of undesired cases ii and iii, and a substantial increase in the probability of occurrence of the desired case i.

To summarize, based on these results in the ensuing experiments, we use the set of hyper-parameters $\lambda_1 = 3$, $\lambda_2 = 5$, $\lambda_3 = 2$, and $\lambda_4 = 0.3$. \footnote{We note the optimal hyper-parameter values may depend slightly on the exact datasets, but we have found that these reported values tend to lead to very good performance across a wide range of datasets.}

\subsection { Experiments with Synthetically Mixed X-ray Data }

\subsubsection{Set-up}

\begin{figure*}[h]
    \centering
    \subfigure[]{\label{fig2a}\includegraphics[width=0.18\textwidth]{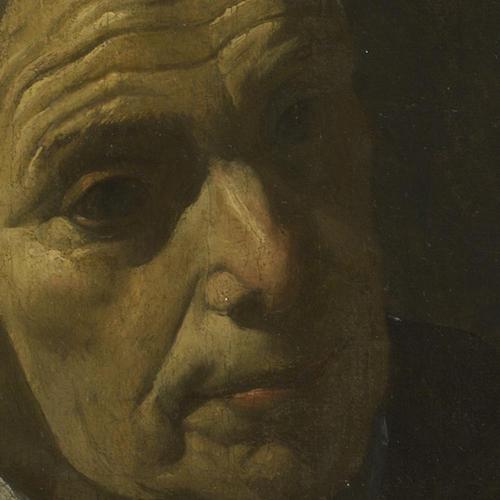}}
    \subfigure[]{\label{fig2b}\includegraphics[width=0.18\textwidth]{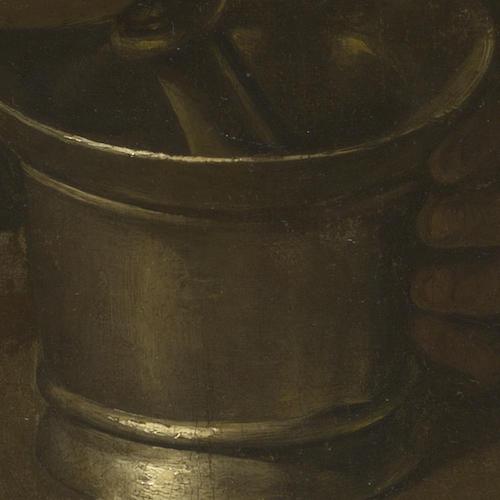}}
    \subfigure[]{\label{fig2c}\includegraphics[width=0.18\textwidth]{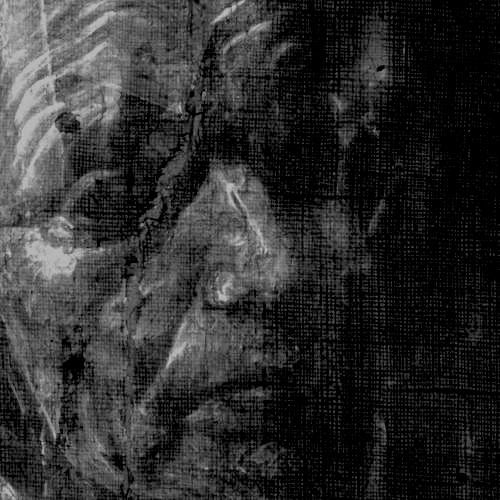}}
    \subfigure[]{\label{fig2d}\includegraphics[width=0.18\textwidth]{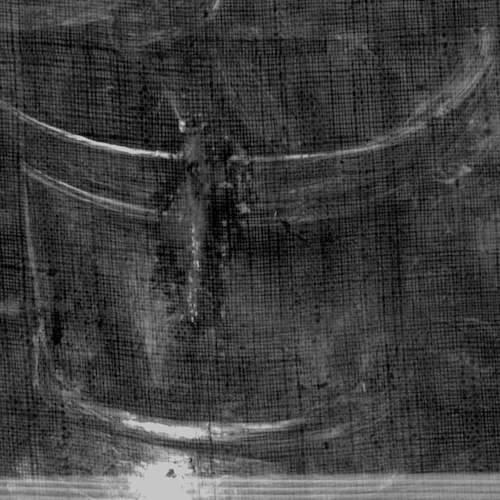}}
    \subfigure[]{\label{fig2e}\includegraphics[width=0.18\textwidth]{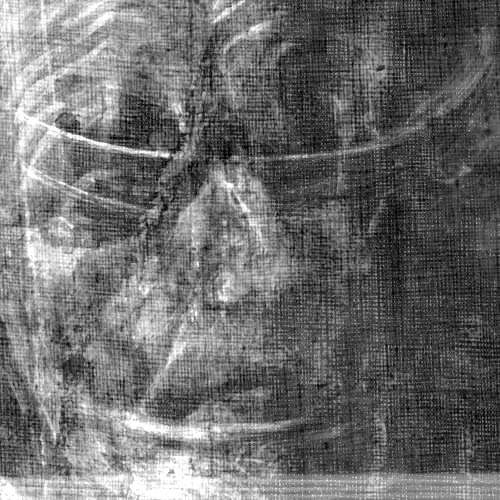}}
    \caption{Images used for synthetic data experiments. (a). First RGB image. (b). Second RGB image. (c). X-ray image corresponding to first RGB image. (d). X-ray image corresponding to second RGB image. (e). Synthetically mixed X-ray image.}\label{figs2}
\end{figure*}

In these experiments, we used two small areas with the same size from another oil painting \textsl{Kitchen Scene with Christ in the House of Martha and Mary} by Diego Velázquez to create a synthetically mixed X-ray image (see Fig. \ref{figs2}).

The previous procedure was again followed: the images – which are of size $1000 \times 1000$ pixels – were divided into patches of size $64 \times 64$ pixels with 56 pixels overlap (both in the horizontal and vertical direction), resulting in 13,924 patches. The patches associated with the synthetically mixed X-ray were  then separated independently. The various patches associated with the individual separated X-rays were  finally put together by placing various patches in the original order and averaging the overlap portions. All patches were utilized in the training of the auto-encoders by randomly shuffling their order.

As mentioned previously, we adopted the hyper-parameter values $\lambda_1 = 3$, $\lambda_2 = 5$, $\lambda_3 = 2$, and $\lambda_4 = 0.3$.

\subsubsection{Results}

In this section, some insights into the operation of the algorithm are provided. Fig. \ref{figs3} depicts the evolution of the overall loss function along with the individual ones as a function of the number of epochs. Various trends can be observed:

\begin{figure*}[h]
    \centering
    \subfigure[]{\label{fig2a}\includegraphics[width=0.32\textwidth]{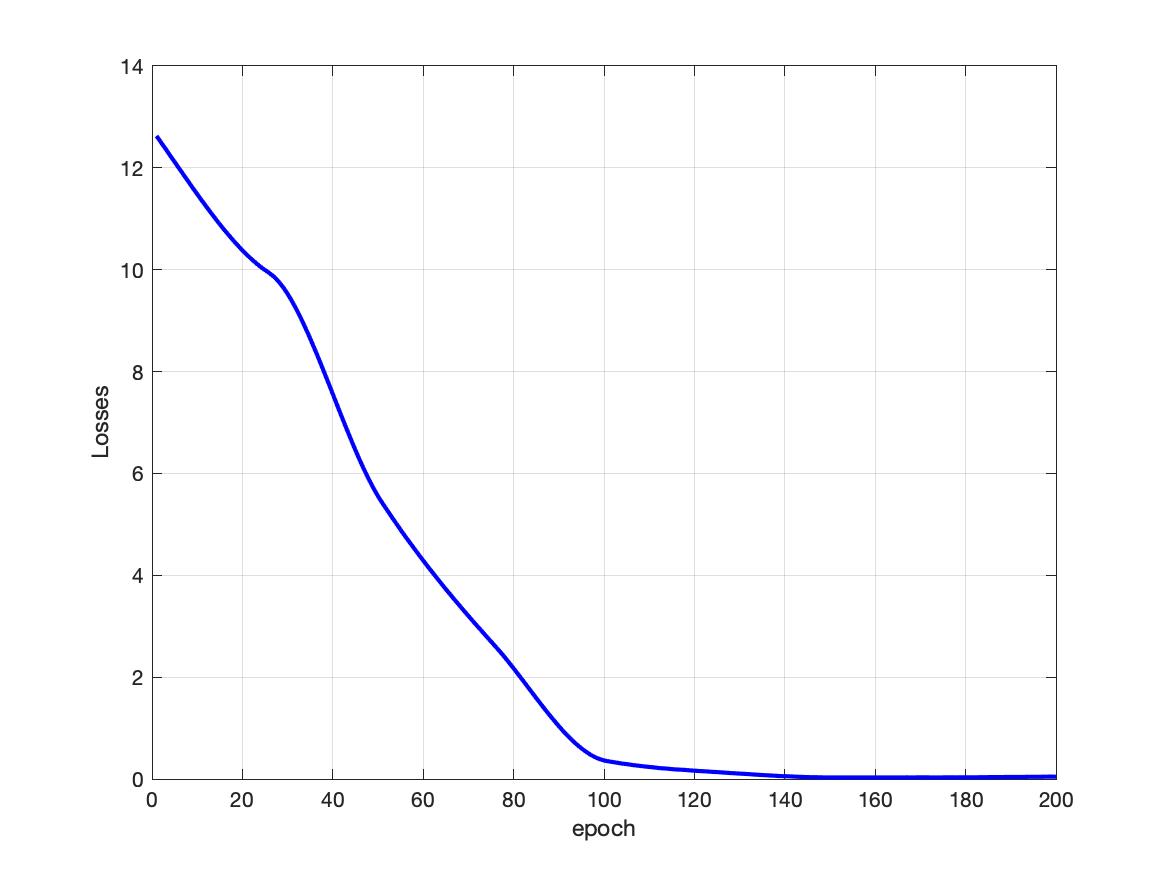}}
    \hfil
    \subfigure[]{\label{fig2b}\includegraphics[width=0.32\textwidth]{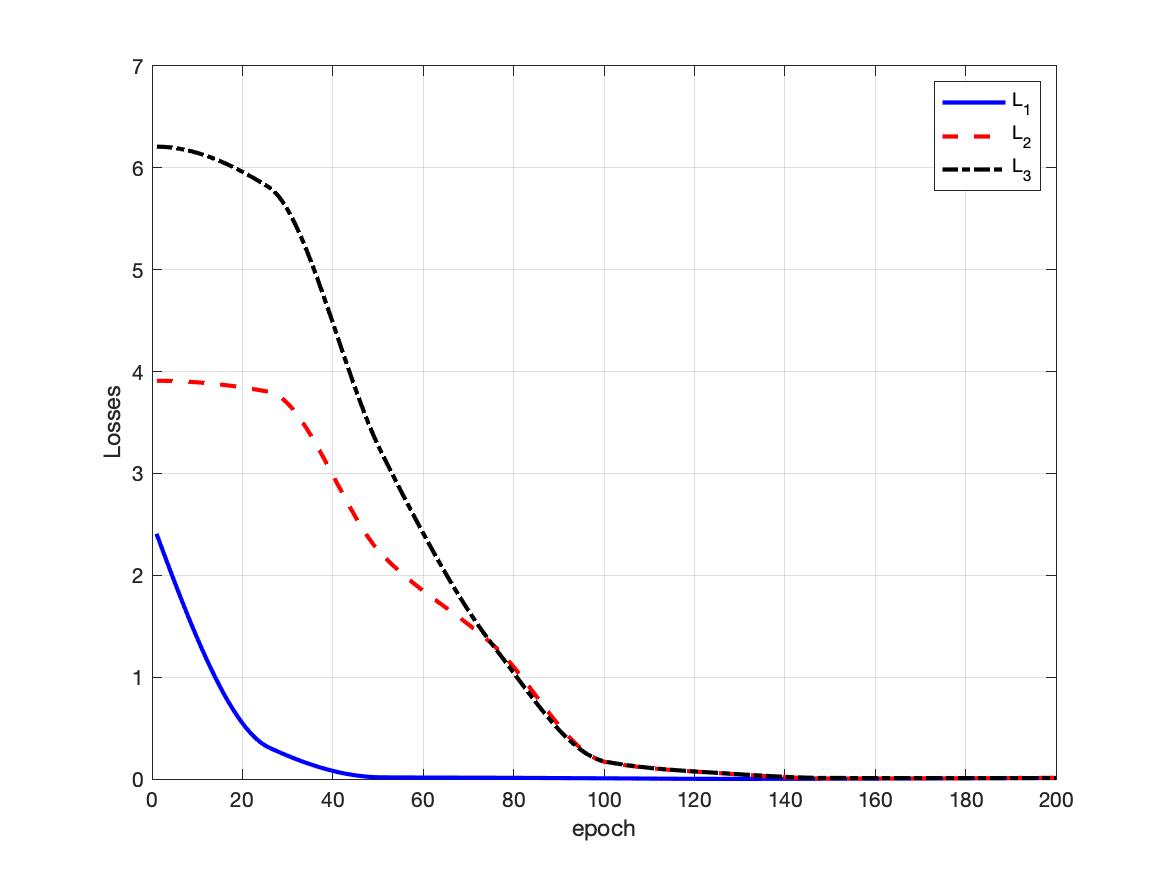}}
    \hfil
    \subfigure[]{\label{fig2c}\includegraphics[width=0.32\textwidth]{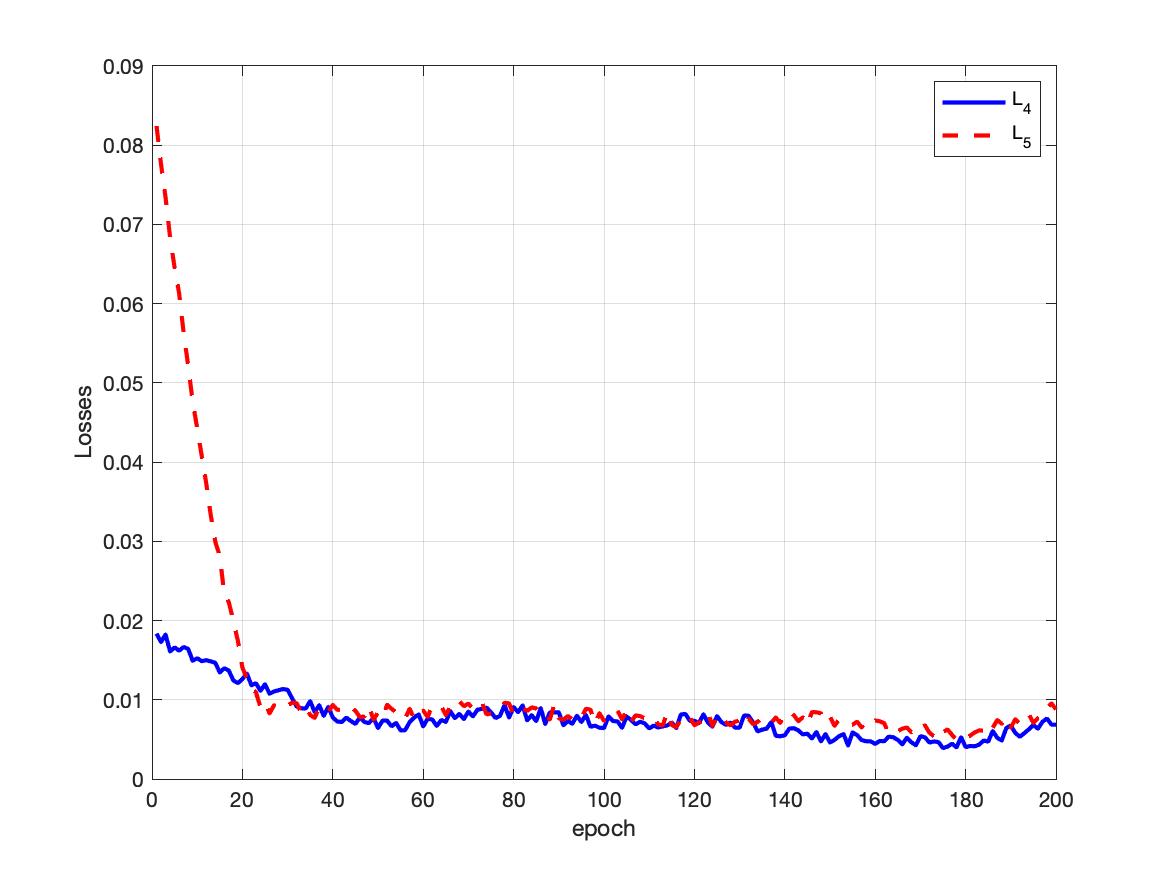}}
    \caption{Losses vs. number of epochs on synthetic data. (a). $L_{total}$. (b). $L_{1}$, $L_{2}$ and $L_{3}$. (c). $L_{4}$ and $L_{5}$.}\label{figs3}
\end{figure*}

\begin{figure*}[t]
    \centering
    \subfigure{\label{fig2a}\includegraphics[width=0.136\textwidth]{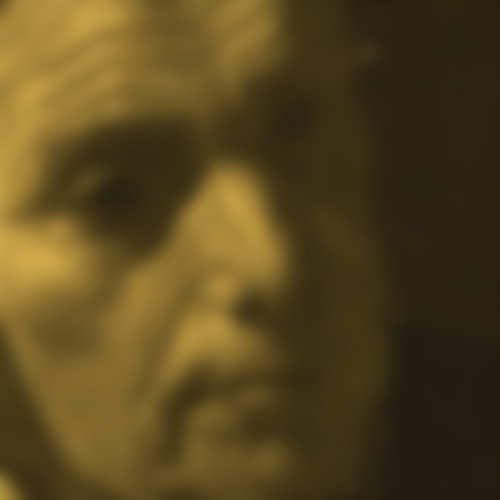}}
    \hfil
    \subfigure{\label{fig2b}\includegraphics[width=0.136\textwidth]{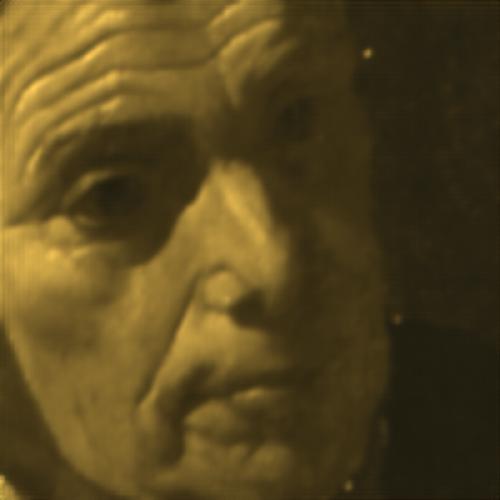}}
    \hfil
    \subfigure{\label{fig2c}\includegraphics[width=0.136\textwidth]{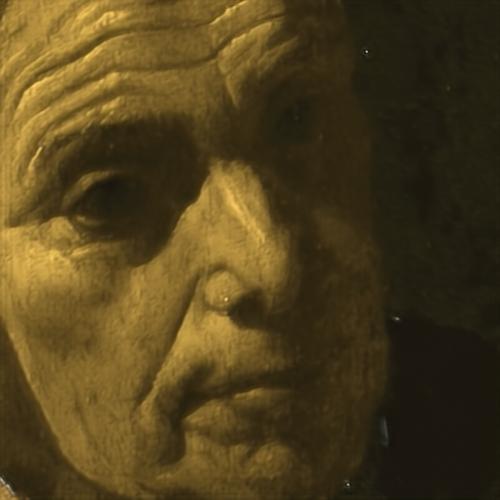}}
    \hfil
    \subfigure{\label{fig2a}\includegraphics[width=0.136\textwidth]{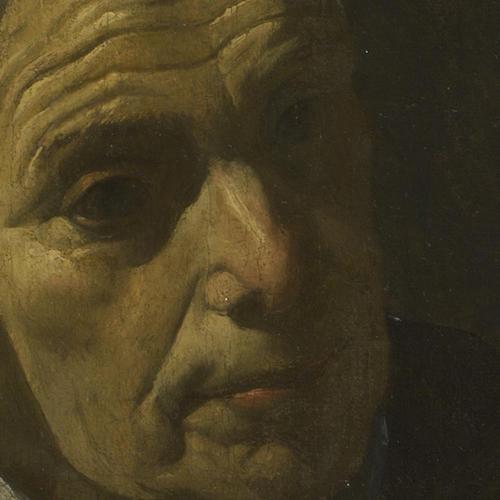}}
    \hfil
    \subfigure{\label{fig2b}\includegraphics[width=0.136\textwidth]{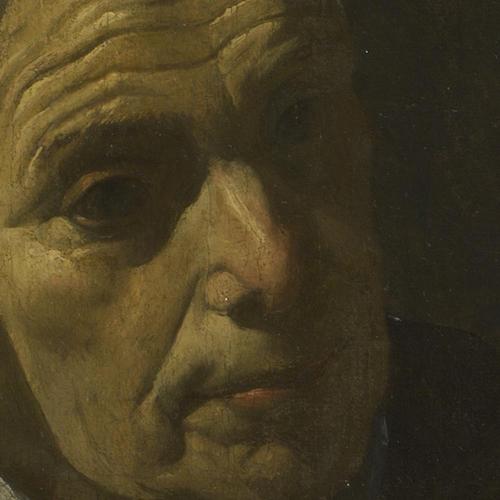}}
    \hfil
    \subfigure{\label{fig2c}\includegraphics[width=0.136\textwidth]{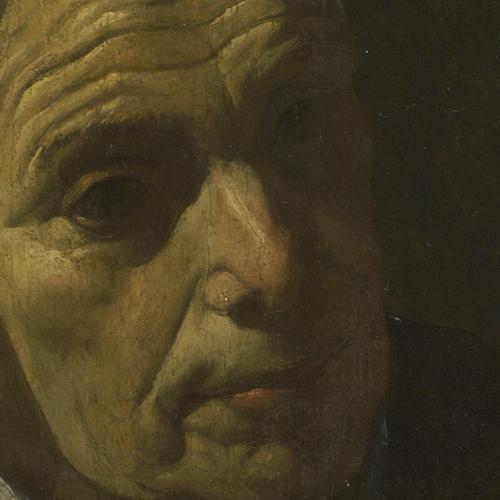}}
    \hfil
    \subfigure{\label{fig2c}\includegraphics[width=0.136\textwidth]{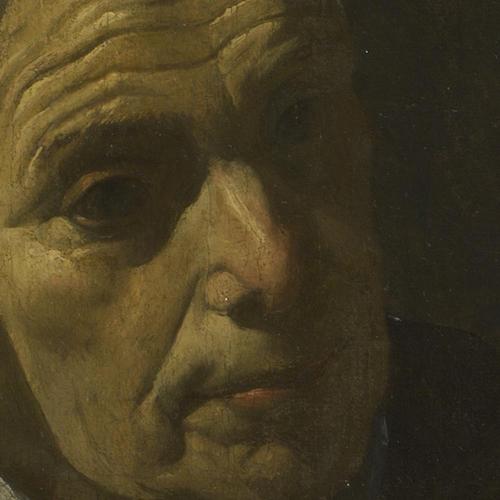}}    
    
    \subfigure{\label{fig2a}\includegraphics[width=0.136\textwidth]{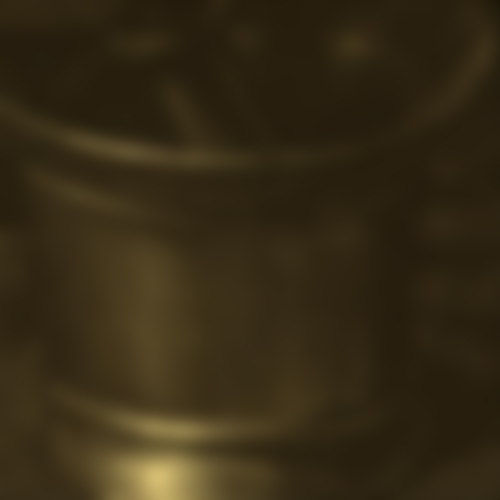}}
    \hfil
    \subfigure{\label{fig2b}\includegraphics[width=0.136\textwidth]{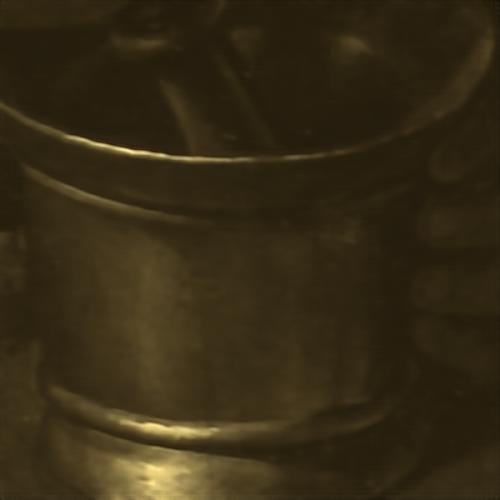}}
    \hfil
    \subfigure{\label{fig2c}\includegraphics[width=0.136\textwidth]{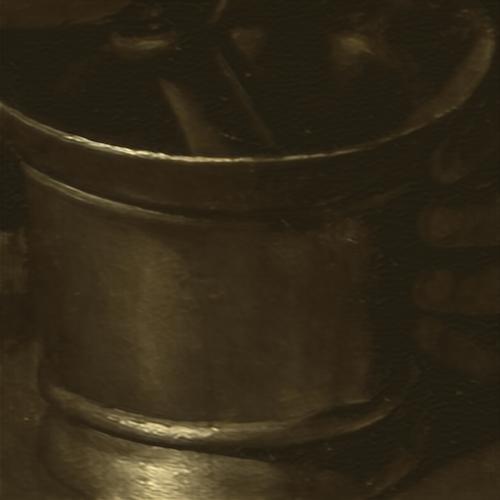}}
    \hfil
    \subfigure{\label{fig2a}\includegraphics[width=0.136\textwidth]{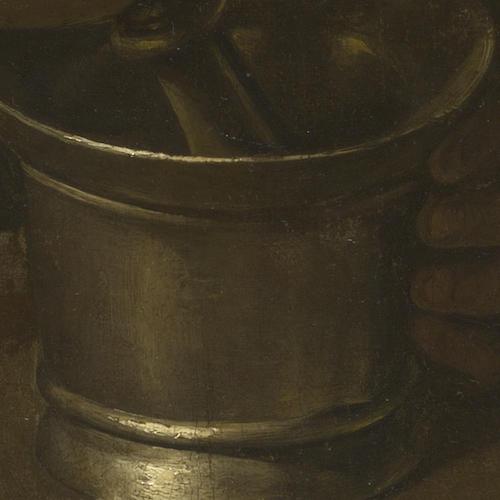}}
    \hfil
    \subfigure{\label{fig2b}\includegraphics[width=0.136\textwidth]{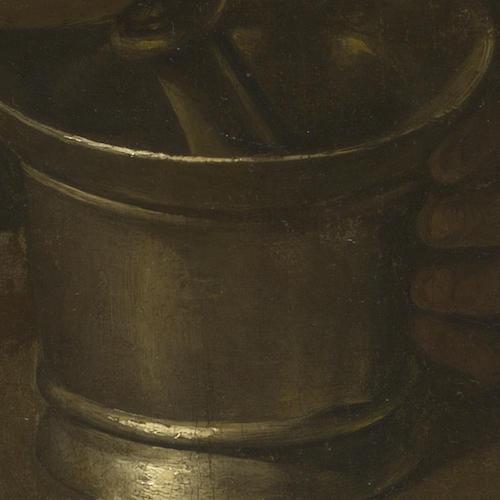}}
    \hfil
    \subfigure{\label{fig2c}\includegraphics[width=0.136\textwidth]{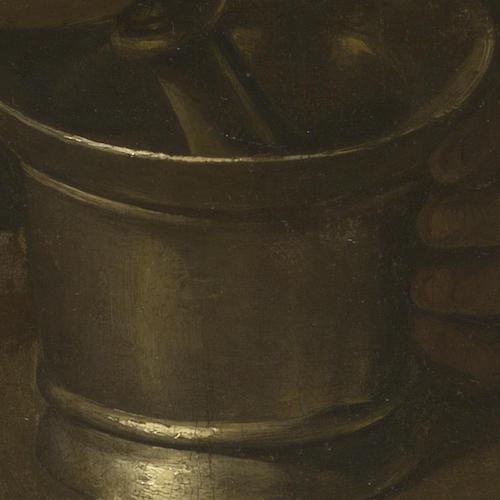}}
    \hfil
    \subfigure{\label{fig2c}\includegraphics[width=0.136\textwidth]{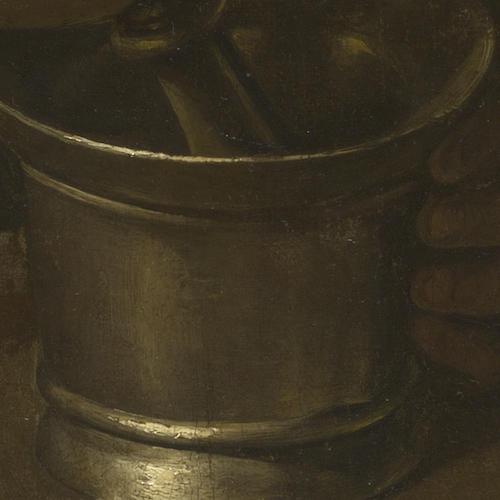}}

    \subfigure{\label{fig2a}\includegraphics[width=0.136\textwidth]{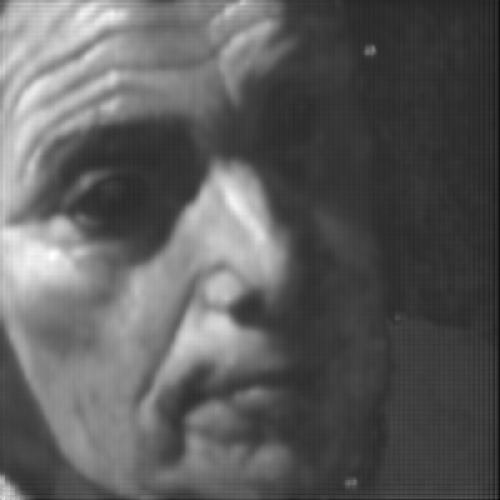}}
    \hfil
    \subfigure{\label{fig2b}\includegraphics[width=0.136\textwidth]{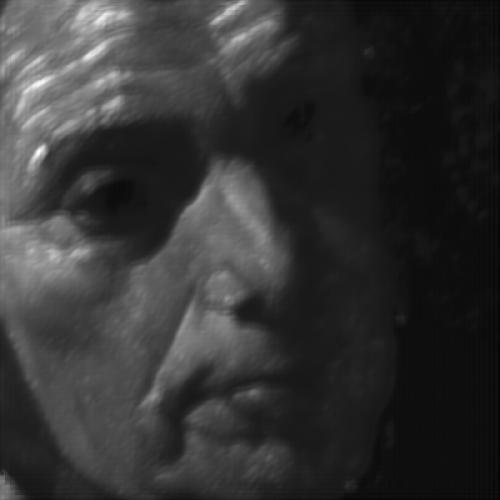}}
    \hfil
    \subfigure{\label{fig2c}\includegraphics[width=0.136\textwidth]{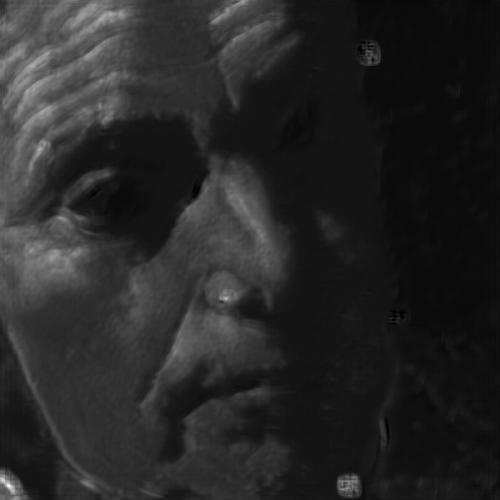}}
    \hfil
    \subfigure{\label{fig2a}\includegraphics[width=0.136\textwidth]{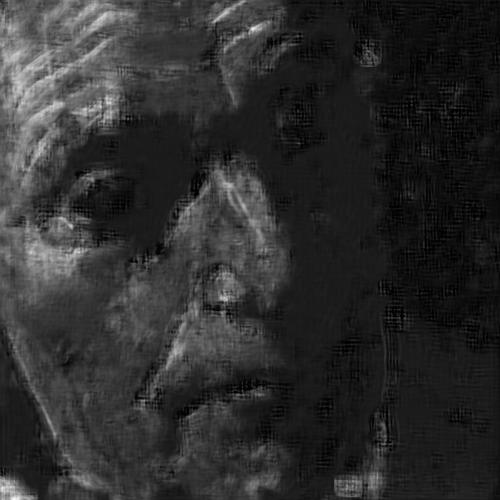}}
    \hfil
    \subfigure{\label{fig2b}\includegraphics[width=0.136\textwidth]{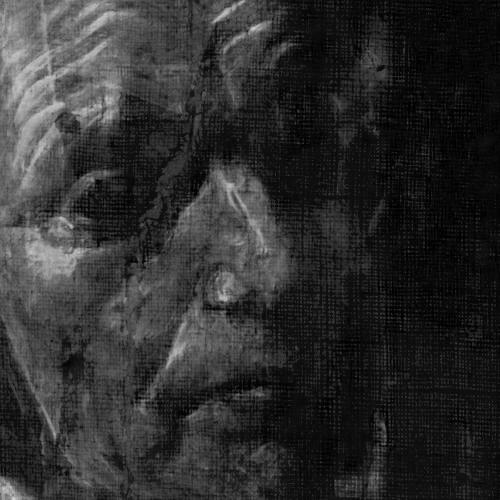}}
    \hfil
    \subfigure{\label{fig2c}\includegraphics[width=0.136\textwidth]{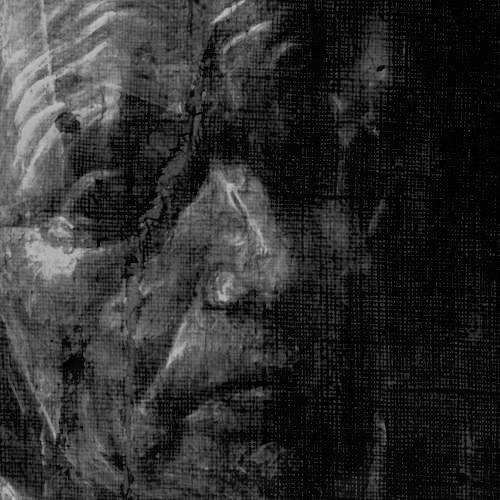}}
    \hfil
    \subfigure{\label{fig2c}\includegraphics[width=0.136\textwidth]{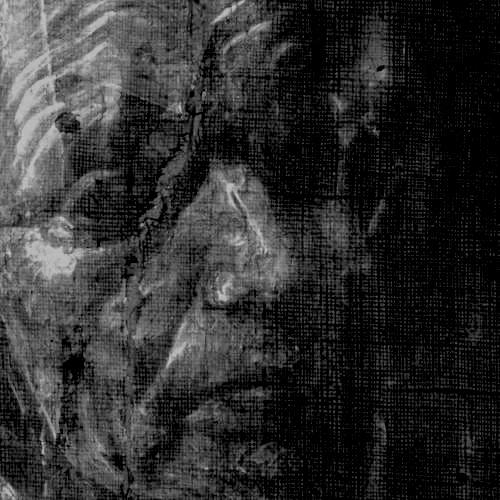}}    
    
    \subfigure{\label{fig2a}\includegraphics[width=0.136\textwidth]{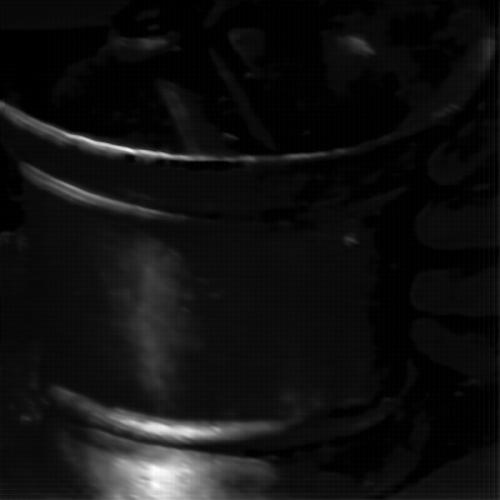}}
    \hfil
    \subfigure{\label{fig2b}\includegraphics[width=0.136\textwidth]{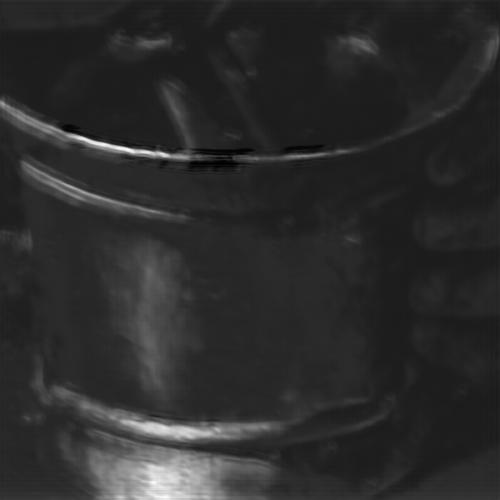}}
    \hfil
    \subfigure{\label{fig2c}\includegraphics[width=0.136\textwidth]{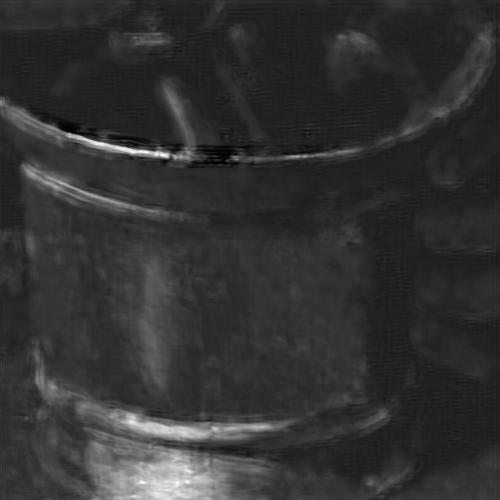}}
    \hfil
    \subfigure{\label{fig2a}\includegraphics[width=0.136\textwidth]{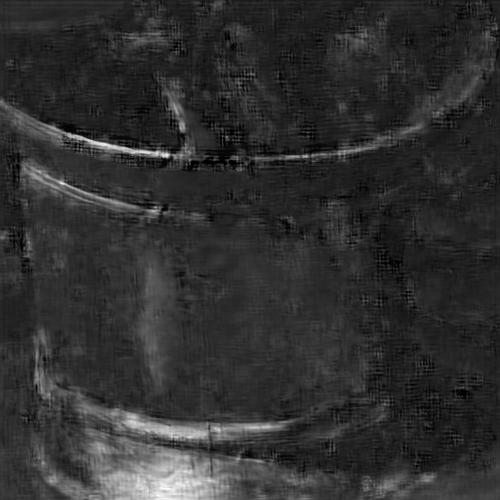}}
    \hfil
    \subfigure{\label{fig2b}\includegraphics[width=0.136\textwidth]{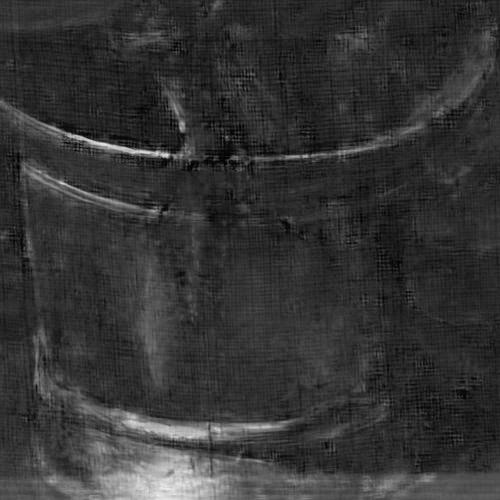}}
    \hfil
    \subfigure{\label{fig2c}\includegraphics[width=0.136\textwidth]{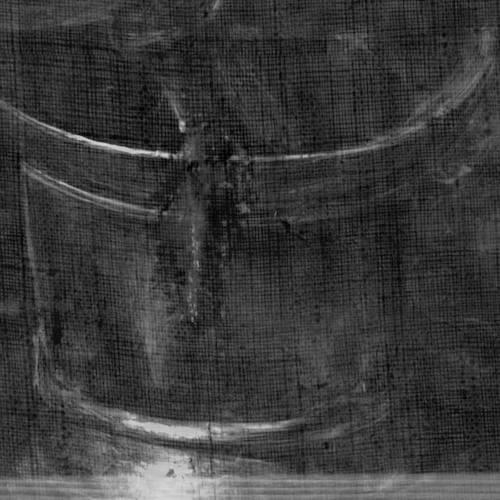}}
    \hfil
    \subfigure{\label{fig2c}\includegraphics[width=0.136\textwidth]{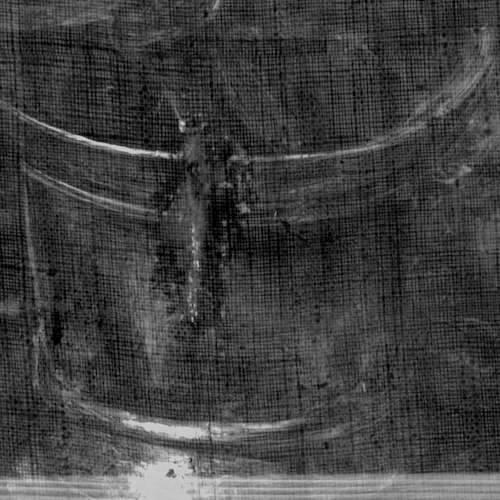}}    
      
    \caption{Reconstructed images vs. number of epochs on synthetic data experiments. Columns 1 to 7 correspond to reconstructed result under 1st, 4th, 10th, 50th, 100th, 150th and 200th epoch, respectively. Rows 1 to 2 correspond to the reconstructed RGB images. Rows 3 to 4 correspond to the reconstructed X-ray images.}\label{figs4}
\end{figure*}

\begin{itemize}

\item $L_{total}$: The overall loss function gradually decreases as the number of epochs increases. This suggests our method will eventually reconstruct perfectly the individual RGB images, the mixed X-ray image, and -- as a by-product -- the individual X-ray images (See Fig. \ref{figs4}).

\item $L_1$: The loss component $L_1$ decreases very rapidly during the initial 40 epochs but decreases less dramatically then onwards. This implies that the encoder $E_r$ and the decoder $D_r$ are learnt during the initial epochs, suggesting that the method can reconstruct very well the individual RGB images during this initial phase (See Fig. \ref{figs4} rows 1 and 2).

\item $L_2$ $\&$ $L_3$: The loss components $L_2$ and $L_3$ only decrease rapidly after epoch 30. This implies that the decoder $D_x$ is only learnt after epoch 30, suggesting in turn that the method can only reconstruct well the mixed X-ray image and the individual X-ray images well after epoch 30 (See Fig. \ref{figs4} rows 3 and 4).

\item $L_4$ $\&$ $L_5$: Fig. \ref{figs3} also suggests that these loss components indeed function to prevent the algorithm from converging to unwanted local minima, by playing a role during the initial learning stages. These losses rapidly converge to zero during the initial epochs, and in doing so do not affect much further the overall loss function after epoch 30.

\end{itemize}

Interestingly, in line with these observations, Fig. \ref{figs4} shows that the evolution of the reconstruction of the individual X-ray images ranges from a grayscale version of the corresponding RGB images (during the initial learning stages) to the true X-ray images (during the later learning stages). Once again, this is due to the fact that the decoder $D_x$ is learnt during a later learning phase.

\begin{figure*}[h]
    \centering
    \subfigure{\label{fig2a}\includegraphics[width=0.16\textwidth]{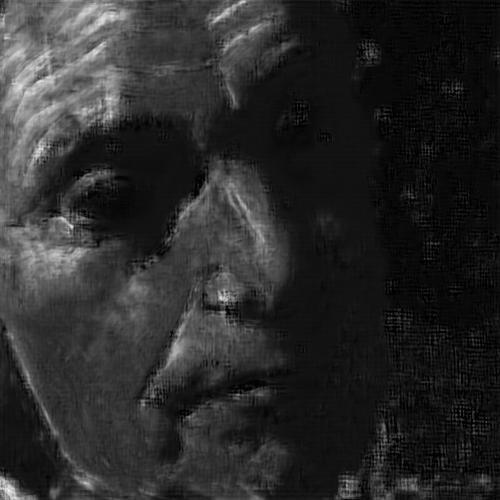}}
    \hfil    
    \subfigure{\label{fig2a}\includegraphics[width=0.16\textwidth]{S_x1.jpg}}
    \hfil
    \subfigure{\label{fig2b}\includegraphics[width=0.16\textwidth]{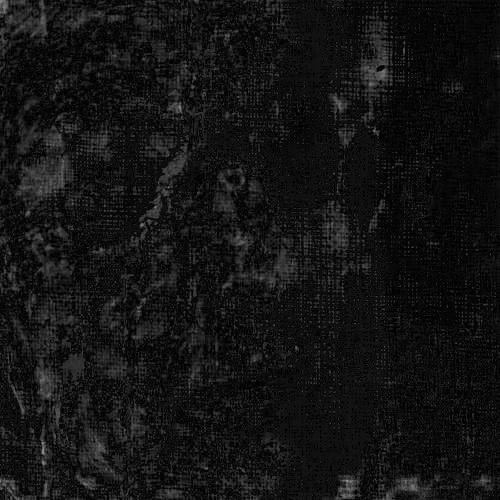}}
    \hfil  
    \subfigure{\label{fig2a}\includegraphics[width=0.16\textwidth]{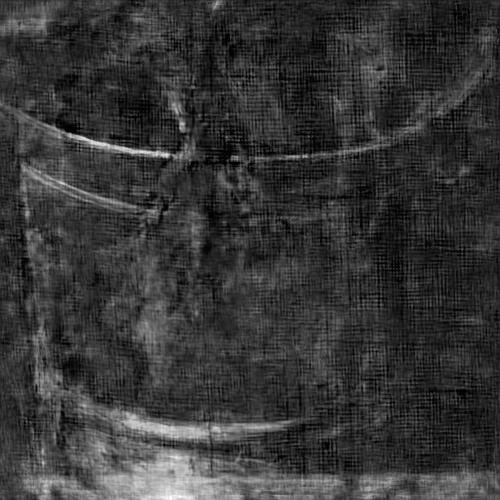}}
    \hfil
    \subfigure{\label{fig2a}\includegraphics[width=0.16\textwidth]{S_x2.jpg}}
    \hfil
    \subfigure{\label{fig2b}\includegraphics[width=0.16\textwidth]{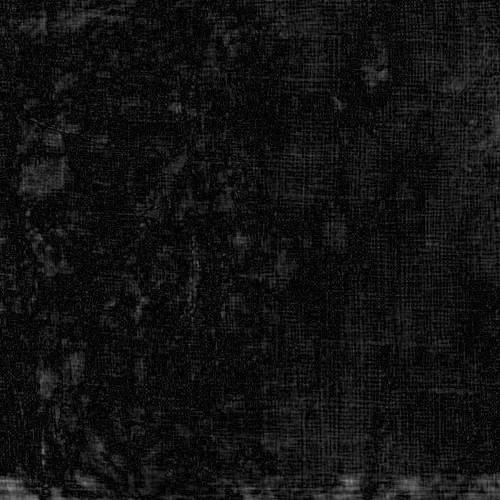}}
    
    \subfigure{\label{fig2a}\includegraphics[width=0.16\textwidth]{S_x1_pre_200.jpg}}
    \hfil
    \subfigure{\label{fig2a}\includegraphics[width=0.16\textwidth]{S_x1.jpg}}
    \hfil
    \subfigure{\label{fig2b}\includegraphics[width=0.16\textwidth]{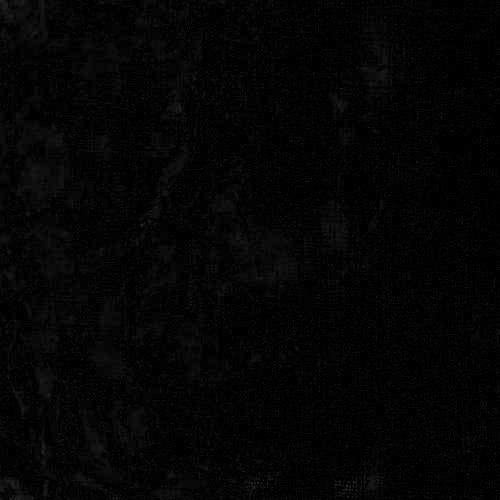}}
    \hfil
    \subfigure{\label{fig2a}\includegraphics[width=0.16\textwidth]{S_x2_pre_200.jpg}}
    \hfil
    \subfigure{\label{fig2a}\includegraphics[width=0.16\textwidth]{S_x2.jpg}}
    \hfil
    \subfigure{\label{fig2b}\includegraphics[width=0.16\textwidth]{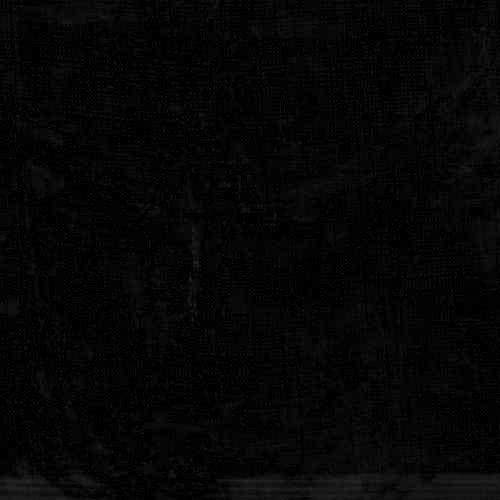}}
    \caption{Comparison of different X-ray separation methods. Rows 1 and 2 correspond to the results yielded by the method in \cite{IS2} and our proposed method, respectively. Columns 1 and 4 correspond to the reconstructed X-ray images, Columns 2 and 5 correspond to the true X-ray images, and Columns 3 and 5 correspond to the error maps.}\label{Figs6}
\end{figure*}

In Fig. \ref{Figs6}, we compare our proposed image separation algorithm to a recently reported state-of-the-art one\cite{IS2}, demonstrating that the proposed algorithm produces much better separations than the algorithm in \cite{IS2}. In particular, the MSE associated with the reconstruction of the first X-ray image (column 1 and 2 in Fig. \ref{Figs6}) is 0.00062 with our method and 0.0016 with the method in \cite{IS2}; in turn, the MSE associated with the reconstruction of the second X-ray image (column 4 and 5 in Fig. \ref{Figs6}) is 0.00057 with our method and 0.0021 with the method in \cite{IS2}.

\subsection { Experiments with Real Mixed X-ray Data }

\subsubsection { Set-up }

\begin{figure}[h]
    \centering
    \subfigure[]{\label{fig2a}\includegraphics[width=0.157\textwidth]{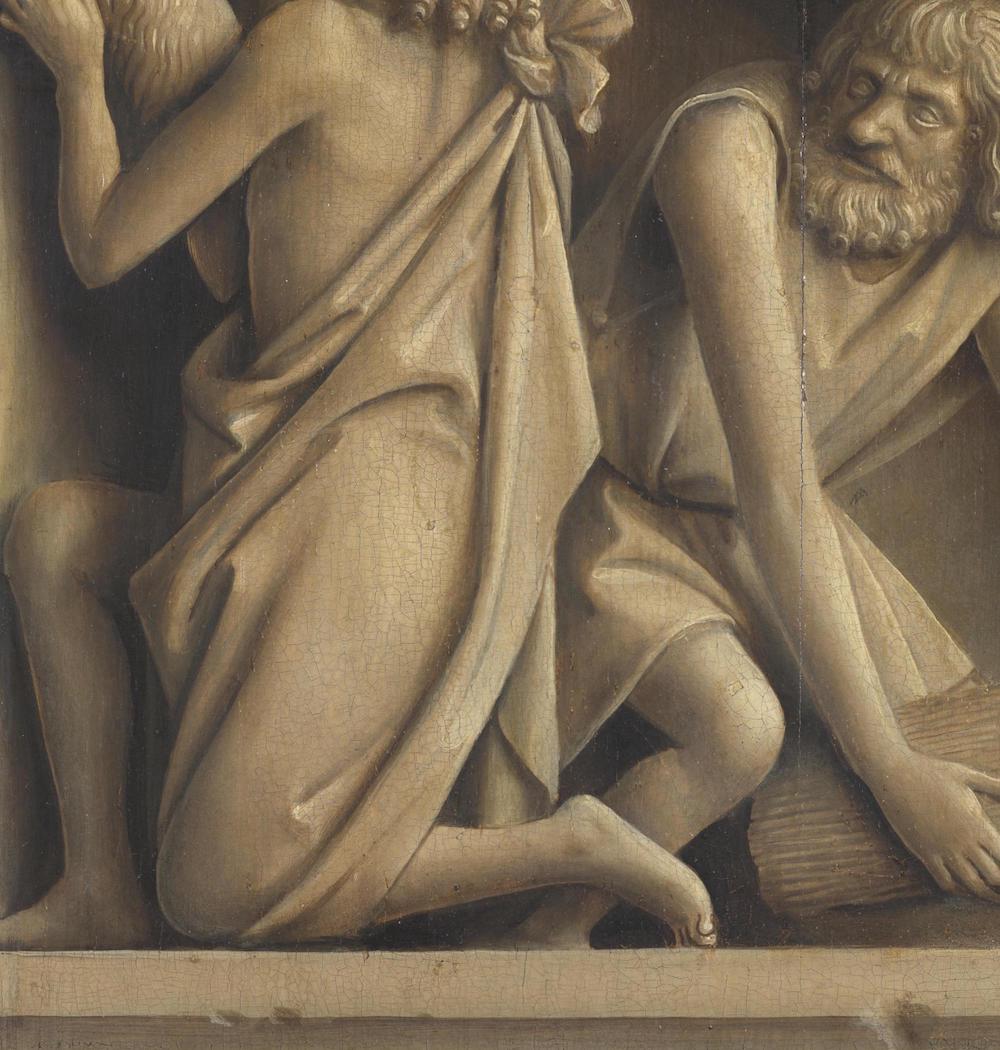}}
    \hfil
    \subfigure[]{\label{fig2b}\includegraphics[width=0.157\textwidth]{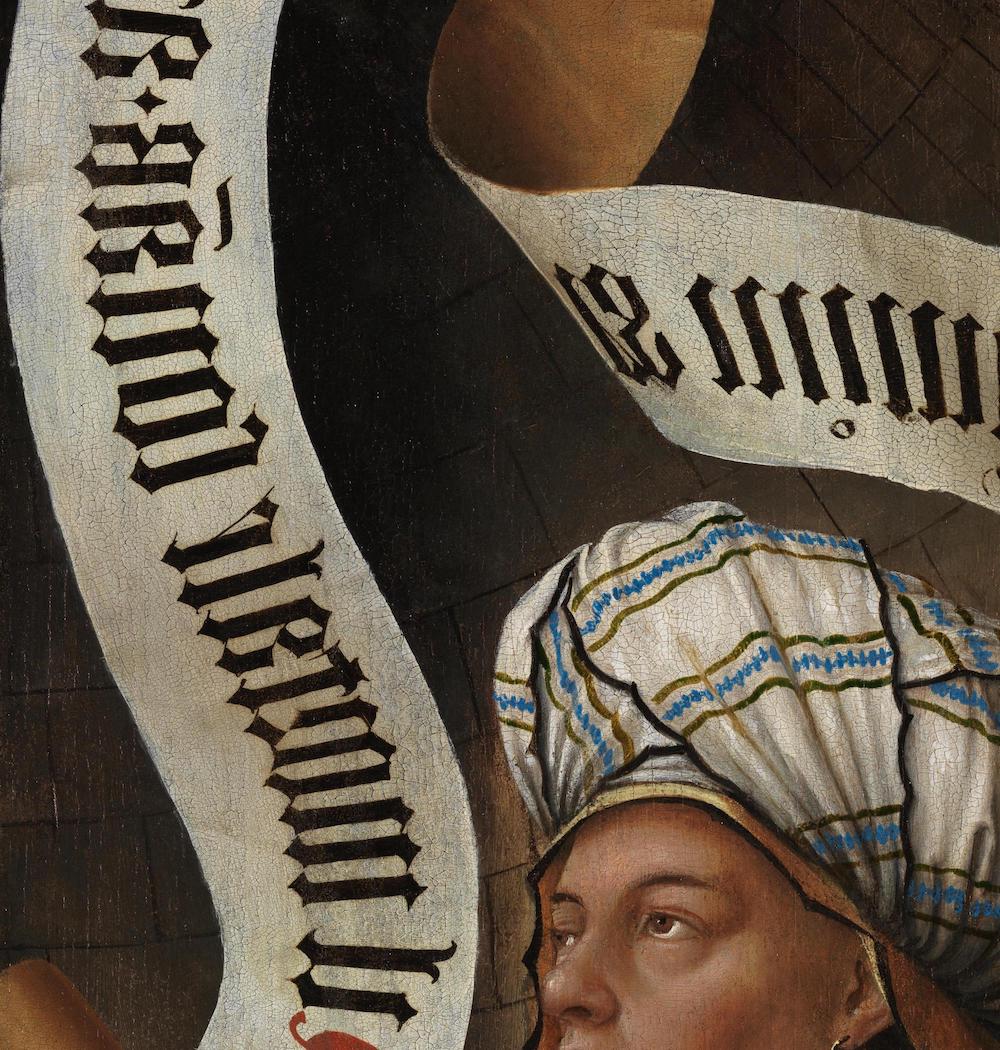}}
    \hfil
    \subfigure[]{\label{fig2c}\includegraphics[width=0.157\textwidth]{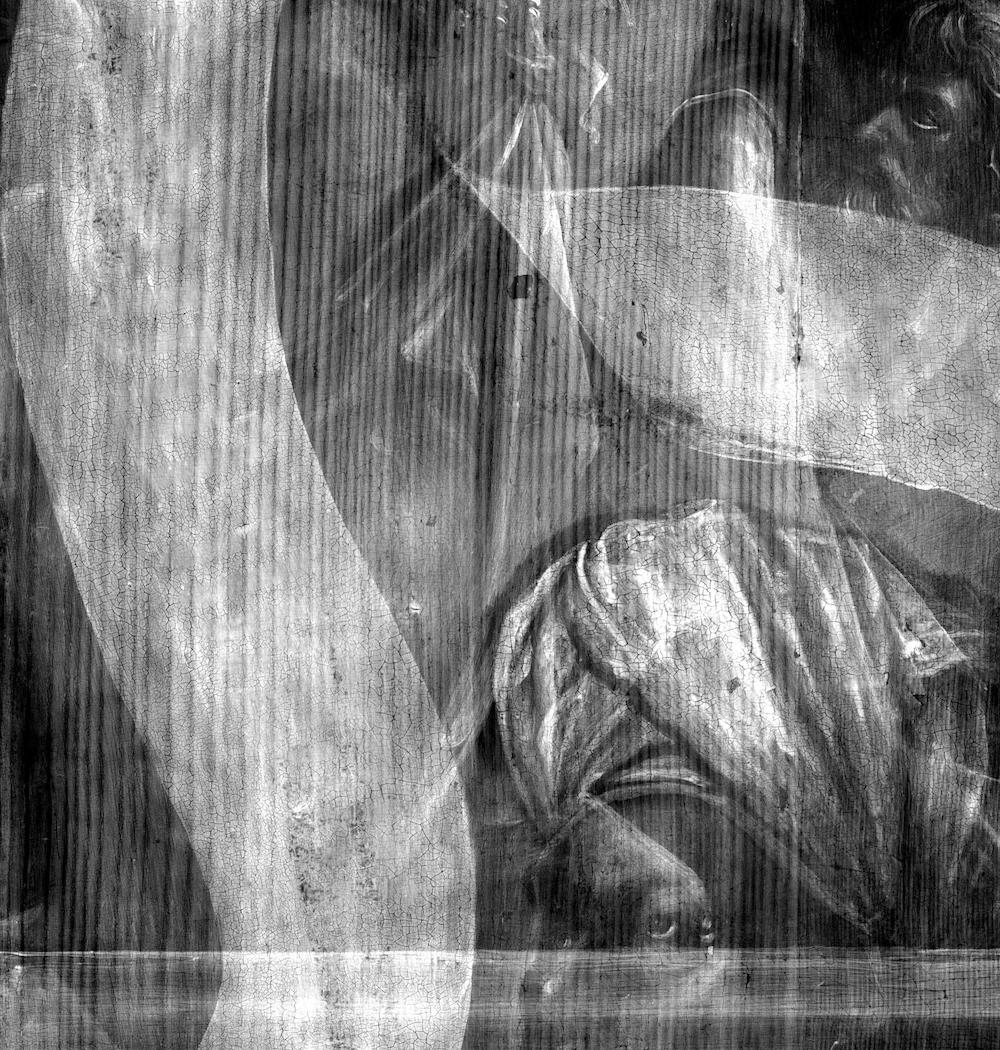}}
    \caption{Images used for real data experiments. (a). RGB image of the front side. (b). RGB image of the back of the panel (image reversed). (c). mixed X-ray image.}\label{figr1}
\end{figure}

In this experiment, we use a small area of size $1000 \times 1000$ pixels from the \textsl{Ghent Altarpiece} (see Fig. \ref{figr1}).

The previous procedure was again followed: the two RGB images and the corresponding mixed X-ray image were  divided into patches of size 64$\times$64 pixels with 56 pixels over- lap (both in the horizontal and vertical direction), resulting in 13,924 patches. The patches associated with the mixed X-ray were separated independently. The various patches associated with the individual separated X-rays were finally put together by placing various patches in the original order and averaging the overlap portions. All patches were also used in the training of the auto-encoders by randomly shuffling their order.

Once again, we adopted the hyper-parameter values $\lambda_1 = 3$, $\lambda_2 = 5$, $\lambda_3 = 2$, and $\lambda_4 = 0.3$.

\subsubsection{Results}

\begin{figure*}[h]
    \centering
    \subfigure{\label{fig2a}\includegraphics[width=0.32\textwidth]{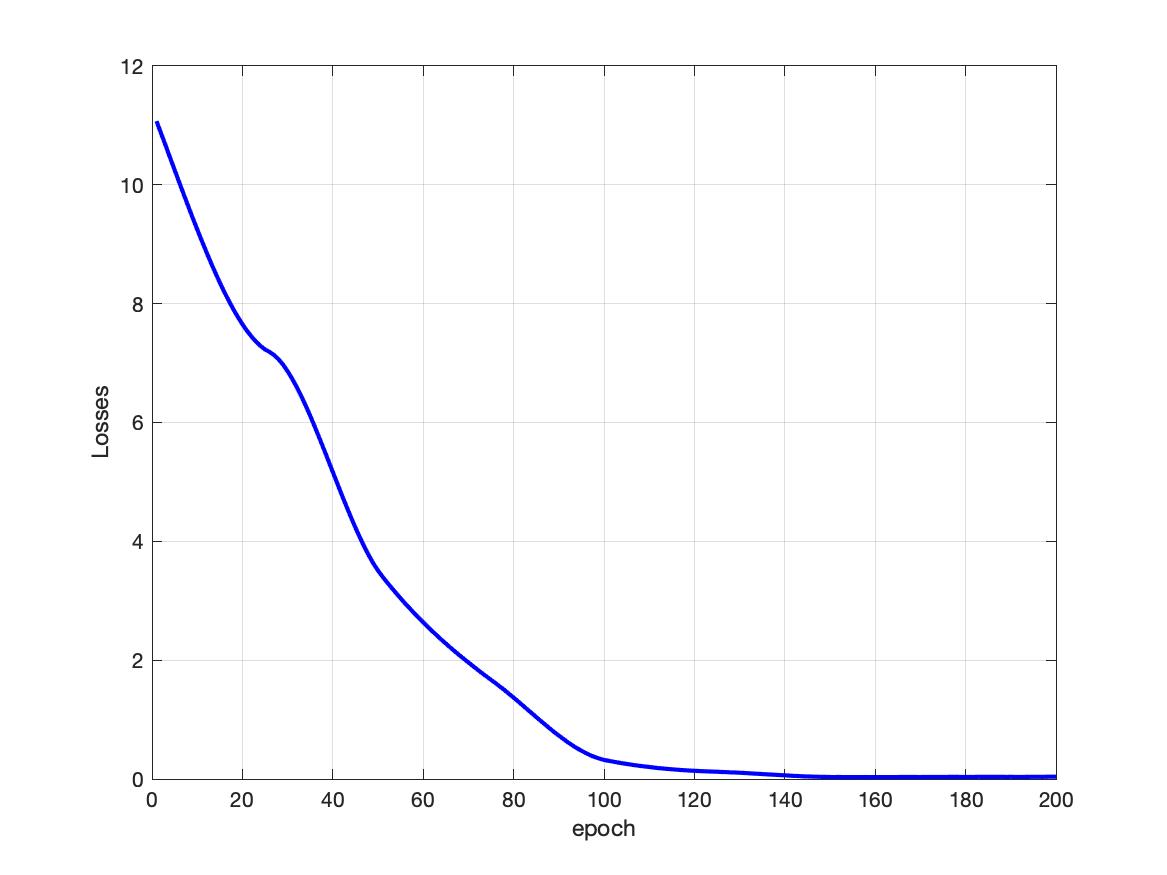}}
    \hfil
    \subfigure{\label{fig2b}\includegraphics[width=0.32\textwidth]{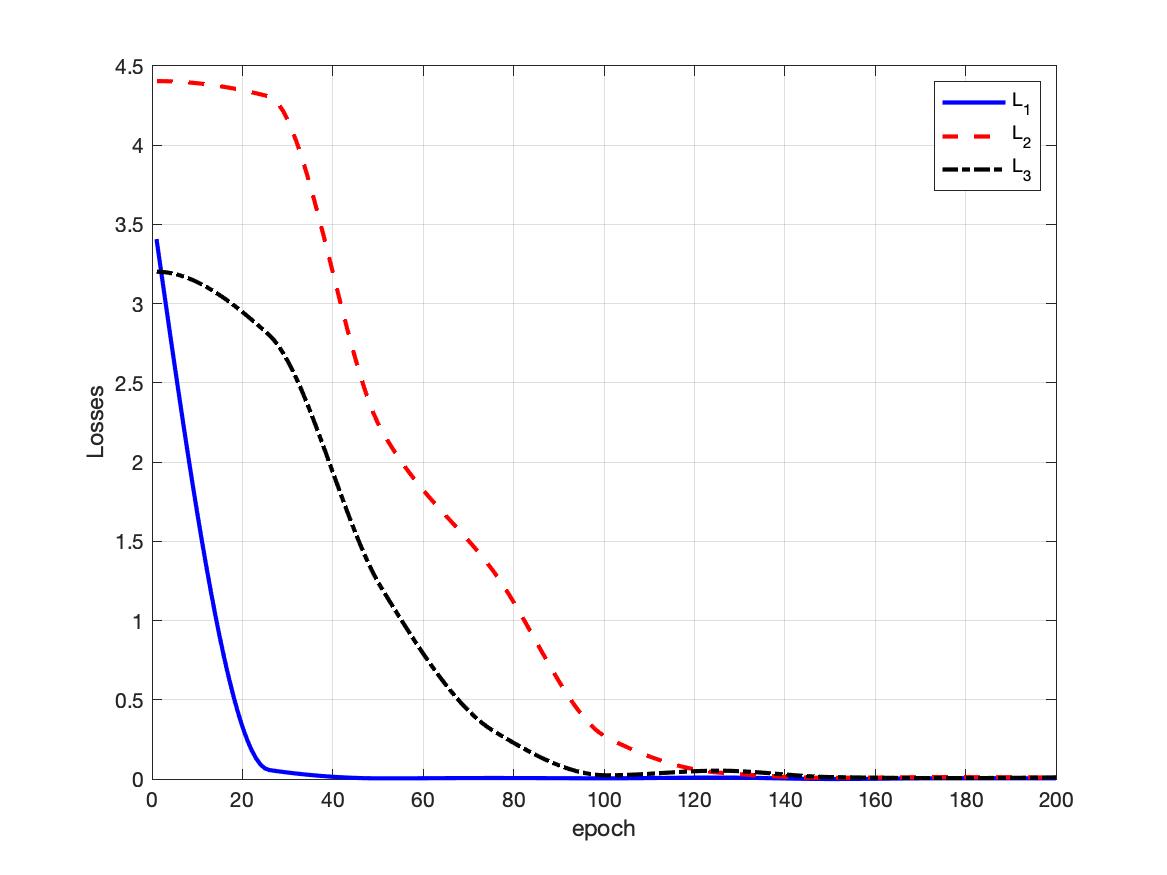}}
    \hfil
    \subfigure{\label{fig2c}\includegraphics[width=0.32\textwidth]{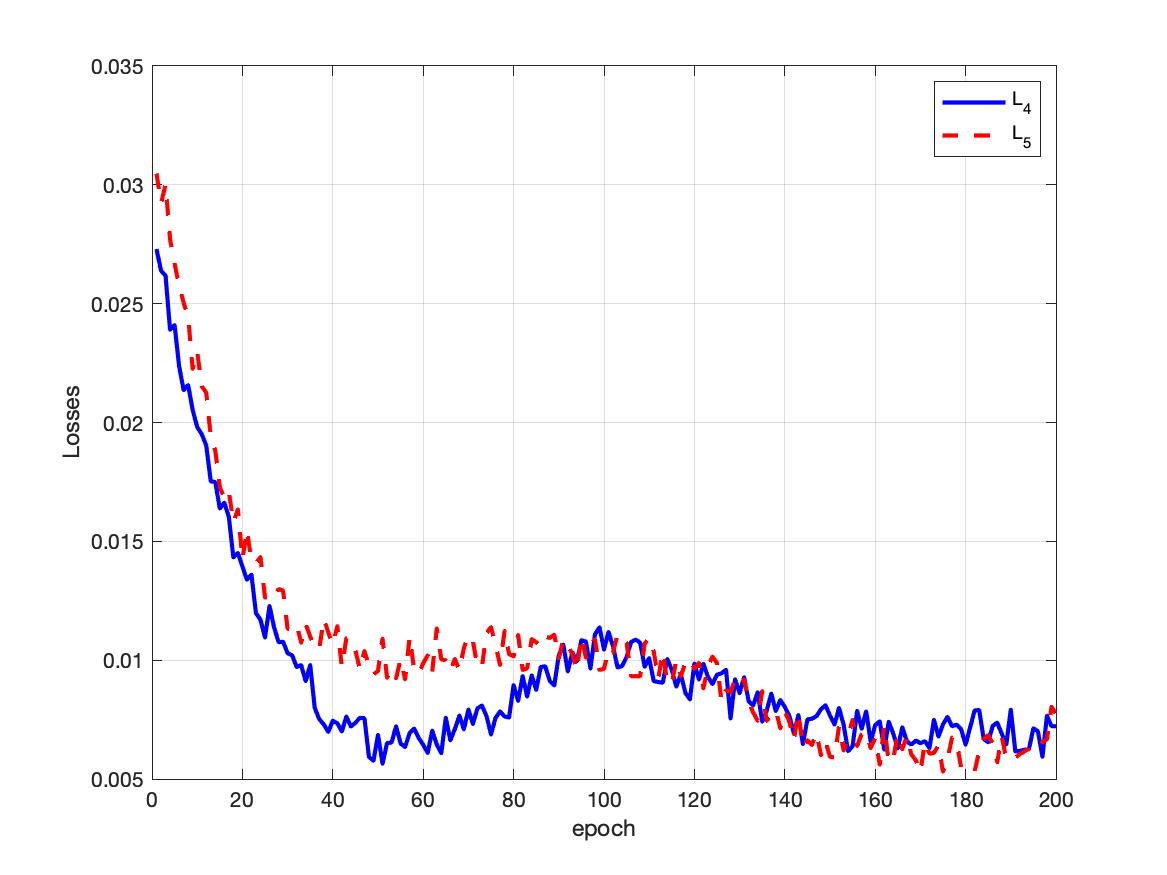}}
    \caption{Losses vs. number of epochs on real data experiments. (a). $L_{total}$. (b). $L_{1}$, $L_{2}$ and $L_{3}$. (a). $L_{4}$ and $L_{5}$.}\label{figr2}
\end{figure*}

\begin{figure*}[h]
    \centering
    \subfigure{\label{fig2a}\includegraphics[width=0.136\textwidth]{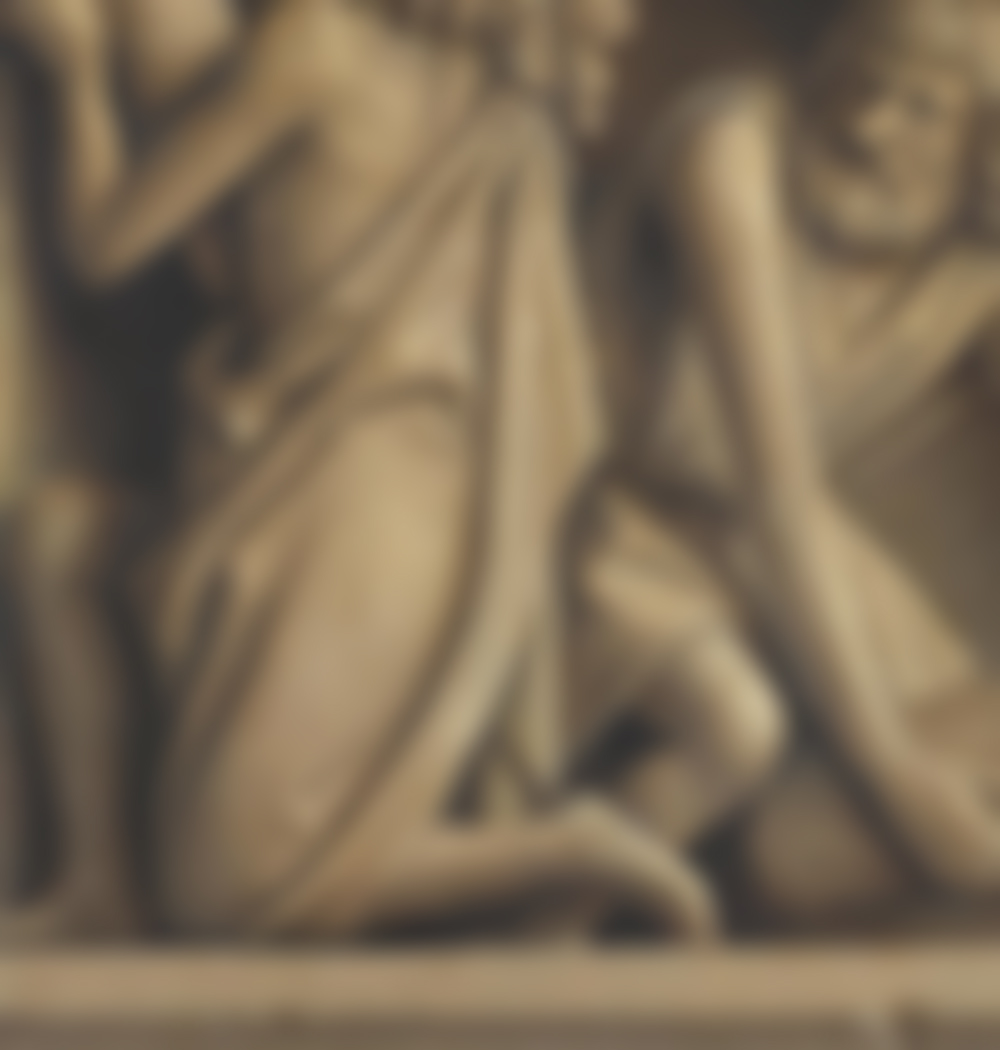}}
    \hfil
    \subfigure{\label{fig2b}\includegraphics[width=0.136\textwidth]{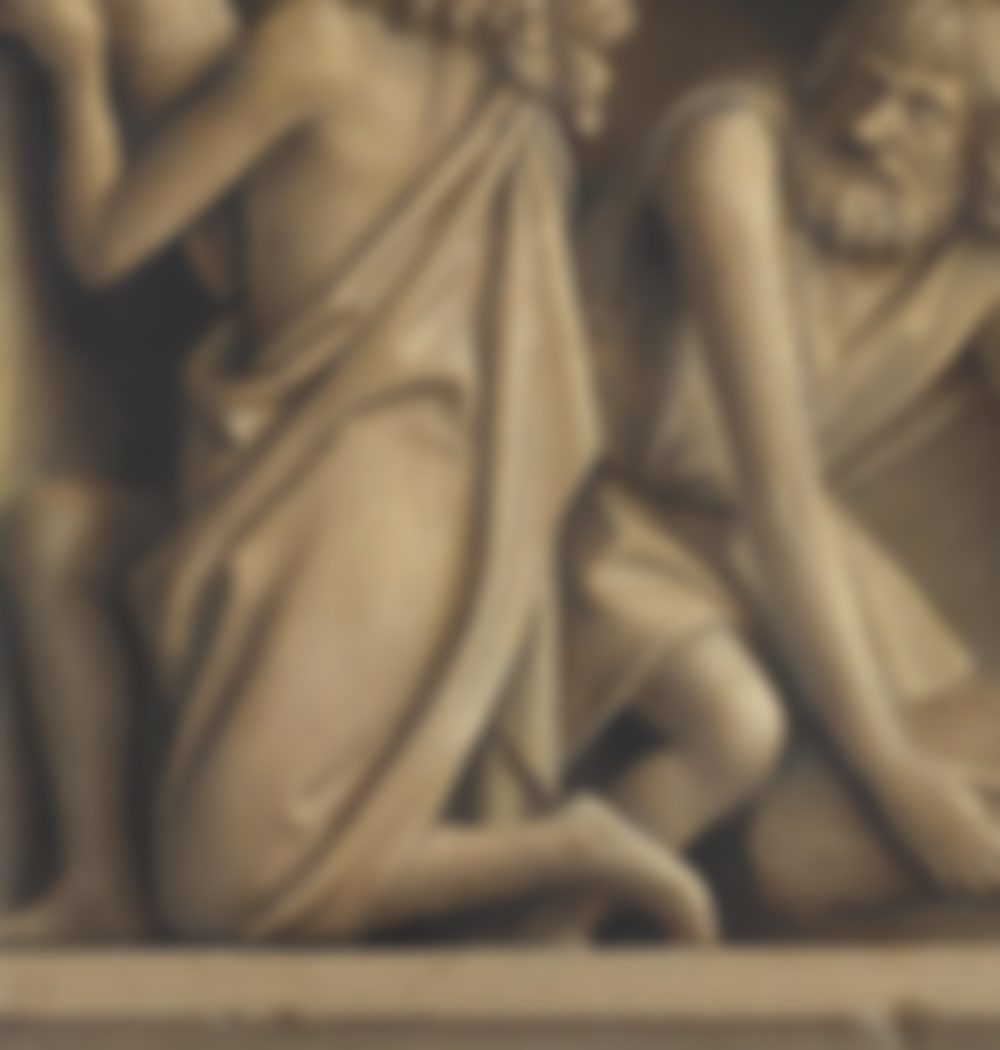}}
    \hfil
    \subfigure{\label{fig2c}\includegraphics[width=0.136\textwidth]{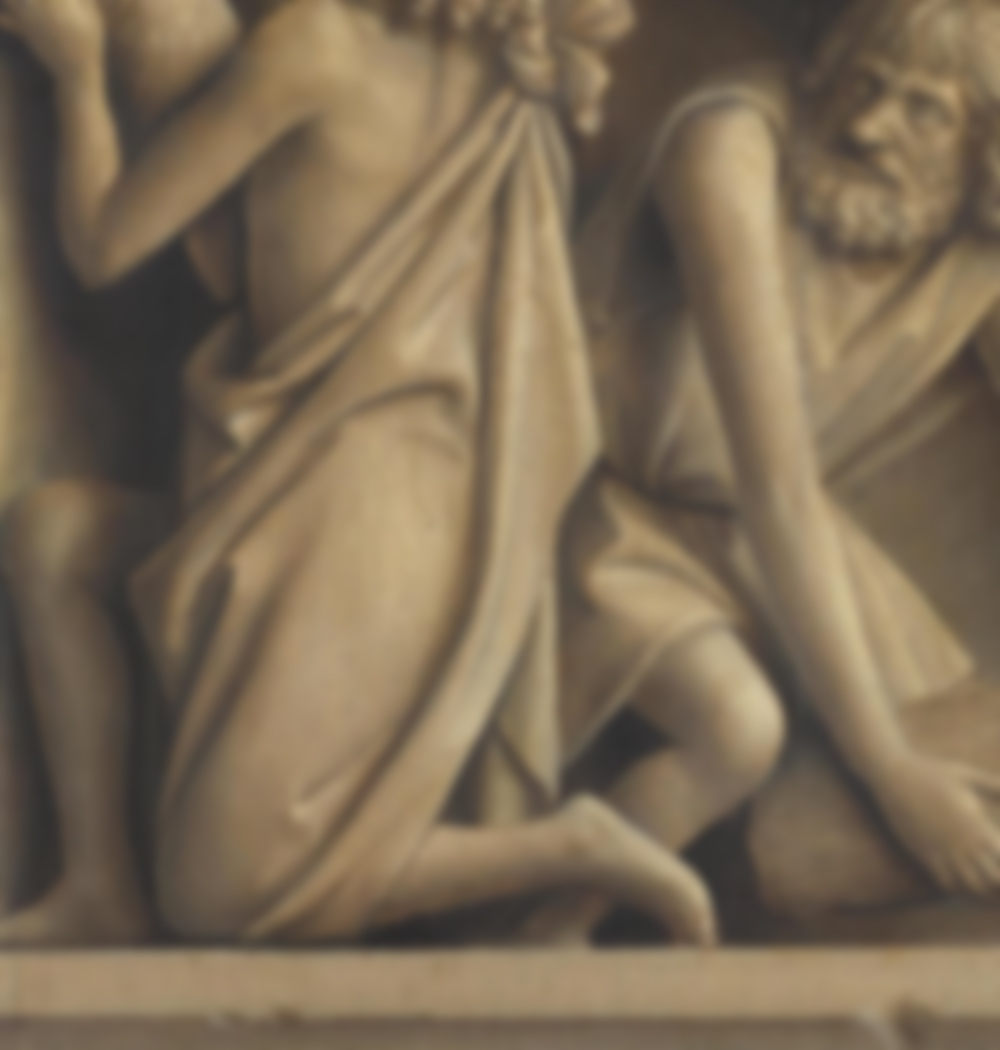}}
    \hfil
    \subfigure{\label{fig2a}\includegraphics[width=0.136\textwidth]{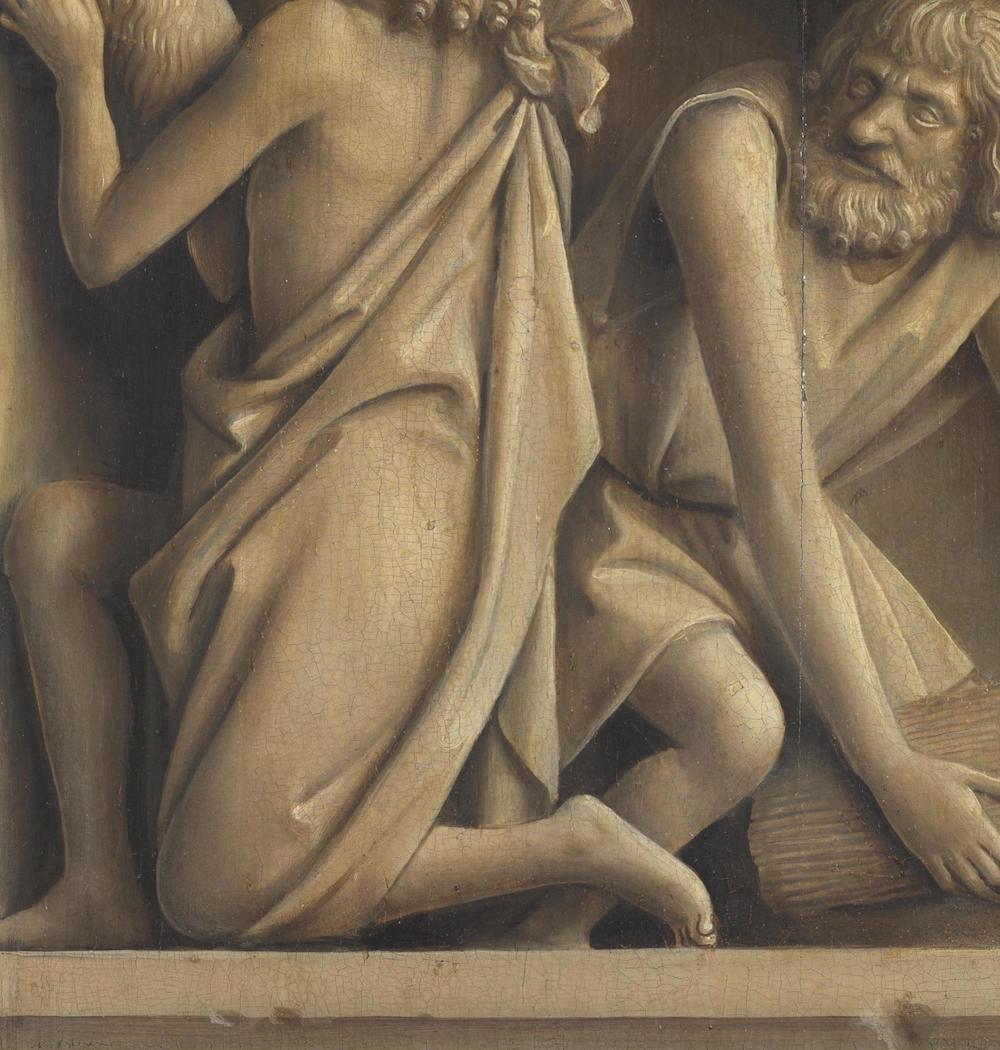}}
    \hfil
    \subfigure{\label{fig2b}\includegraphics[width=0.136\textwidth]{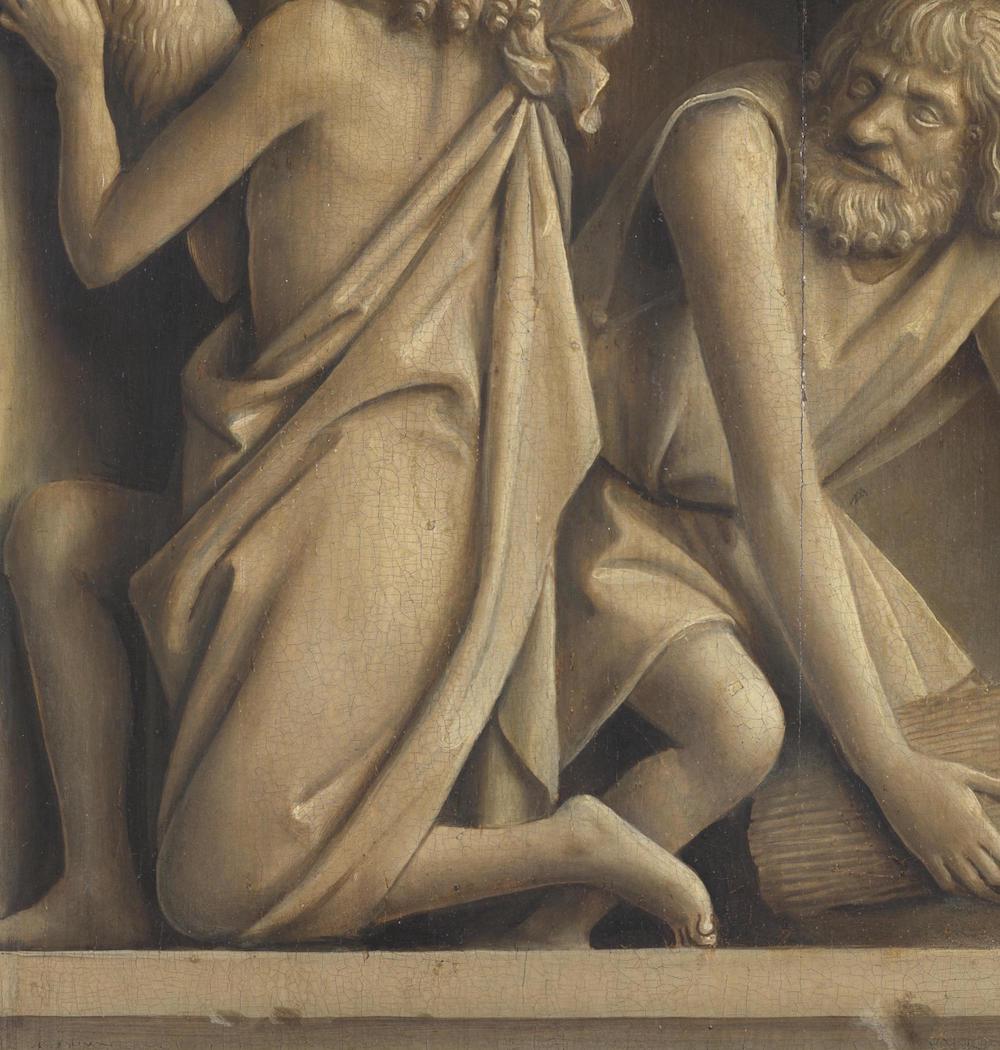}}
    \hfil
    \subfigure{\label{fig2c}\includegraphics[width=0.136\textwidth]{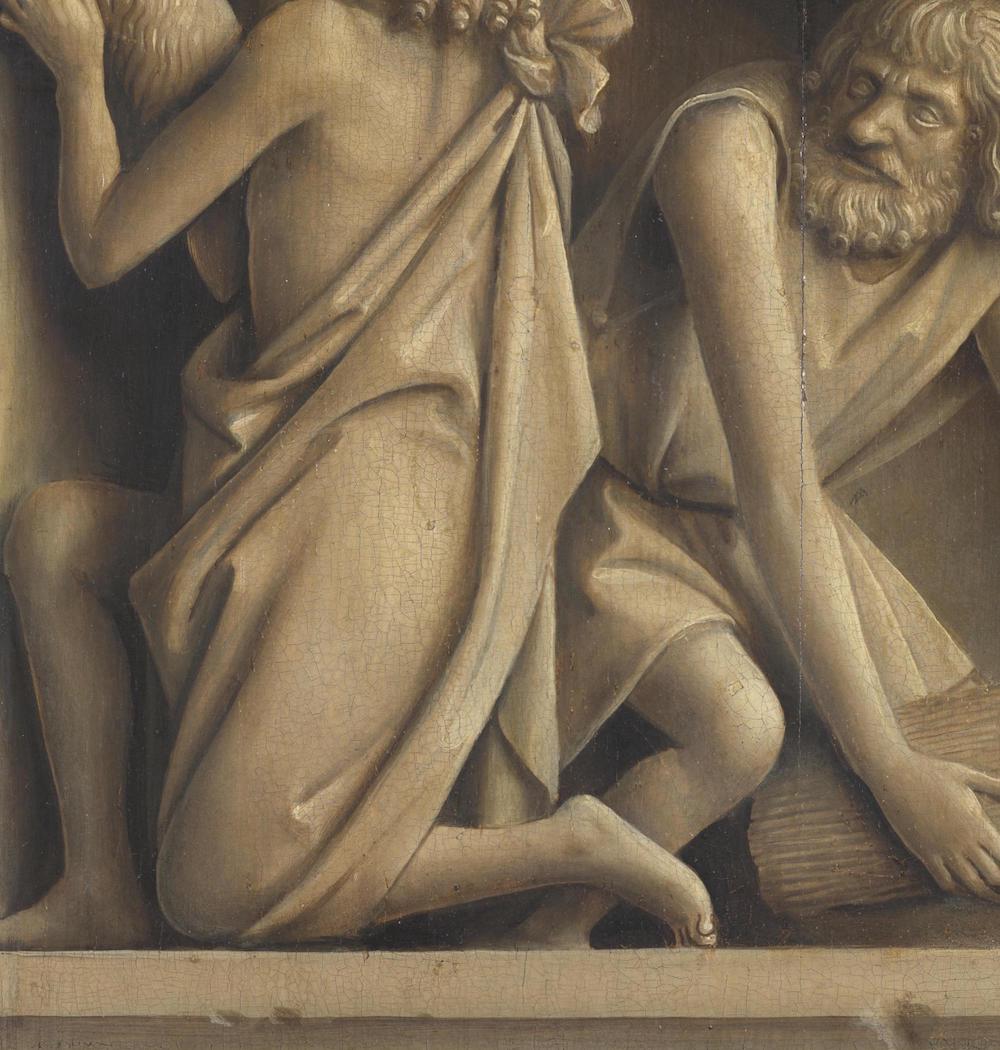}}
    \hfil
    \subfigure{\label{fig2c}\includegraphics[width=0.136\textwidth]{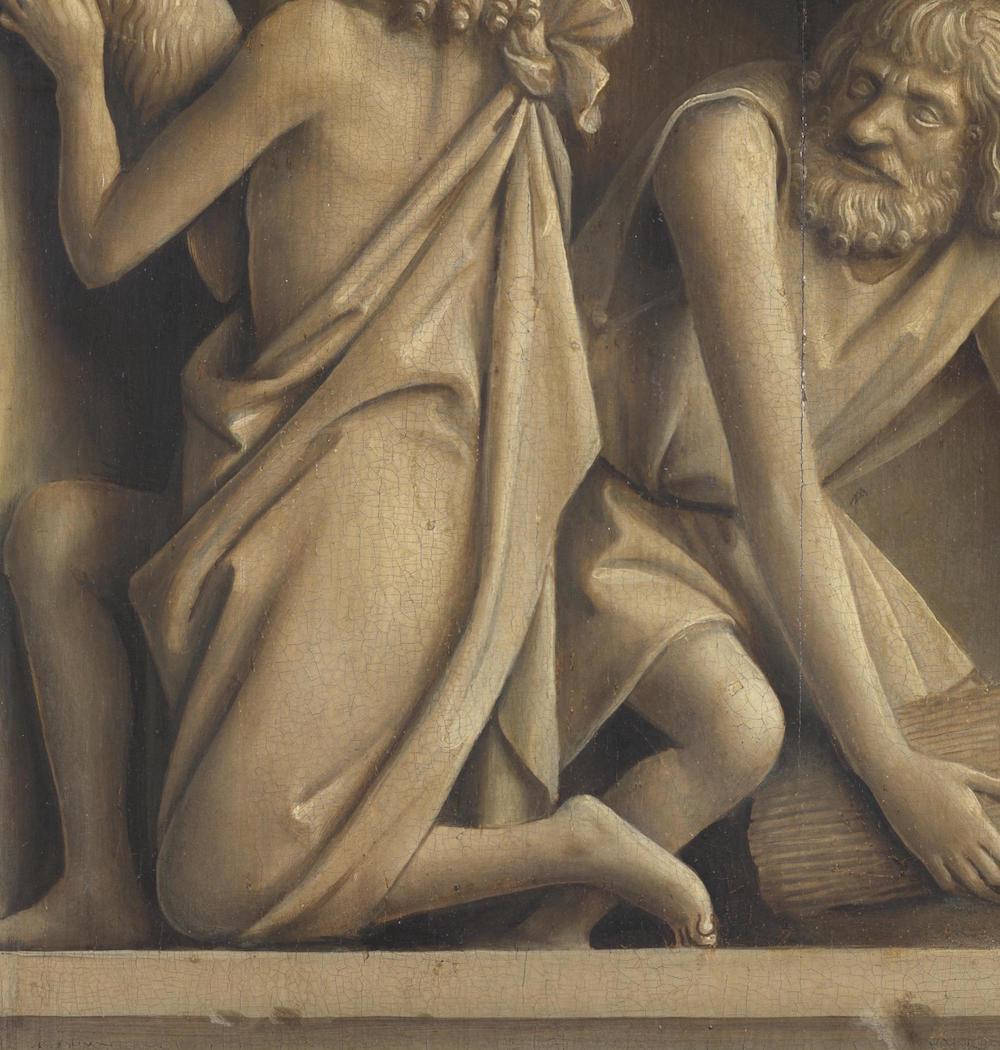}}    
    
    \subfigure{\label{fig2a}\includegraphics[width=0.136\textwidth]{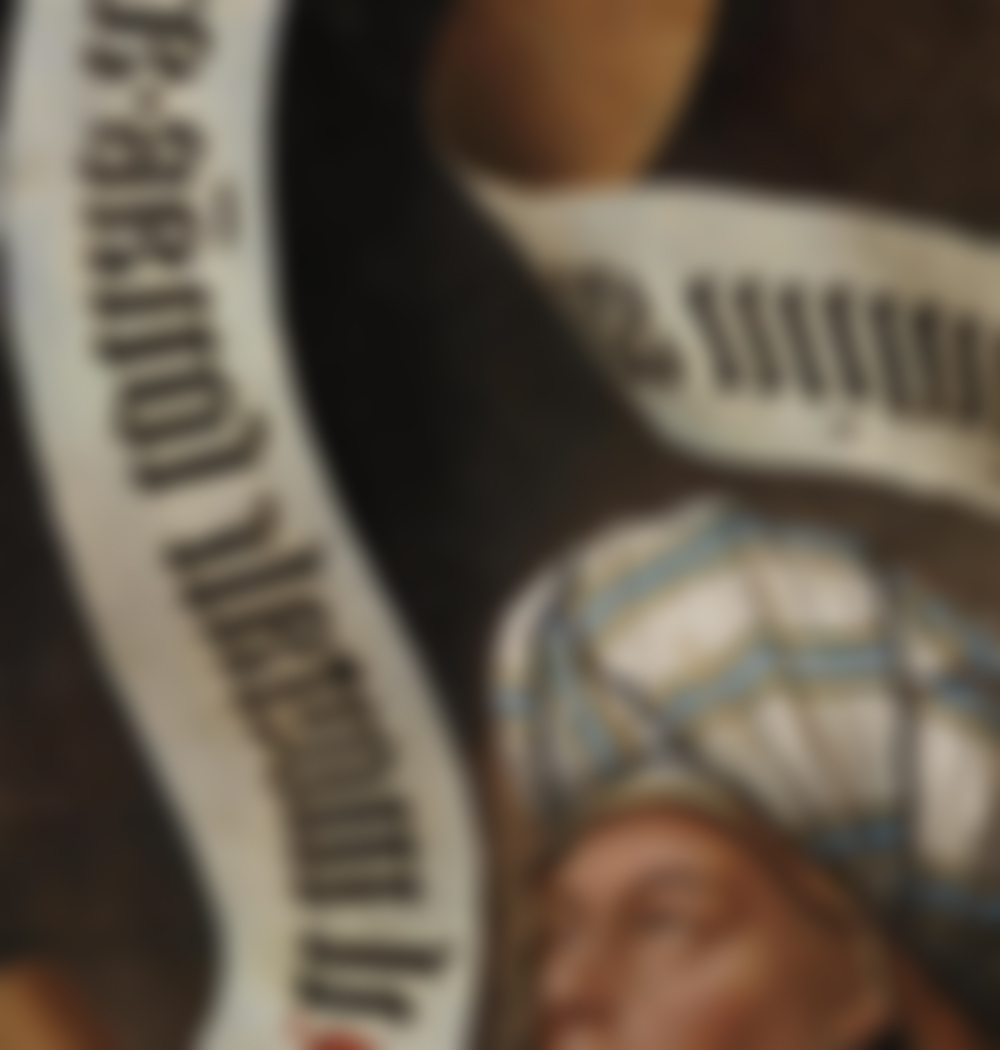}}
    \hfil
    \subfigure{\label{fig2b}\includegraphics[width=0.136\textwidth]{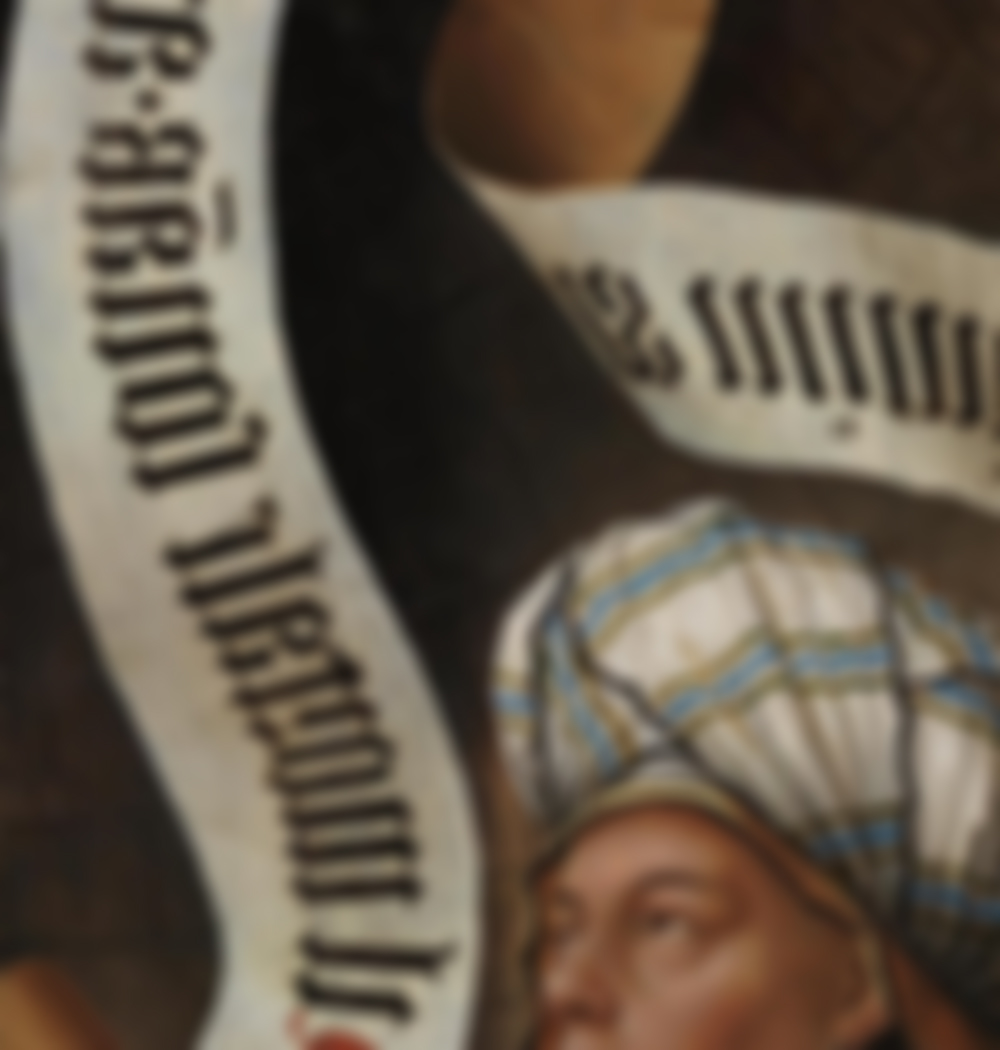}}
    \hfil
    \subfigure{\label{fig2c}\includegraphics[width=0.136\textwidth]{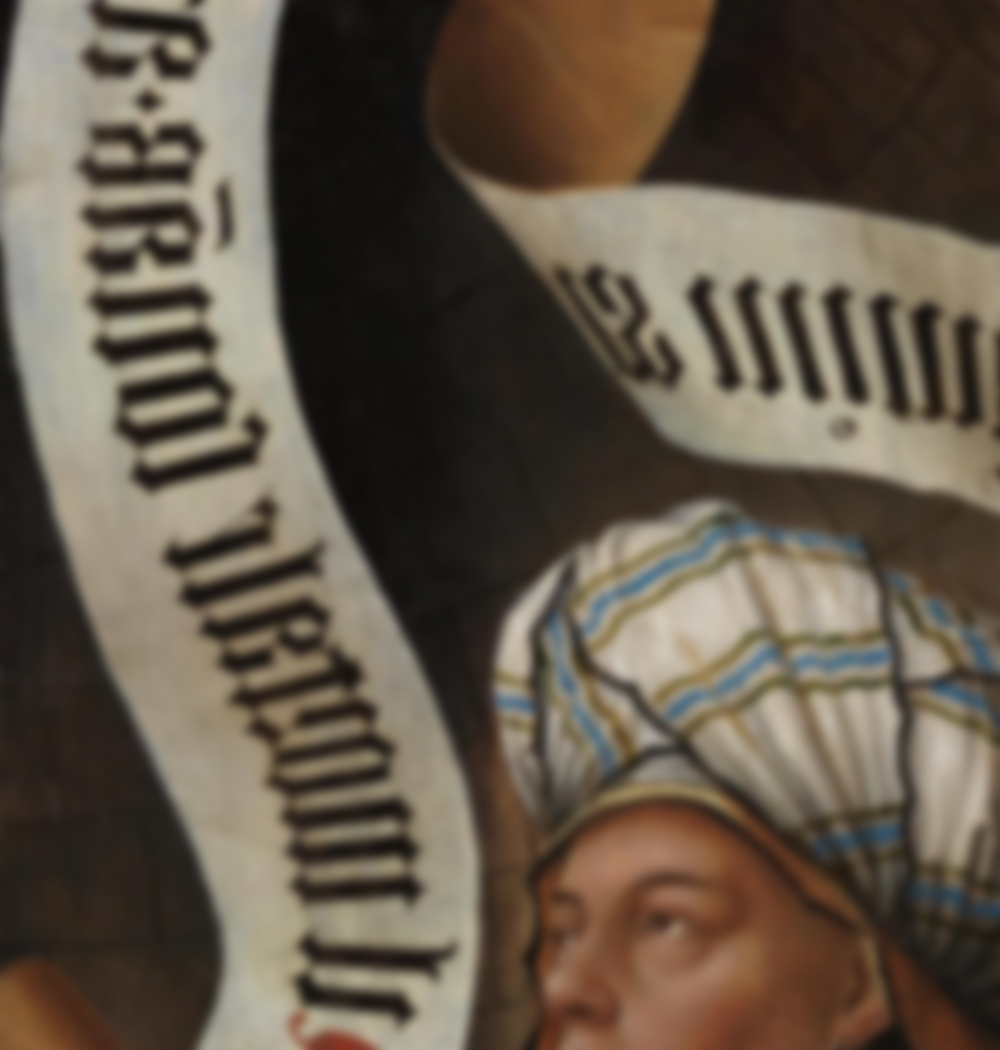}}
    \hfil
    \subfigure{\label{fig2a}\includegraphics[width=0.136\textwidth]{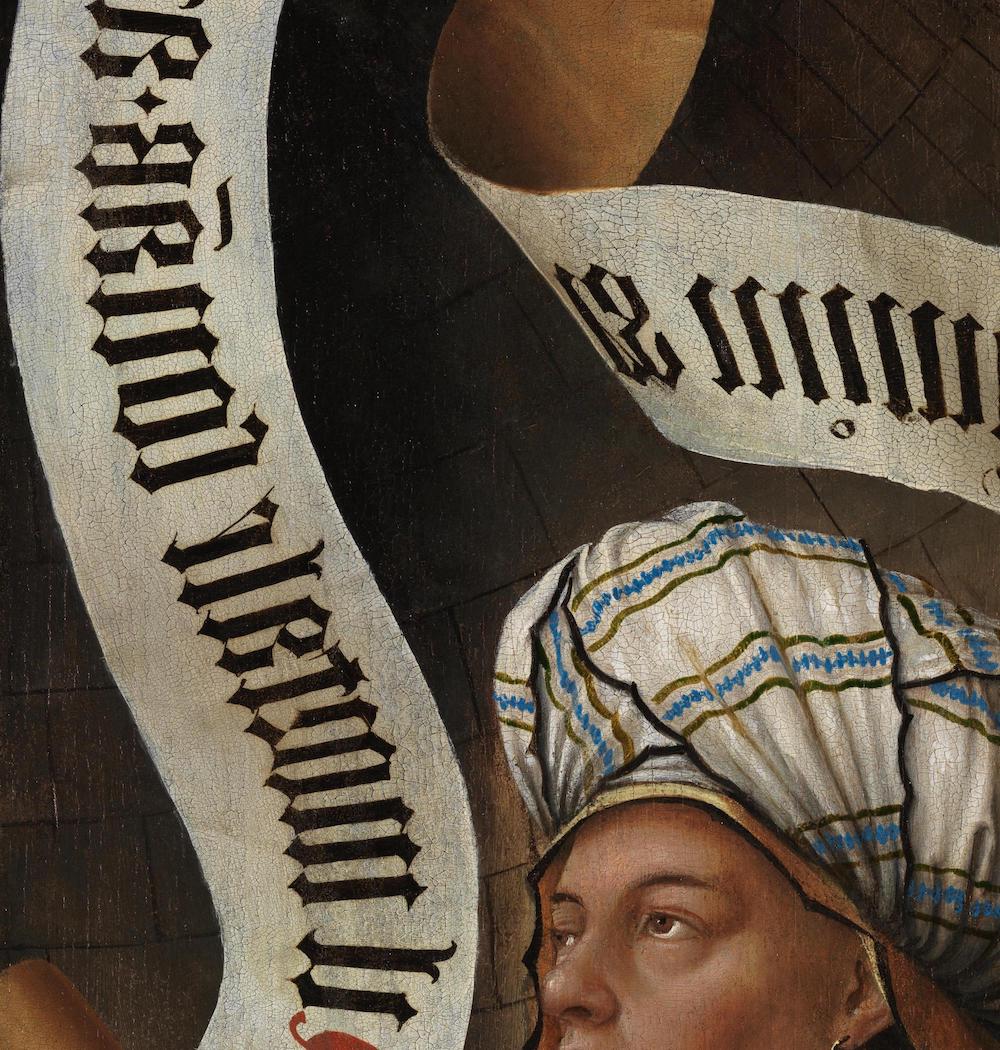}}
    \hfil
    \subfigure{\label{fig2b}\includegraphics[width=0.136\textwidth]{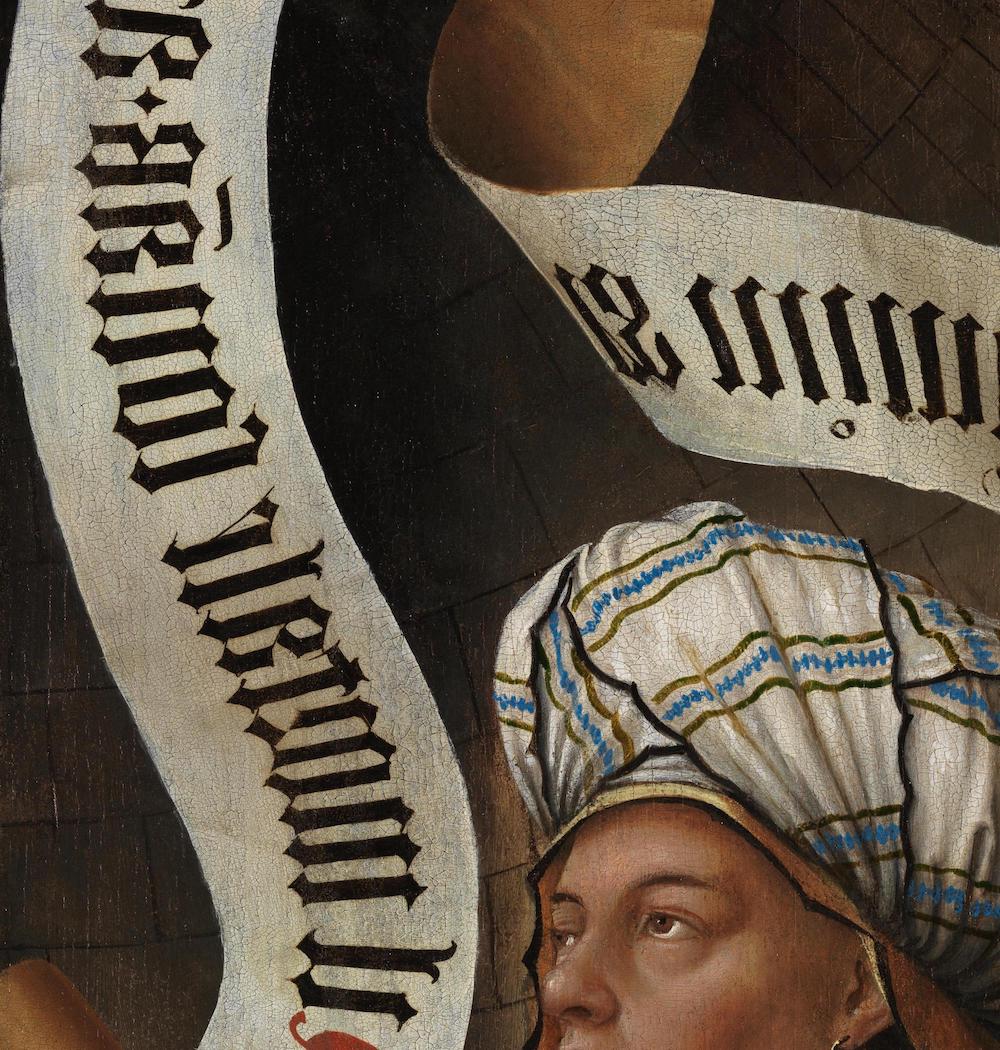}}
    \hfil
    \subfigure{\label{fig2c}\includegraphics[width=0.136\textwidth]{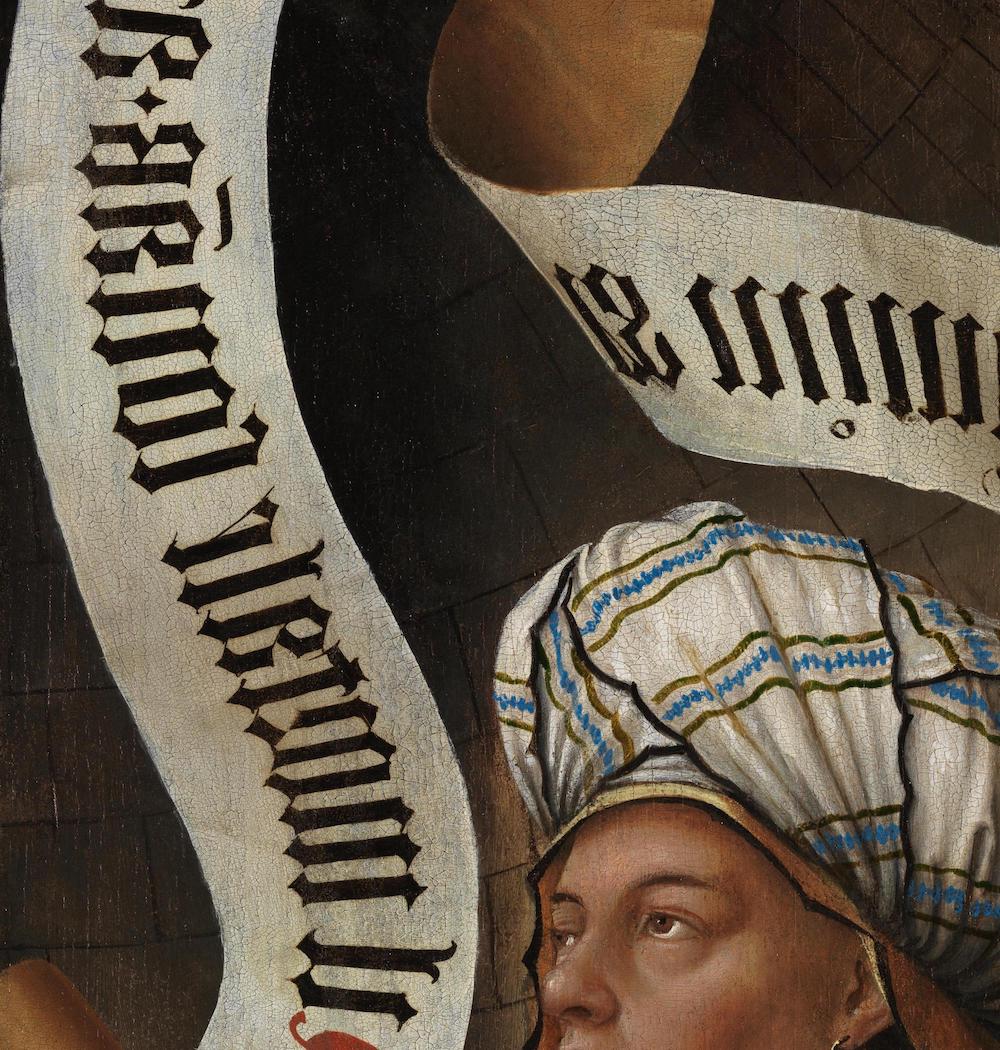}}
    \hfil
    \subfigure{\label{fig2c}\includegraphics[width=0.136\textwidth]{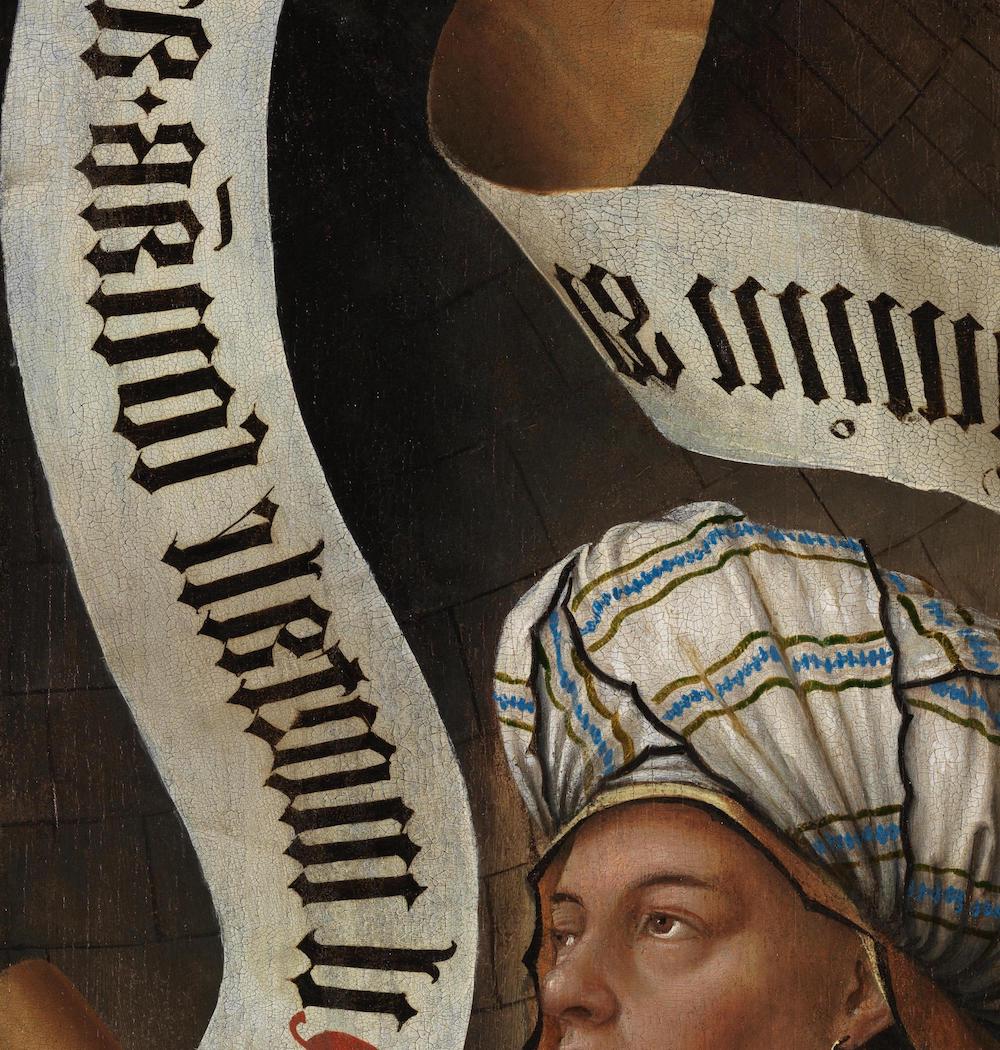}}    
    
    \subfigure{\label{fig2a}\includegraphics[width=0.136\textwidth]{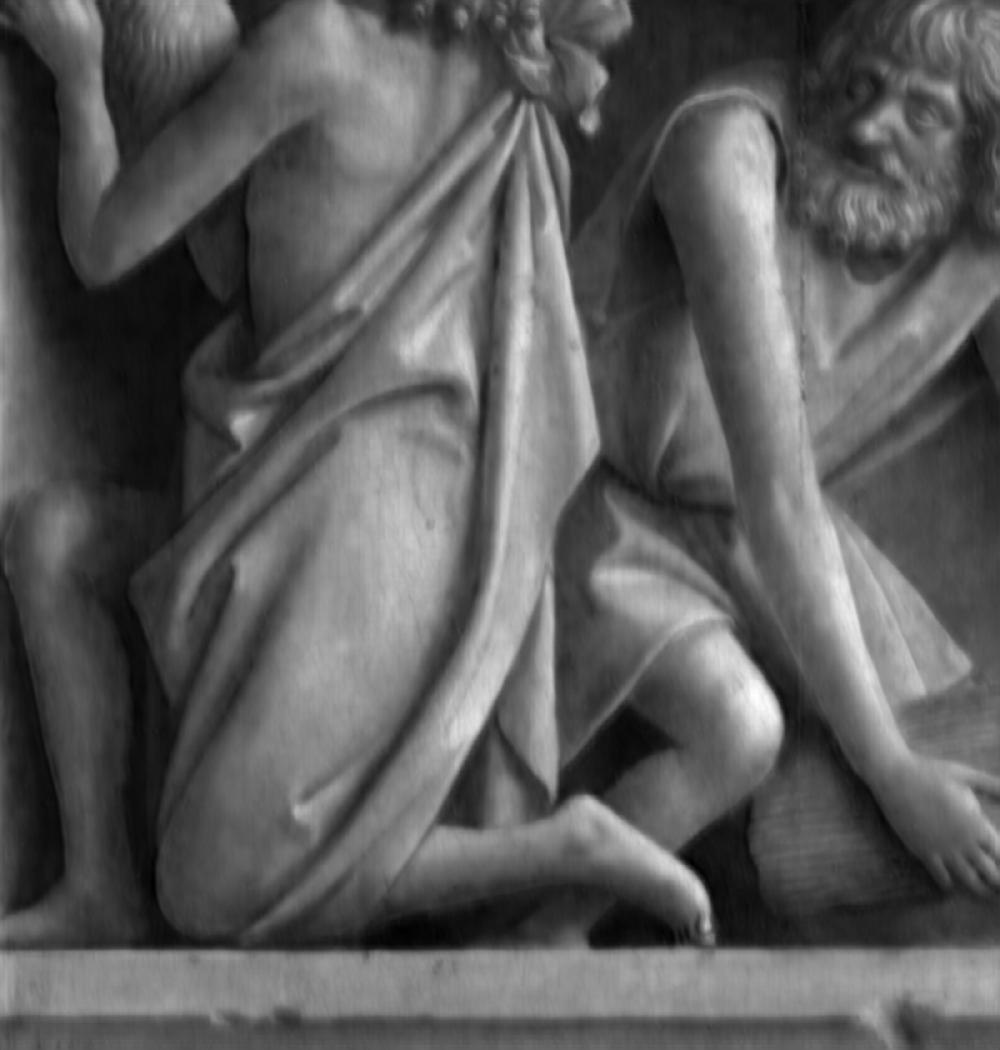}}
    \hfil
    \subfigure{\label{fig2b}\includegraphics[width=0.136\textwidth]{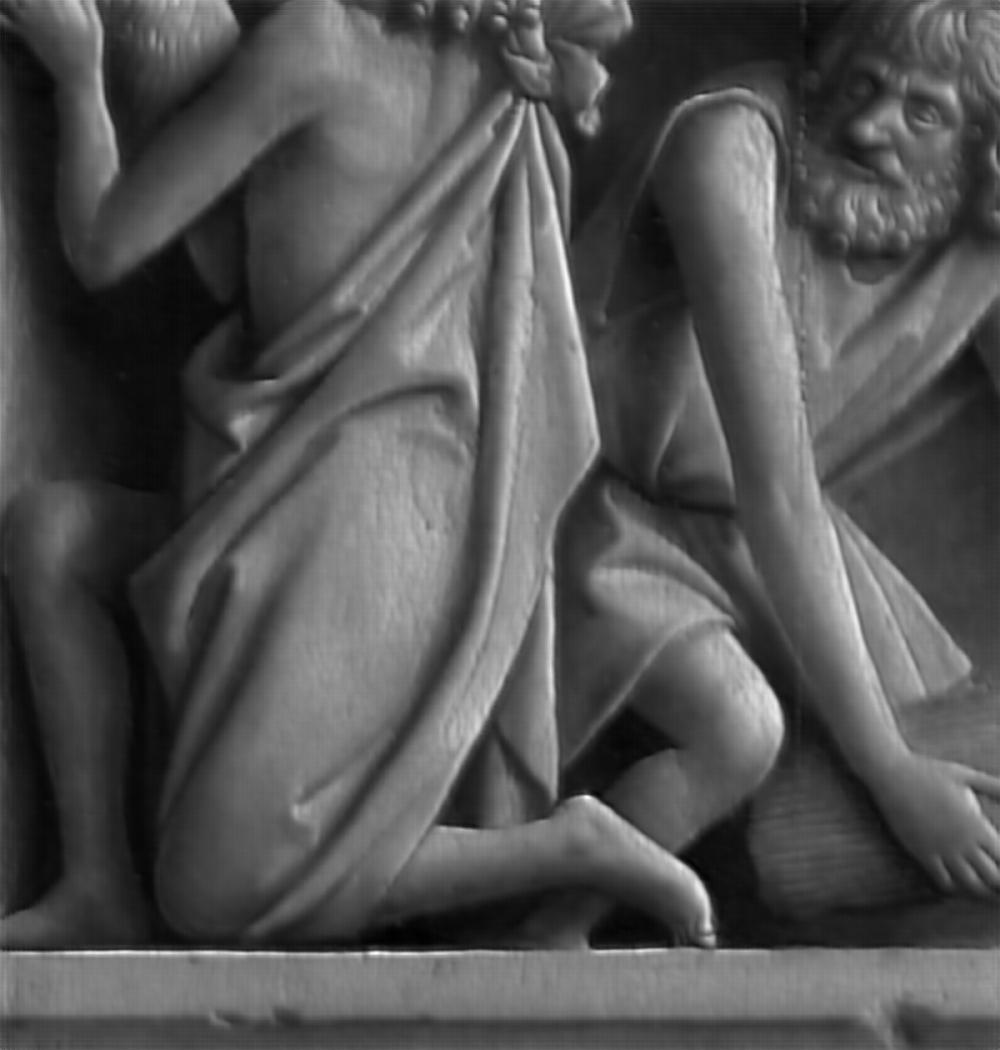}}
    \hfil
    \subfigure{\label{fig2c}\includegraphics[width=0.136\textwidth]{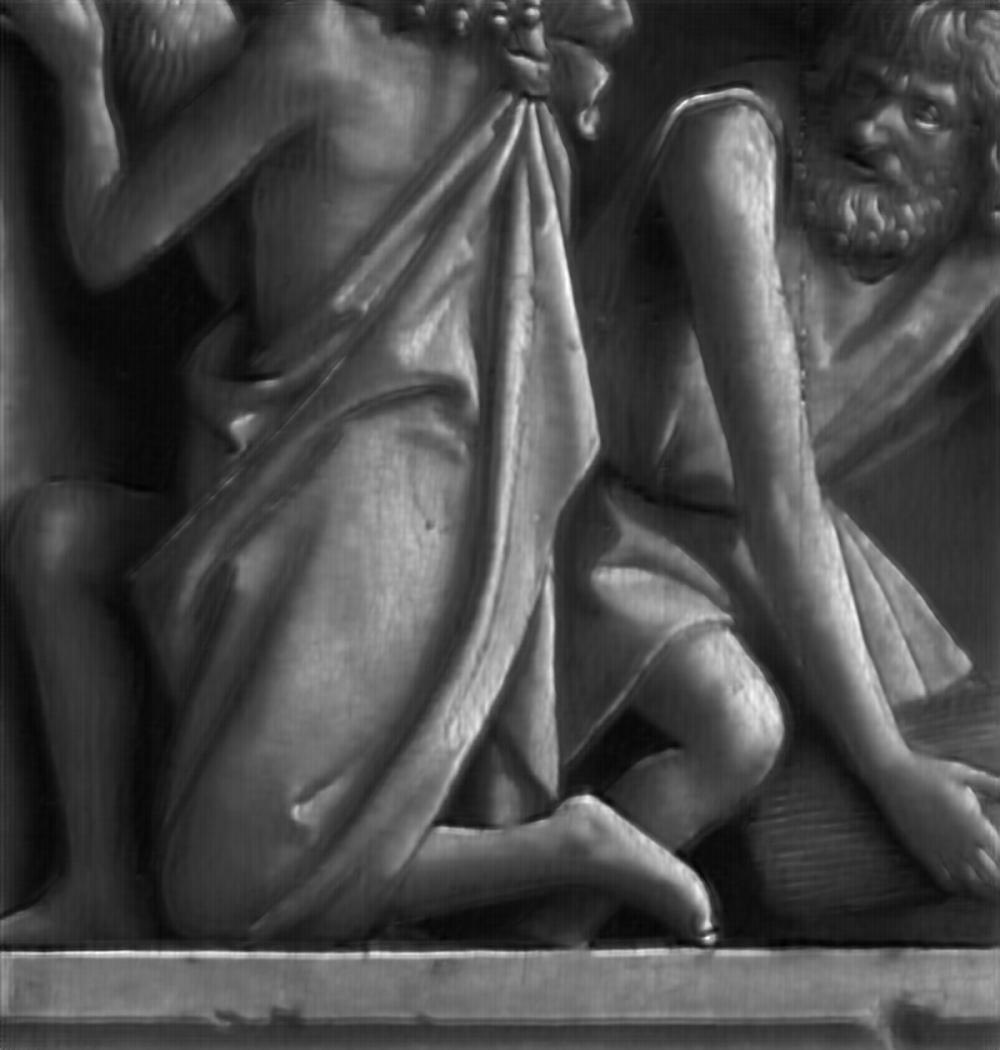}}
    \hfil
    \subfigure{\label{fig2a}\includegraphics[width=0.136\textwidth]{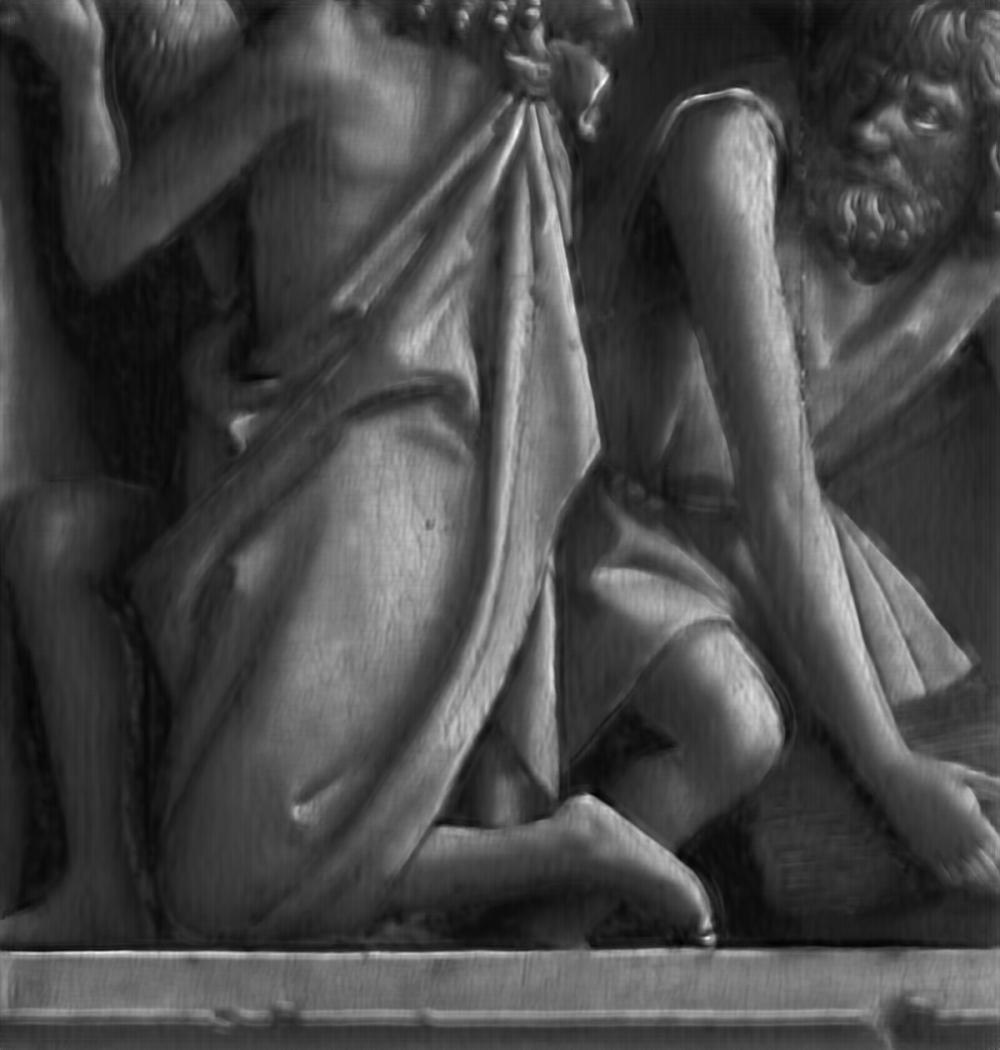}}
    \hfil
    \subfigure{\label{fig2b}\includegraphics[width=0.136\textwidth]{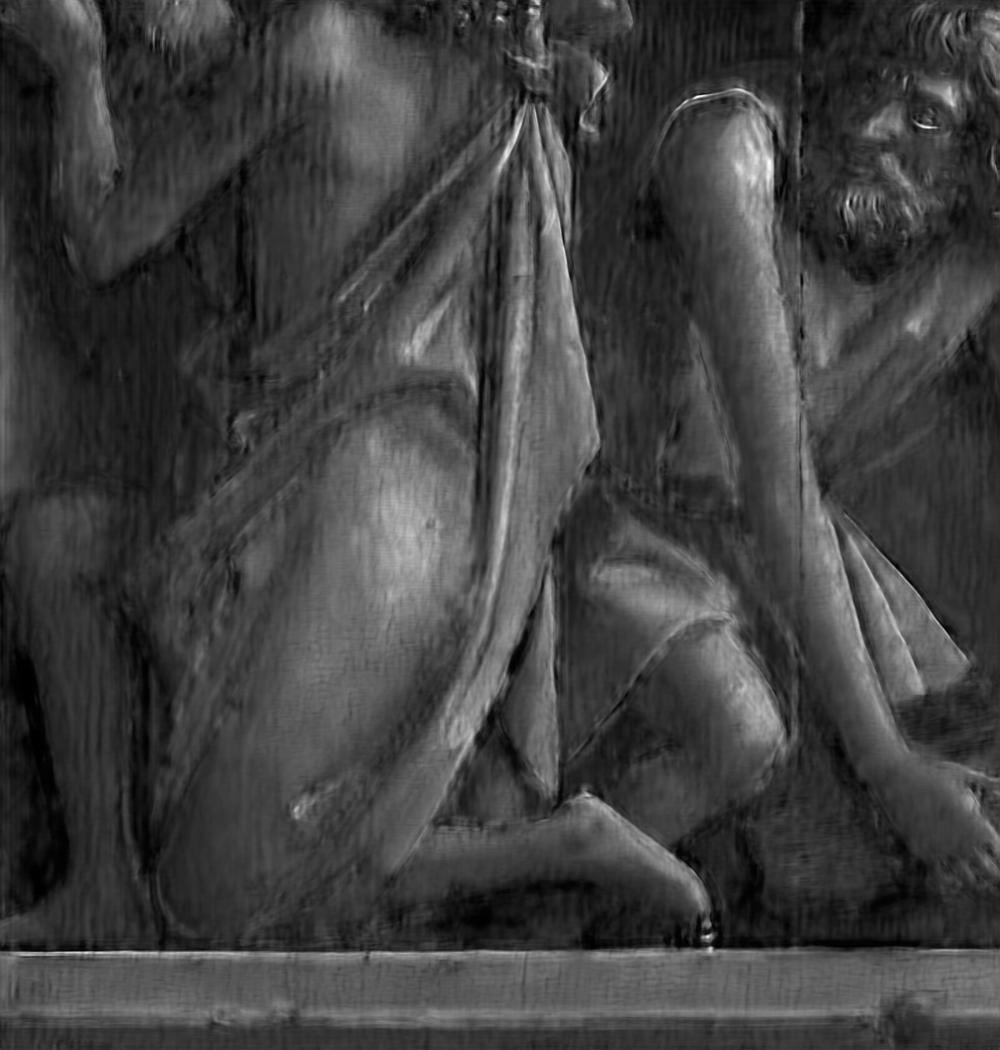}}
    \hfil
    \subfigure{\label{fig2c}\includegraphics[width=0.136\textwidth]{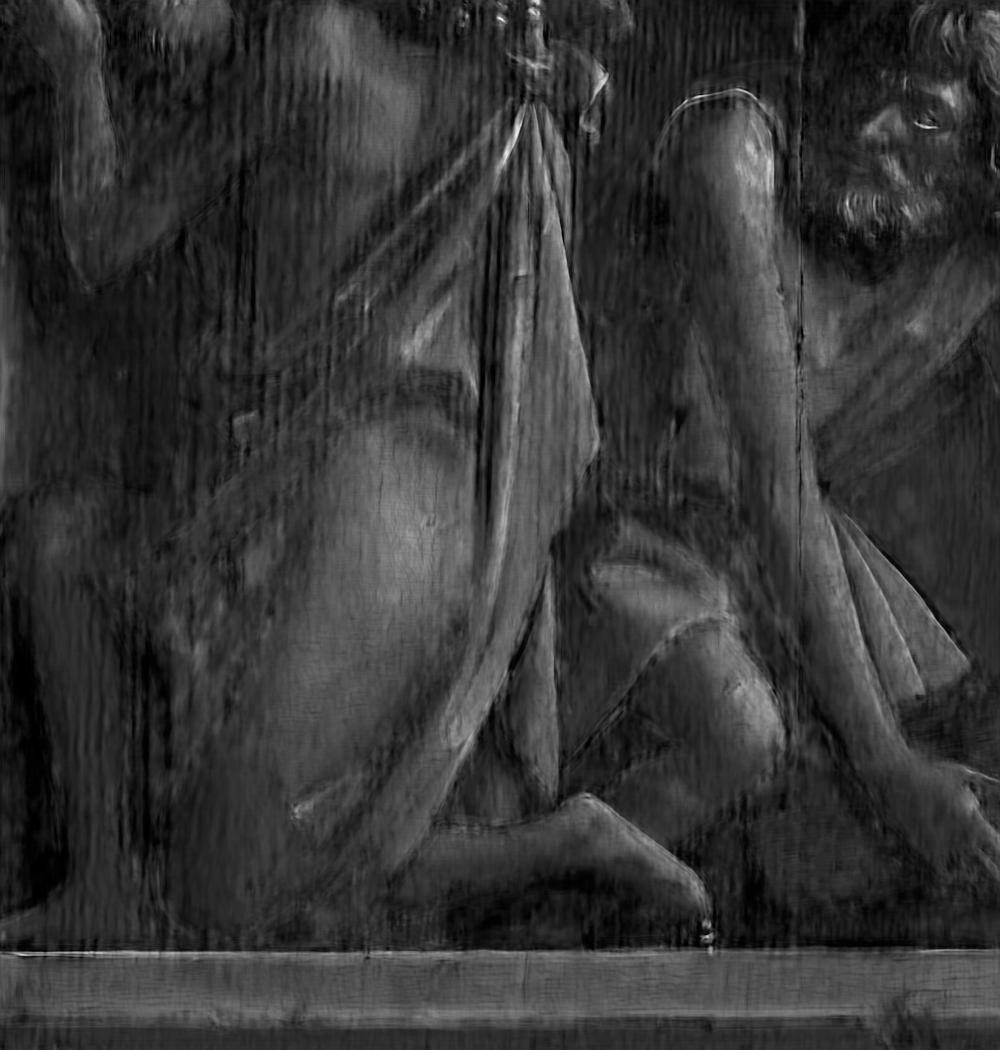}}
    \hfil
    \subfigure{\label{fig2c}\includegraphics[width=0.136\textwidth]{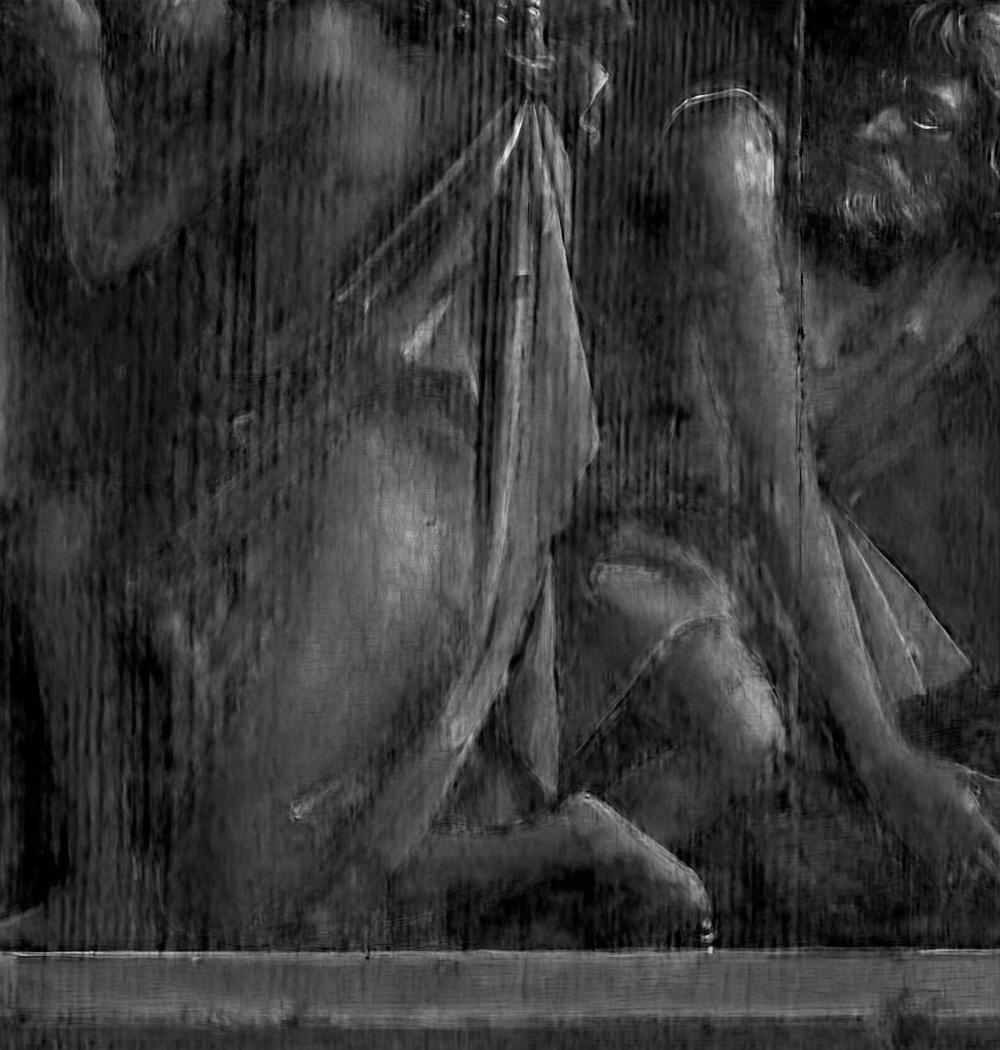}}    
    
    \subfigure{\label{fig2a}\includegraphics[width=0.136\textwidth]{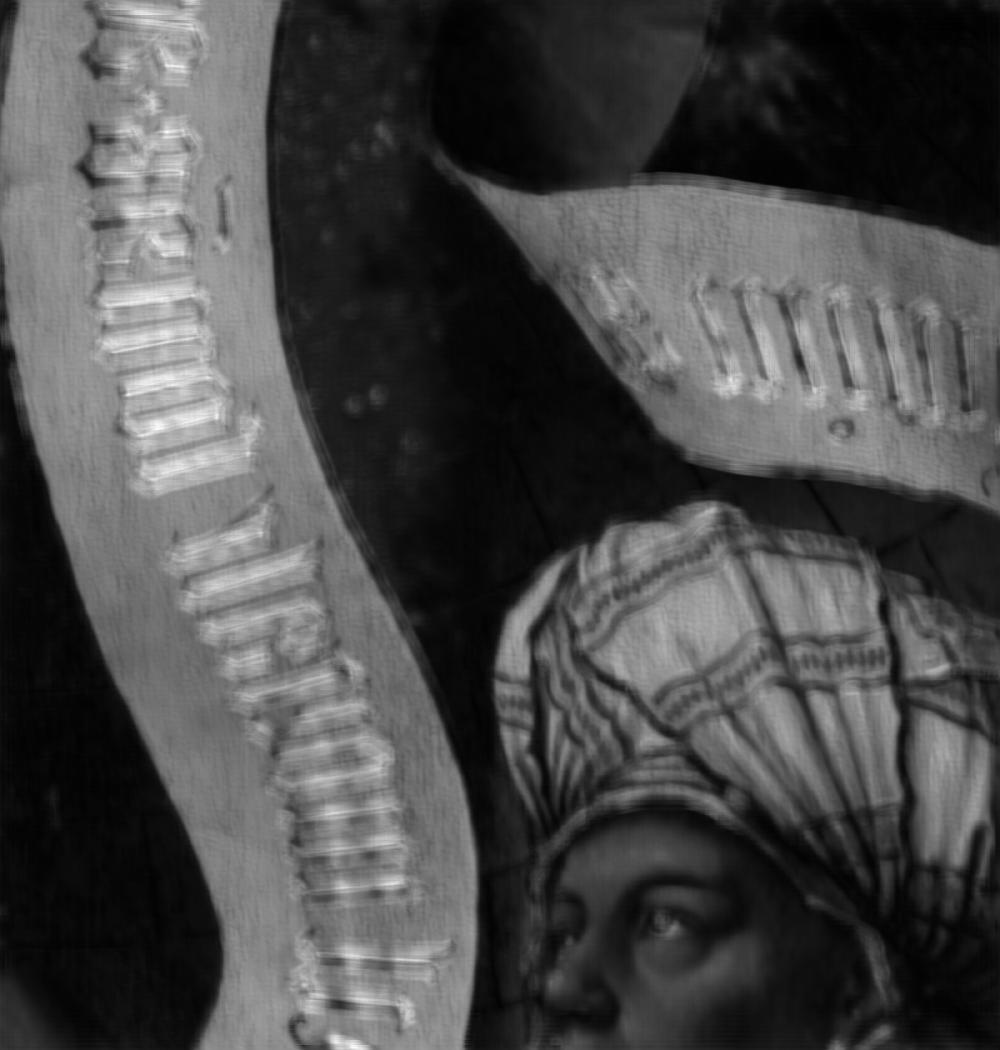}}
    \hfil
    \subfigure{\label{fig2b}\includegraphics[width=0.136\textwidth]{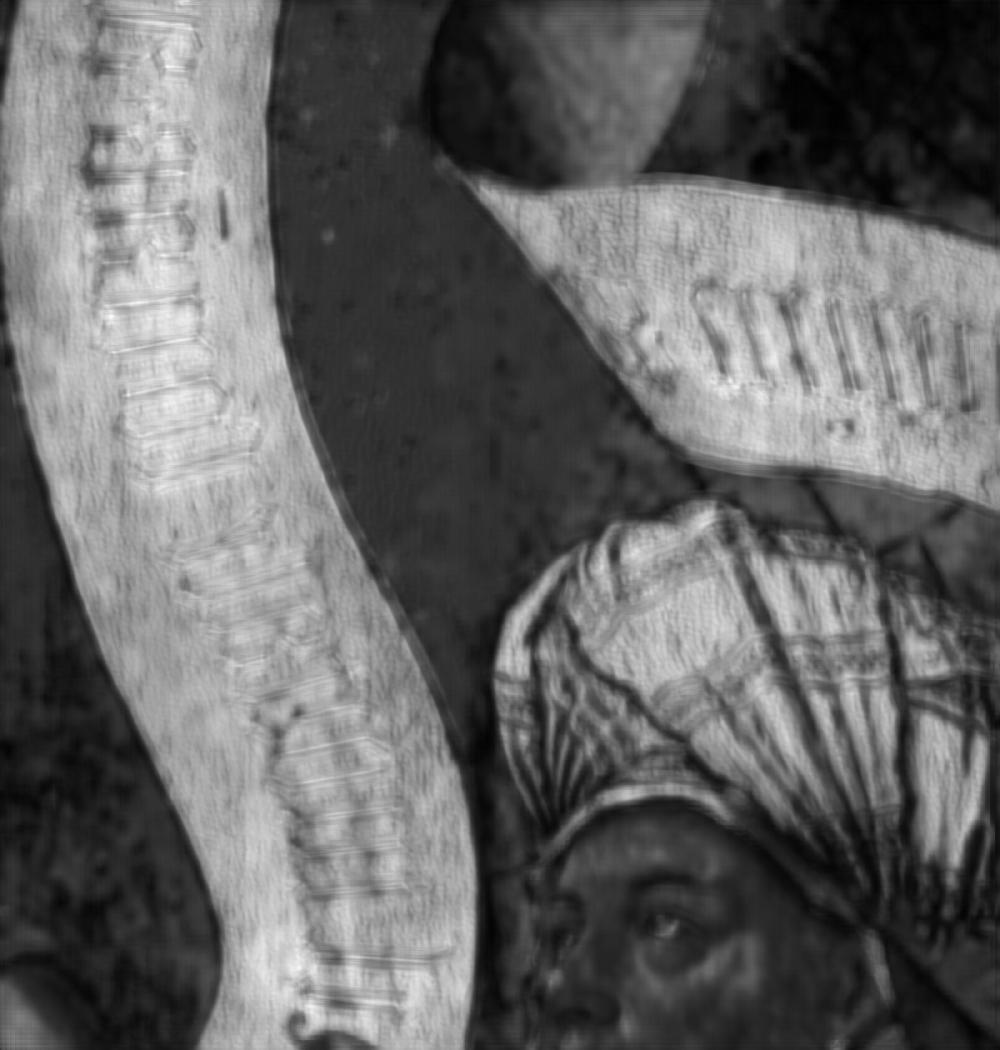}}
    \hfil
    \subfigure{\label{fig2c}\includegraphics[width=0.136\textwidth]{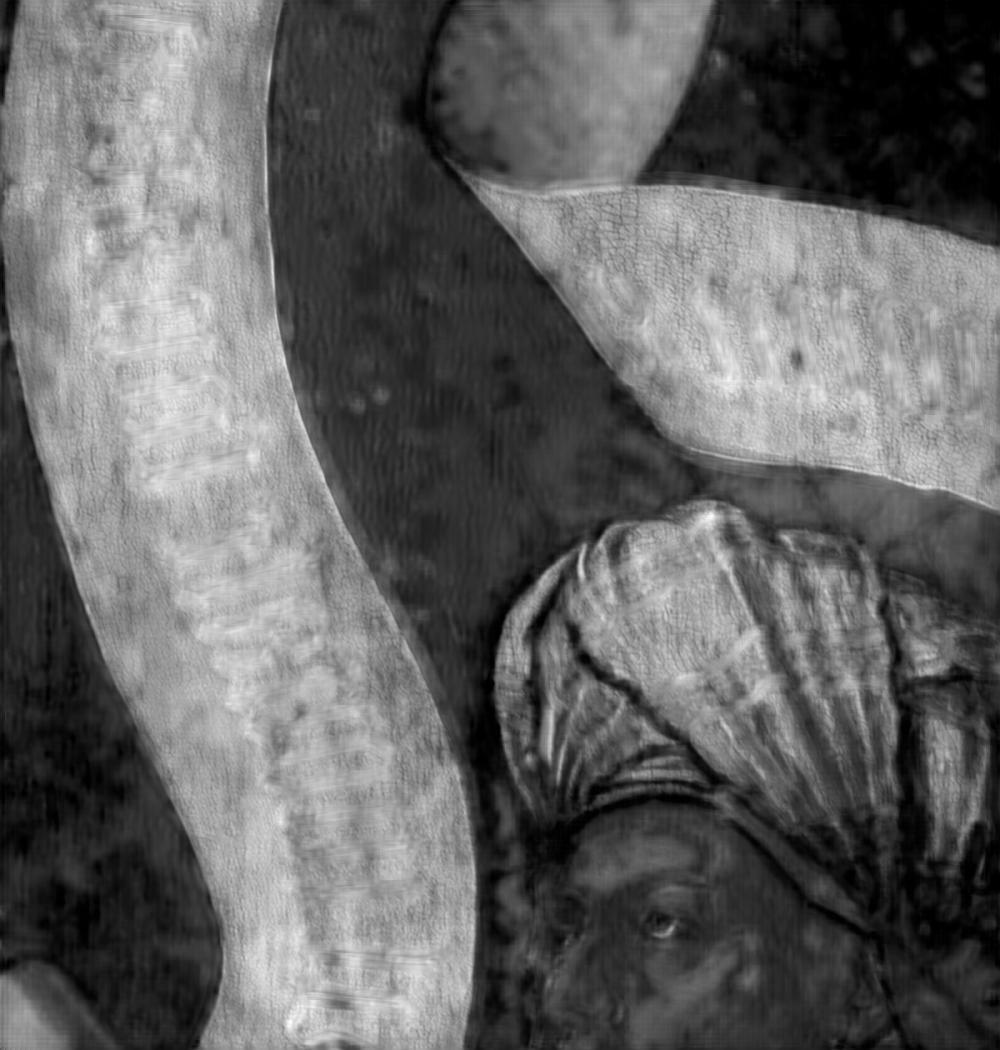}}
    \hfil
    \subfigure{\label{fig2a}\includegraphics[width=0.136\textwidth]{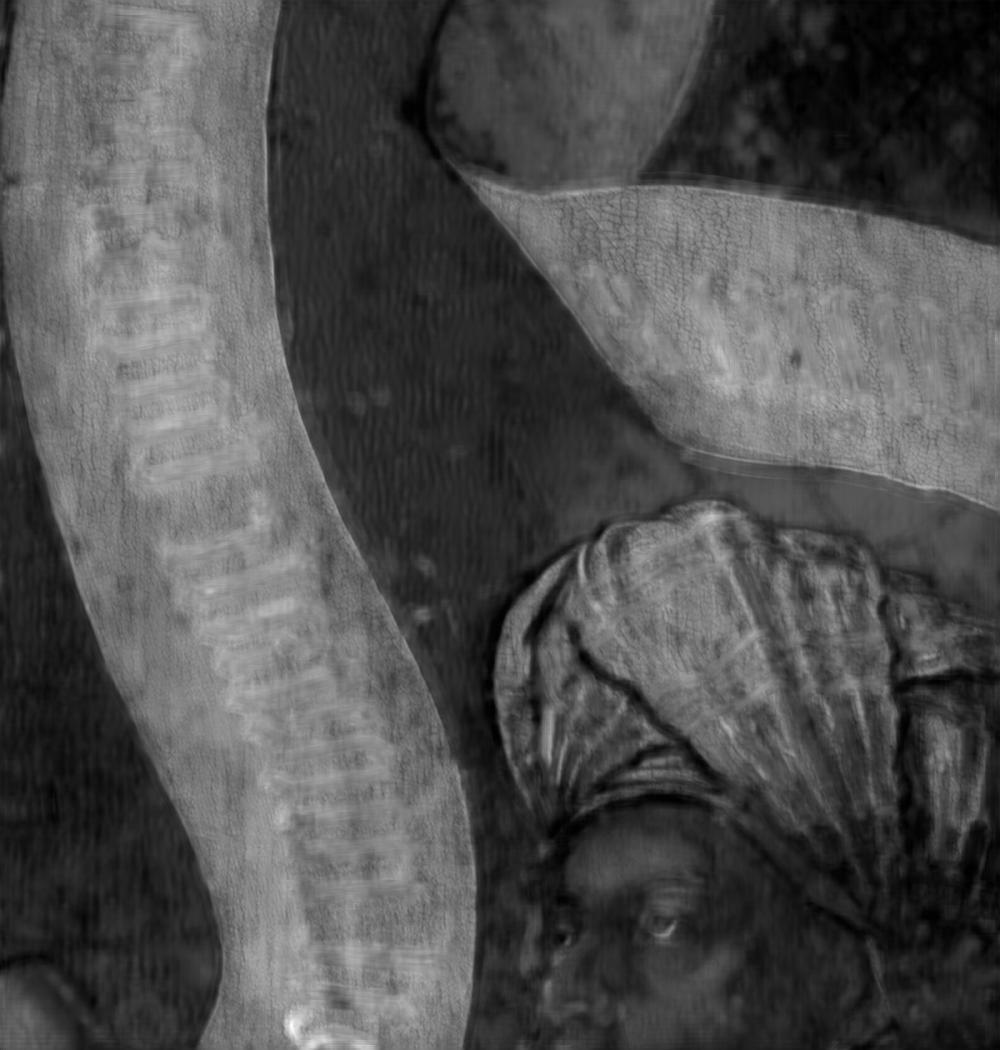}}
    \hfil
    \subfigure{\label{fig2b}\includegraphics[width=0.136\textwidth]{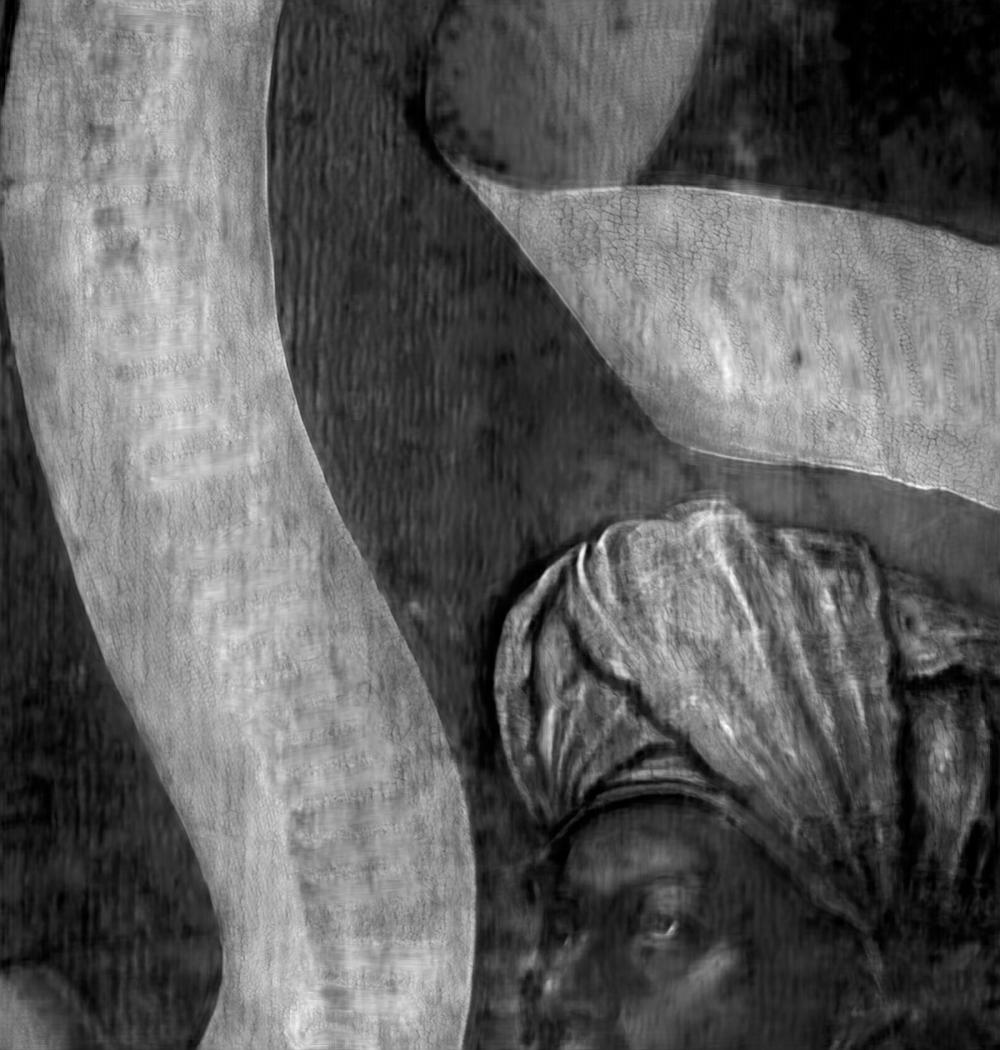}}
    \hfil
    \subfigure{\label{fig2c}\includegraphics[width=0.136\textwidth]{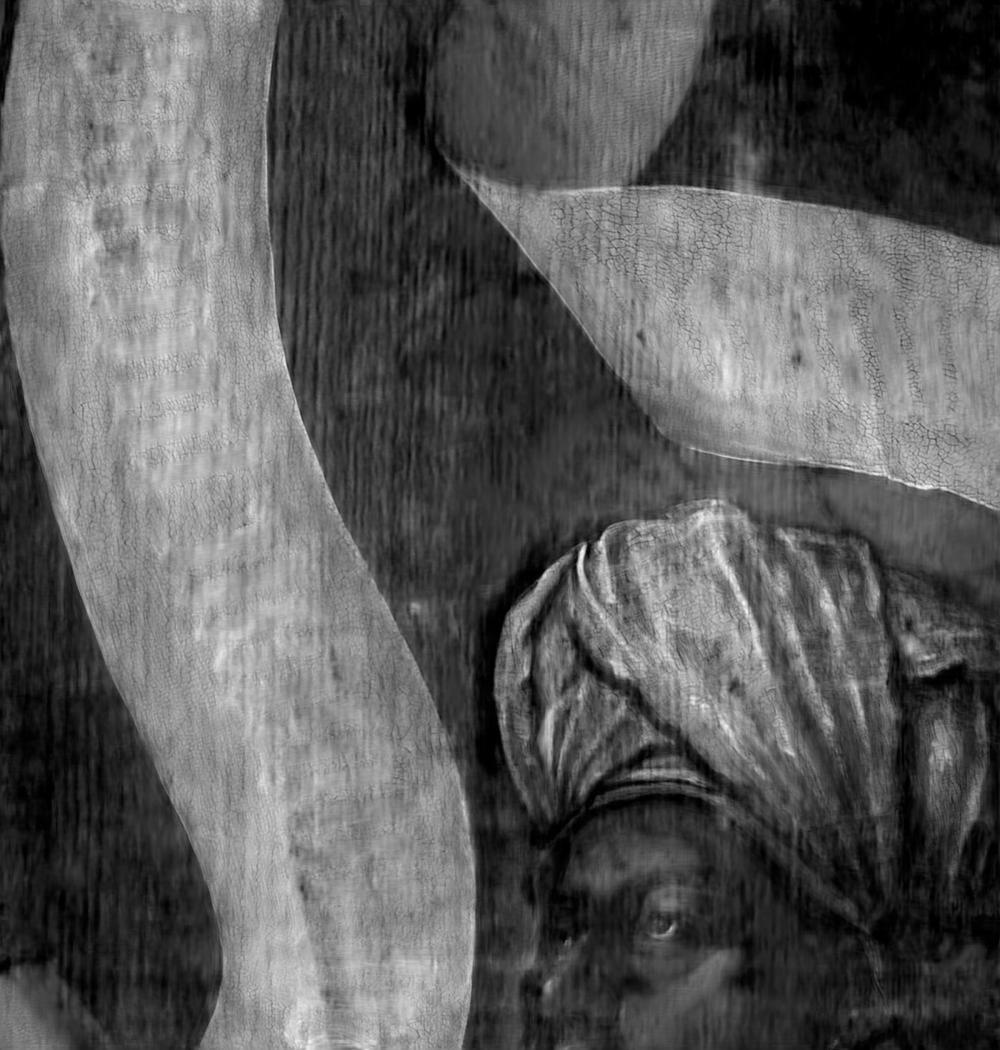}}
    \hfil
    \subfigure{\label{fig2c}\includegraphics[width=0.136\textwidth]{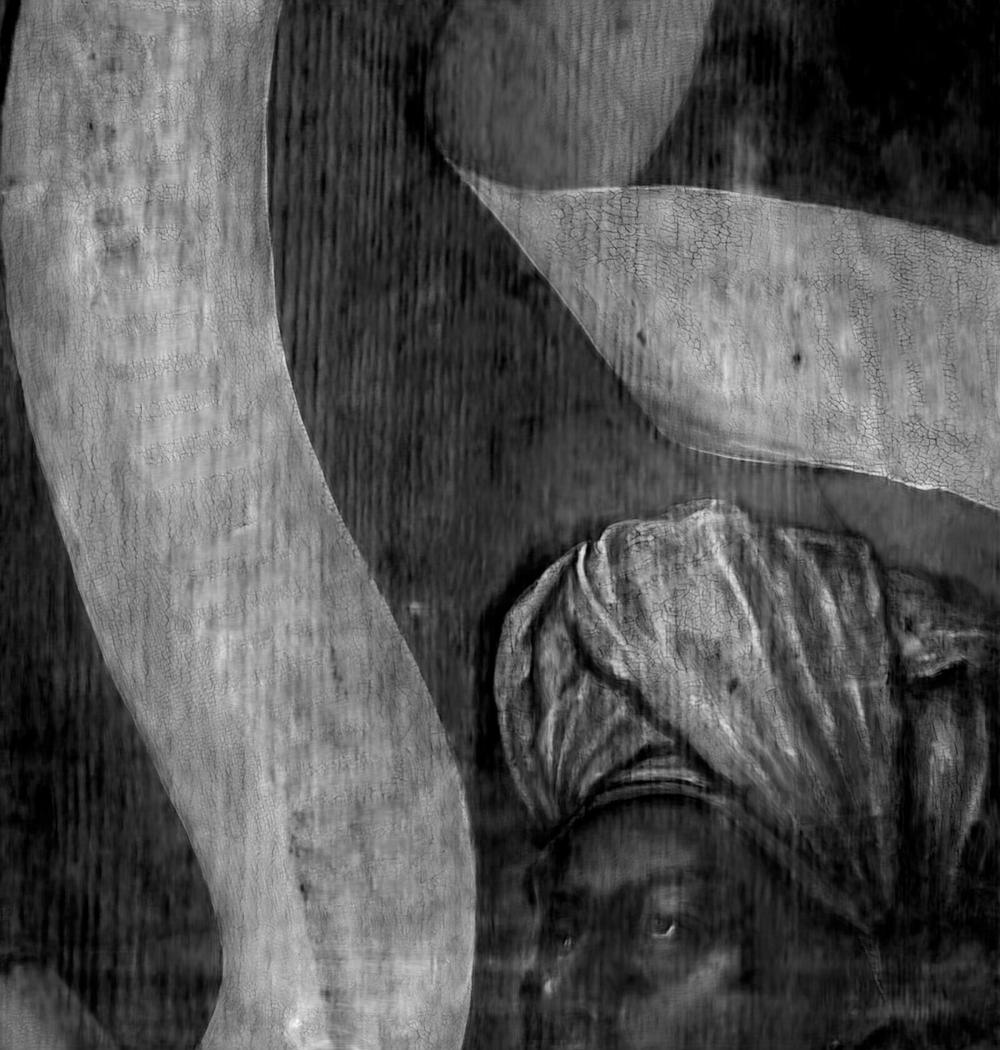}}   
    \caption{Reconstructed images vs. number of epochs on real data experiments. Columns 1 to 7 correspond to reconstructed result under 1st, 4th, 10th, 50th, 100th, 150th and 200th epoch, respectively. Rows 1 to 2 correspond to the reconstructed RGB images. Rows 3 to 4 correspond to the reconstructed X-ray images.}\label{figr3}
\end{figure*}

Fig. \ref{figr2} also depicts the evolution of the overall loss function along with the individual ones as a function of the number of epochs. The trends are similar to the previous ones: the $L_1$ component of the loss function is agressively minimized during the first 30 iterations, implying learning of $E_r$ and $D_r$; the $L_2$ and $L_3$ components of the loss function are minimized during the last 170 iterations, leading to learning of $D_x$; in parallel, the components $L_4$ and $L_5$ guide the convergence of the learning algorithm during the initial stages.

\begin{figure*}[h]
    \centering
    \subfigure{\label{fig2a}\includegraphics[width=0.19\textwidth]{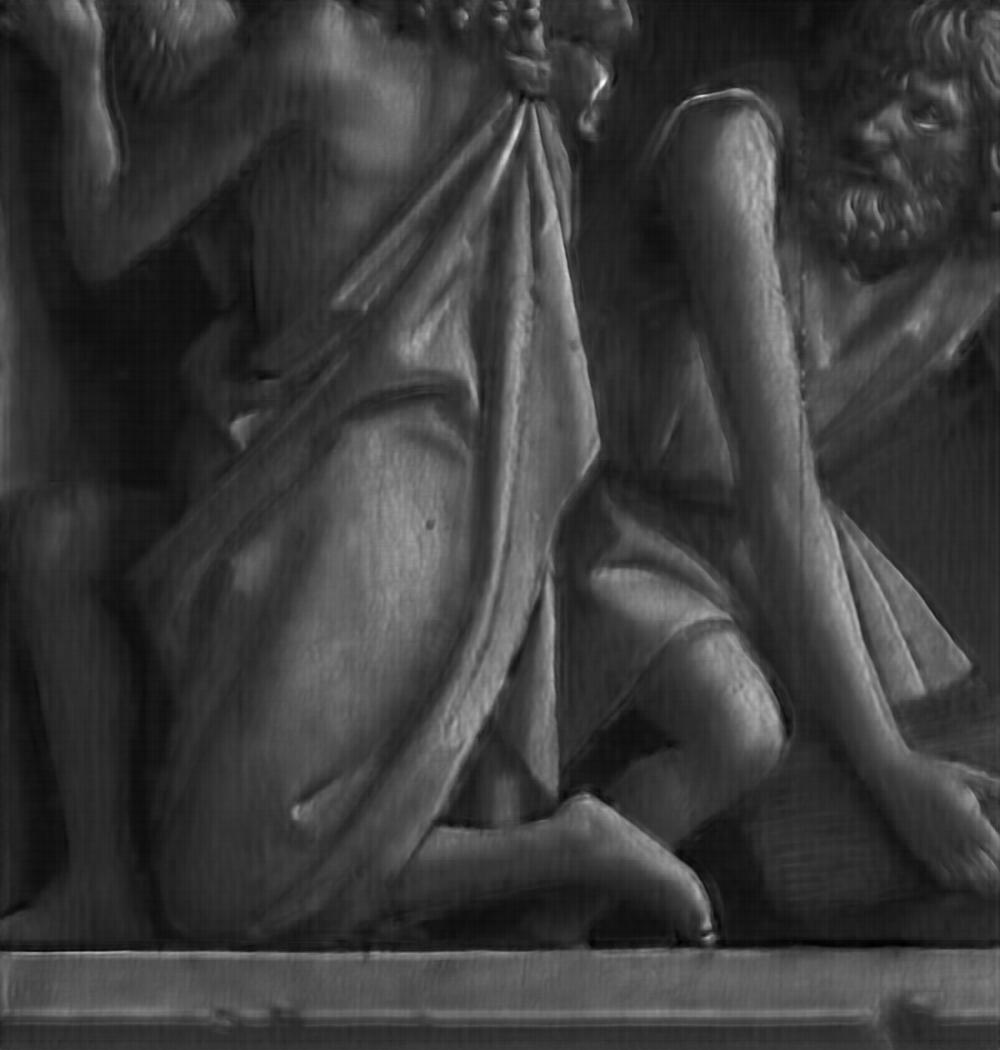}}
    \hfil
    \subfigure{\label{fig2b}\includegraphics[width=0.19\textwidth]{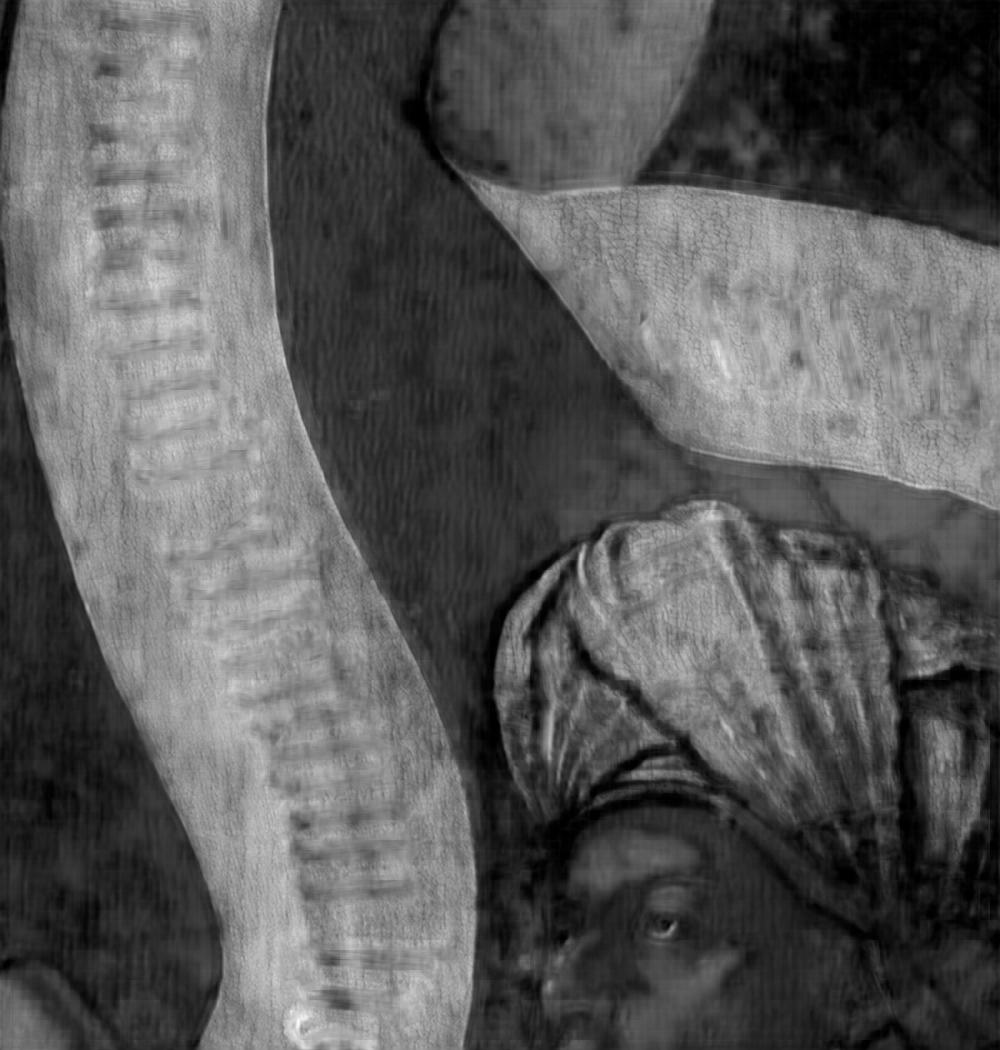}}
    \hfil
    \subfigure{\label{fig2c}\includegraphics[width=0.19\textwidth]{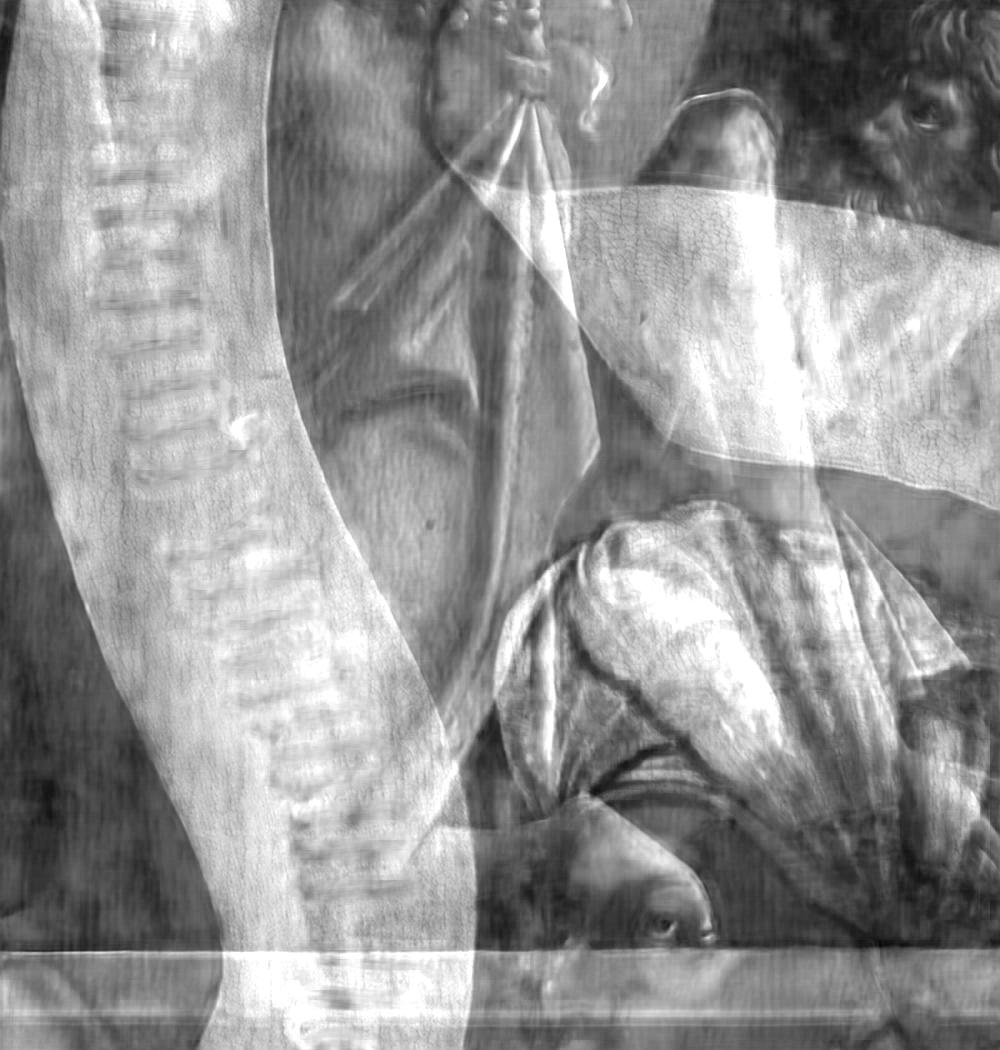}}
    \hfil
    \subfigure{\label{fig2a}\includegraphics[width=0.19\textwidth]{A2_xray_mix.jpg}}
    \hfil
    \subfigure{\label{fig2e}\includegraphics[width=0.19\textwidth]{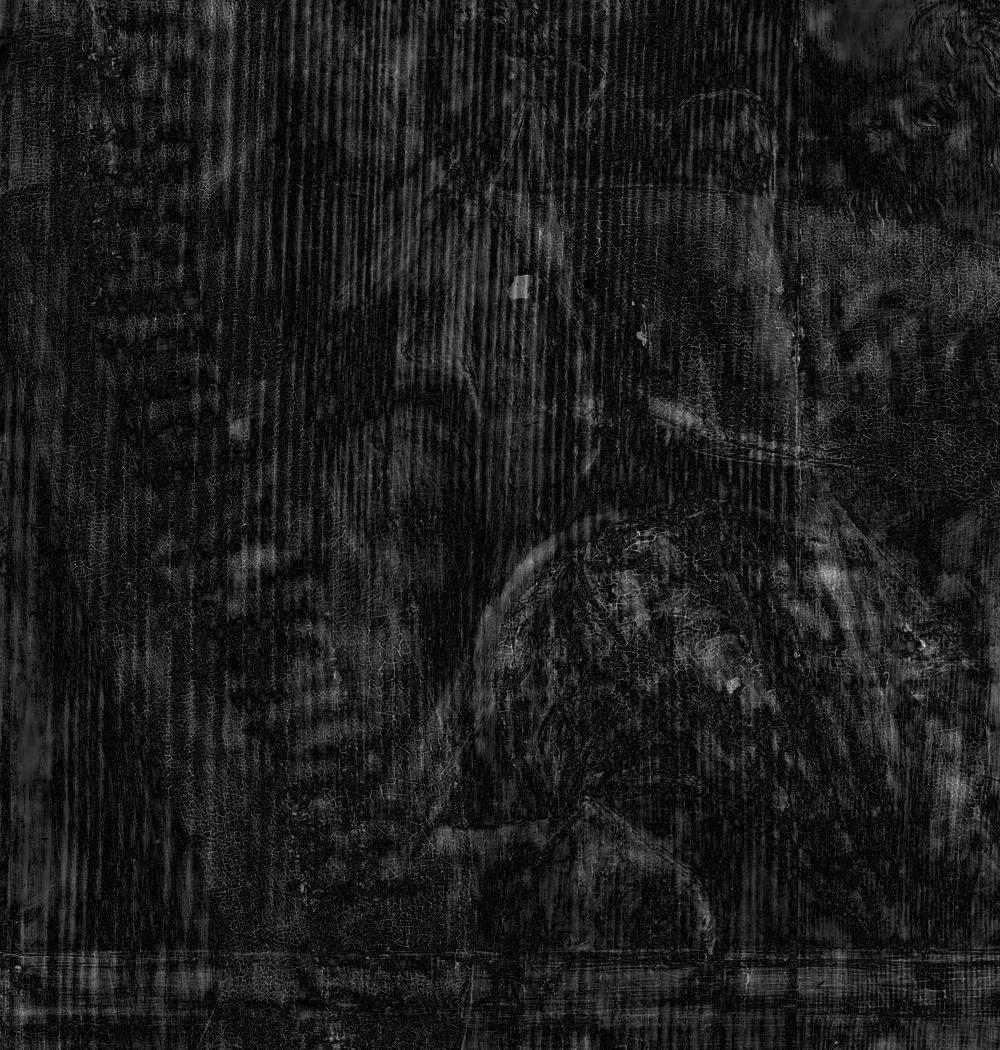}}
    
    \subfigure{\label{fig2a}\includegraphics[width=0.19\textwidth]{A2_xray1_P.jpg}}
    \hfil
    \subfigure{\label{fig2b}\includegraphics[width=0.19\textwidth]{A2_xray2_P.jpg}}
    \hfil
    \subfigure{\label{fig2c}\includegraphics[width=0.19\textwidth]{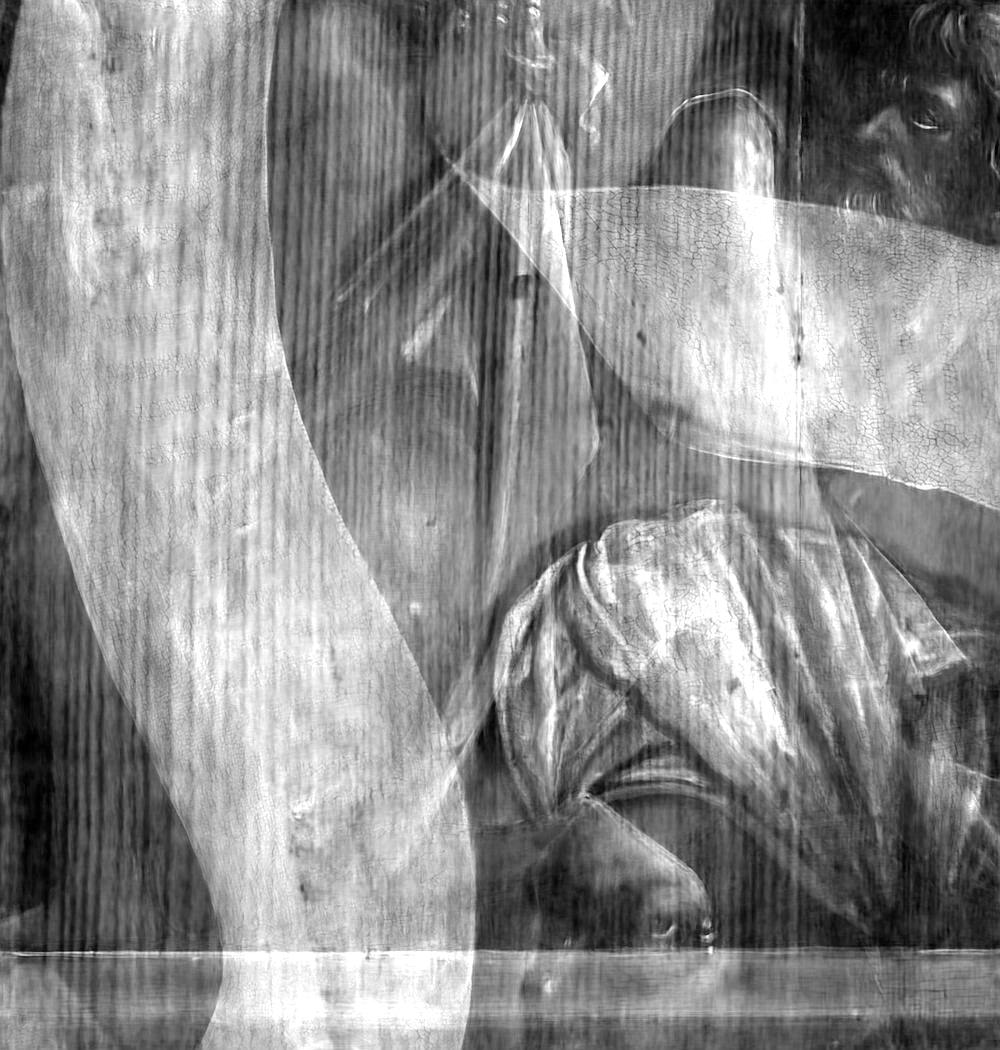}}
    \hfil
    \subfigure{\label{fig2a}\includegraphics[width=0.19\textwidth]{A2_xray_mix.jpg}}
    \hfil
    \subfigure{\label{fig2e}\includegraphics[width=0.19\textwidth]{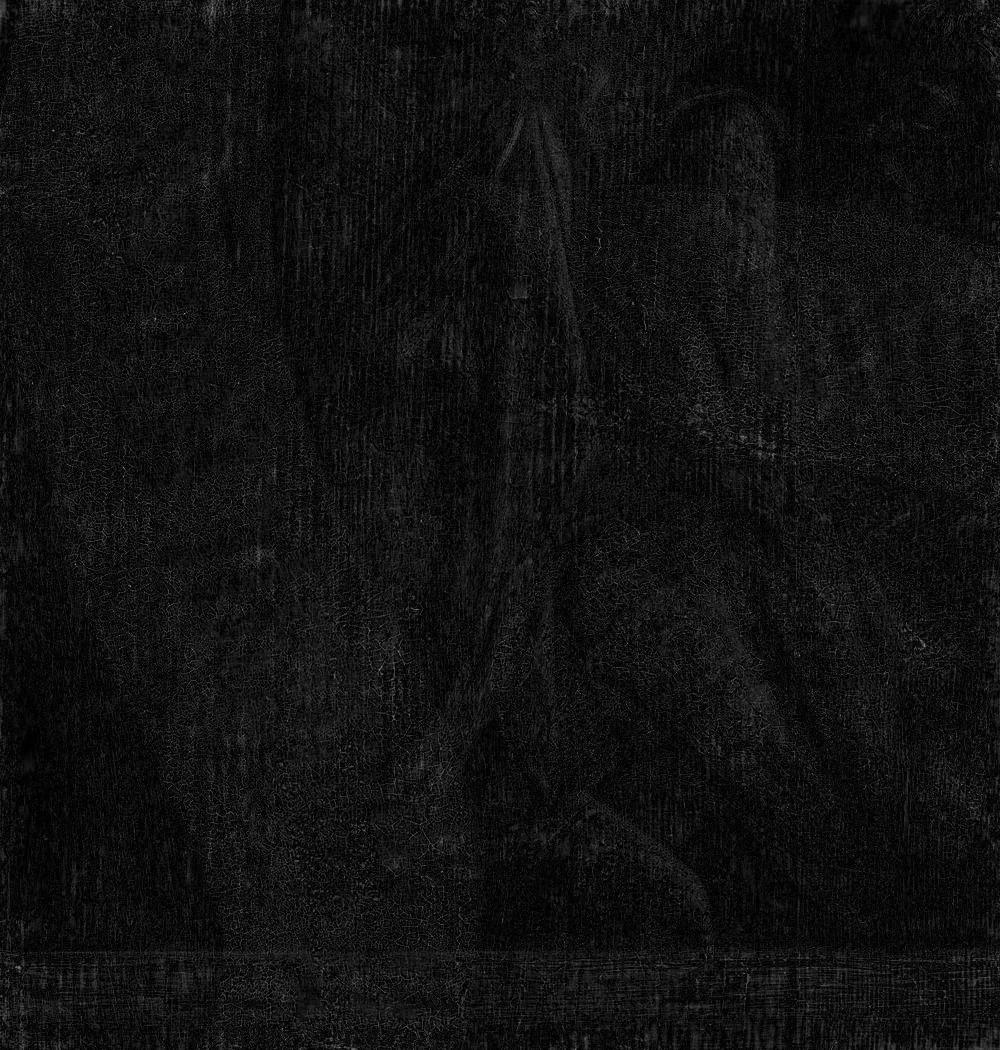}}
    \caption{Comparison of different X-ray separation methods. Rows 1 and 2 correspond to the results yielded by the method in \cite{IS2} and our proposed method, respectively. Columns 1 and 2 correspond to the reconstructed X-ray images, Columns 3 and 4 correspond to the reconstructed and true mixed X-ray images, and Columns 5 corresponds to the error maps.}\label{figr4}
\end{figure*}

Fig. \ref{figr3} depicts the evolution of the reconstruction of the various images as a function of the number of epochs. It is clear that the proposed algorithm has learnt how to reconstruct the RGB images by epoch 40; it is also clear that the algorithm only learns how to reconstruct the individual X-ray images and the mixed one well by epoch 100. Indeed, during the initial learning stages, the individual reconstructed X-ray images are very similar to grayscale versions of the RGB ones (e.g. see the inscription on the banner that is present in the RGB images but should not be present in the X-ray images). By contrast, during the last learning stages,

Finally, Fig. \ref{figr4} shows that the proposed algorithm produces much better separations than the algorithm in \cite{IS2}. Specifically,

\begin{itemize}

\item The proposed algorithm produces two individual X-ray images that recombine almost perfectly to match the original mixed X-ray image. By contrast, the algorithm in \cite{IS2} produces individual X-ray images that do not quite add up to the original mixed X-ray image. In fact, the MSE between the reconstructed mixed X-ray image – corresponding to the sum of the reconstructed individual X-ray images -- and the true mixed X-ray image is 0.0017 for our algorithm and 0.0050 for the algorithm in \cite{IS2}. \footnote{Note that -- in contrast to the synthetic data experiments --  it is not possible to compare the individual reconstructions to a ground truth.}

\item Our algorithm -- in contrast to that in \cite{IS2} -- also appears to reconstruct the pattern of cracking in the paint and the woodgrain of the panel better, including more fine detail such as the crack on the right hand side of the reconstructed X-ray image for the front side.

\item Our algorithm -- again in contrast to that in \cite{IS2} -- also  produces X-ray images without the addition of RGB image information not present in the true mixed X-ray image.

\end{itemize}

We also note that similar results, demonstrating the supe- riority of the proposed algorithm, have also been verified with other double-sided paintings.

\section{Conclusion}
X-ray images of polyptych wings, or other artworks painted on both sides of their support, contain in a single image content from both paintings, making them difficult for experts to interpret. It is therefore desirable to conceive approaches to separate the composite X-ray image into individual X-ray images containing content pertaining to one side only, in order to improve the utility of X-ray images in studying and conserving artworks.

This paper proposes a new approach to X-ray image separation that outperforms the state-of-the-art separation algorithms, as verified by side-by-side experiments on images from multiple paintings. The proposed architecture is a self-supervised learning approach based on the use of “connected” auto-encoders that extract features from available side information, i.e. RGB images of the front and back of the artwork in order to (1) reproduce both of the original RGB images, (2) reconstruct the two simulated X-ray images corresponding to each side, and (3) regenerate the mixed X-ray image. This approach allows image separation without the need for labelled data. A composite loss function is introduced to improve image separation and offset convergence issues of the learning algorithm that can otherwise occur, with the selected values of the associated hyper-parameters varying little between datasets.

The proposed method is robust, having successfully separated both real X-ray image data from a double-sided panel of the \textsl{Ghent Altarpiece}, as well as synthetically-mixed X-ray images from canvas paintings, into two images that maintain fidelity to the mixed X-ray image while more clearly showing features from each side of the painting. Importantly, the results from this image separation also maintain features of the support, such as wood grain and canvas weave, that are not readily apparent in the RGB images used as side information. In the future, it will be important to further assess this image separation approach on more challenging sets of X-ray imaging data, such as those with large structural components or other features of interest not apparent at the surface of the painting.

\bibliographystyle{IEEETran}

\end{document}